\documentclass[11pt,twoside]{article}
\pdfoutput=1

\usepackage{pslatex}    
\usepackage{amsfonts}
\usepackage{amsmath}
\usepackage{amsthm}
\usepackage{amscd}
\usepackage{cite}
\usepackage{eucal}
\usepackage{graphicx}
\usepackage{a4}
\usepackage{booktabs}
\usepackage{microtype}
\usepackage{appendix}
\usepackage{float}
\usepackage{color}

\let\a=\alpha \let\be=\beta \let\g=\gamma \let\de=\delta
\let\e=\varepsilon   \let\th=\theta
\let\eps=\epsilon
\let\dh=\vartheta \let\k=\kappa \let\la=\lambda 
 \let\x=\xi \let\p=\pi \let\r=\rho \let\s=\sigma
 
\let\ph=\varphi   \let\Ps=\Psi
  \let\Th=\Theta
\let\La=\Lambda \let\G=\Gamma \let\D=\Delta

\let\qd=\quad \let\qqd=\qquad 

\def\epp{\, .}
\def\epc{\, ,}

\def\tst#1{{\textstyle #1}}

\theoremstyle{plain}

\newtheorem{lemma}{Lemma}

\newtheorem*{corollary*}{Corollary}
\newtheorem{conjecture}{Conjecture}
\newtheorem*{conjecture*}{Conjecture}

\theoremstyle{definition}

\newtheorem{Remark}{Remark}

\def\2{\frac{1}{2}} \def\4{\frac{1}{4}}

\def\6{\partial}

\def\+{\dagger}

\def\<{\langle} \def\>{\rangle}

\def\CO{{\cal O}}

\def\i{{\rm i}}

\def\rd{{\rm d}}
\def\re{{\rm e}}

\def\ctg{\, {\rm ctg}\,}

\DeclareMathOperator{\Ln}{Ln}
\DeclareMathOperator{\sh}{sh}
\DeclareMathOperator{\ch}{ch}

\DeclareMathOperator{\Tr}{Tr}
\DeclareMathOperator{\one}{\bf 1}

\DeclareMathOperator{\sn}{sn}
\DeclareMathOperator{\cn}{cn}
\DeclareMathOperator{\dn}{dn}

\DeclareMathOperator{\End}{End}

\DeclareMathOperator{\PV}{PV}

\def\Re{{\rm Re\,}} \def\Im{{\rm Im\,}}

\def\fa{\mathfrak{a}}

\renewcommand{\appendix}{%
   \renewcommand{\section}{
        \secdef\Appendix\sAppendix}%
   \setcounter{section}{0}%
   \renewcommand{\thesection}{\Alph{section}}%
   \renewcommand{\theequation}{\thesection.\arabic{equation}}%
}

\newcommand{\Appendix}[2][?]{%
     \refstepcounter{section}%
     \setcounter{equation}{0}%
     \addcontentsline{toc}{appendix}%
          {\protect\numberline{\appendixname~\thesection} #1}%
     \vspace{\baselineskip}%
     {\noindent\large\bfseries\appendixname\ \thesection: #2\par}%
     \sectionmark{#1}\vspace{\baselineskip}}

\newcommand{\sAppendix}[1]{%
     {\noindent\large\bfseries\appendixname\:: #1\par}%
     \sectionmark{#1}\vspace{\baselineskip}}

\begin{document}

\thispagestyle{empty}

\begin{center}

{\Large {\bf Low-temperature spectrum of correlation lengths of the
XXZ chain in the antiferromagnetic massive regime
\\}}

\vspace{7mm}

{\large
Maxime Dugave,\footnote{e-mail: dugave@uni-wuppertal.de}
Frank G\"{o}hmann\footnote{e-mail: goehmann@uni-wuppertal.de}}%
\\[1ex]
Fachbereich C -- Physik, Bergische Universit\"at Wuppertal,\\
42097 Wuppertal, Germany\\[2.5ex]
{\large Karol K. Kozlowski\footnote{e-mail: karol.kozlowski@u-bourgogne.fr}}%
\\[1ex]
IMB, UMR 5584 du CNRS,
Universit\'e de Bourgogne, France\\[2.5ex]
{\large Junji Suzuki\footnote{e-mail: sjsuzuk@ipc.shizuoka.ac.jp}}%
\\[1ex]
Department of Physics, Faculty of Science, Shizuoka University,\\
Ohya 836, Suruga, Shizuoka, Japan

\vspace{15mm}
{\it Dedicated to Professor R.~J.~ Baxter on the occasion of his 75th birthday}
\vspace{15mm}

{\large {\bf Abstract}}

\end{center}

\begin{list}{}{\addtolength{\rightmargin}{9mm}
               \addtolength{\topsep}{-5mm}}
\item
We consider the spectrum of correlation lengths of the spin-$\2$
XXZ chain in the antiferromagnetic massive regime. These are
given as ratios of eigenvalues of the quantum transfer matrix of
the model. The eigenvalues are determined by integrals over
certain auxiliary functions and by their zeros. The auxiliary
functions satisfy nonlinear integral equations. We analyse these
nonlinear integral equations in the low-temperature limit.
In this limit we can determine the auxiliary functions and 
the expressions for the eigenvalues as functions of a finite
number of parameters which satisfy finite sets of algebraic
equations, the so-called higher-level Bethe Ansatz equations.
The behaviour of these equations, if we send the temperature
$T$ to zero, is different for zero and non-zero magnetic
field $h$. If $h$ is zero the situation is much like in the
case of the usual transfer matrix. Non-trivial higher-level
Bethe Ansatz equations remain which determine certain
complex excitation parameters as functions of hole parameters
which are free on a line segment in the complex plane. If $h$ is
non-zero, on the other hand, a remarkable restructuring occurs,
and all parameters which enter the description of the quantum
transfer matrix eigenvalues can be interpreted entirely in
terms of particles and holes which are freely located on two
curves when $T$ goes to zero.
\\[2ex]
\end{list}

\clearpage

\section{Introduction}
The quantum transfer matrix formalism \cite{Suzuki85,SuIn87}
provides a framework for calculating the thermodynamic
properties \cite{SAW90,Kluemper93} and correlation functions
\cite{GKS04a,GKS05,DGK13a,DGK14a} of integrable lattice
models analytically. It enables, in particular, the calculation
of correlation lengths of integrable Heisenberg chains
\cite{Kluemper92,Kluemper93,Takahashi91,KMSS01,KlSc03}
and related Fermion models \cite{SSSU99}.

The main concern of this work is the calculation of the
full spectrum of correlation lengths of the XXZ chain in
the antiferromagnetic massive regime at finite magnetic
field $h$ and low temperature $T$, \emph{i.e.} for large
ratios $h/T$. The above cited previous works dealt with
the massless regime or with the case that $h/T$ is small
and were mainly restricted to the calculation of a few
largest correlation lengths. Our study is the first
step in the analysis of the low-temperature behaviour
of two-point correlation functions, especially of their
large-distance asymptotics, by means of a form factor
approach, as, in fact, a form factor expansion requires
the summation over a complete set of intermediate states.

\begin{figure}
\begin{center}
\includegraphics[width=.6\textwidth,angle=0,clip=true]{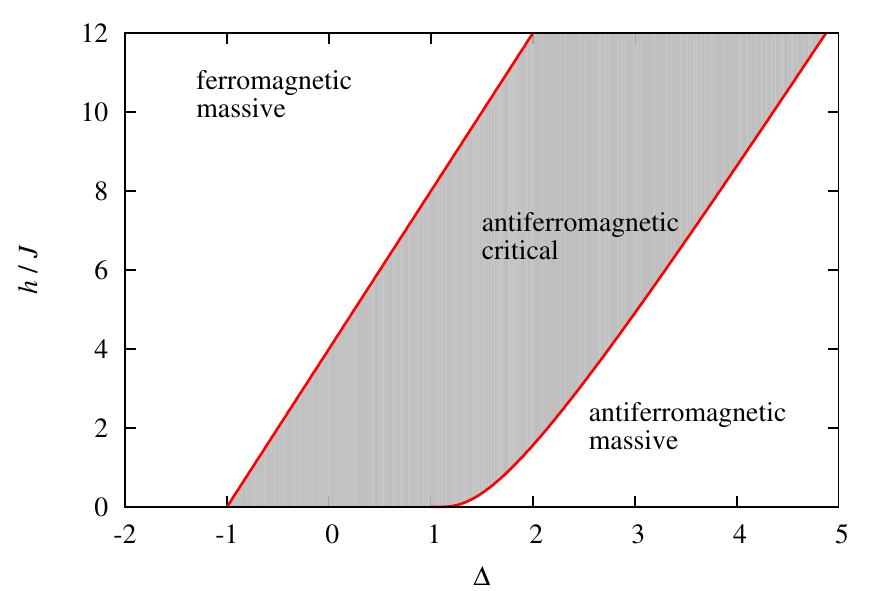}
\caption{\label{fig:phasediagram} The ground state phase diagram of
the XXZ chain in the $\D$-$h$ plane.
}
\end{center}
\end{figure}

A form-factor based analysis of the large-distance asymptotics
of two-point functions at low temperatures was recently completed
for the model in the massless (or critical) regime \cite{DGK13a,%
DGK14a}.\footnote{In fact, the analysis carried out in
\cite{DGK13a,DGK14a} is restricted to the massless regime
at $|\D| < 1$. We have obtained similar results for
$\D > 1$ and the magnetic field between lower and upper
critical field (see Figure~\ref{fig:phasediagram}) which
we hope to publish elsewhere.} In that work, as well as
in the previous analysis of ground-state correlation
functions within a form-factor approach \cite{KKMST11b,%
KKMST12,KKMST11a}, a finite magnetic field turned out
to be an important regularization parameter. As long as
the magnetic field is finite, the low-energy excitations
above the ground state can be classified as particle-hole
excitations about two Fermi points, and a similarly simple
picture holds for the `excitations' of the quantum
transfer matrix as well. When sending first $T$ and then
$h$ to zero we found numerical agreement of our formulae
for the correlation amplitudes \cite{DGK14a} with the
explicit formulae obtained by Lukyanov \cite{Lukyanov99}
for the ground state at vanishing magnetic field.

It is our hope that a similar program can be carried
out in the massive regime and that we will be able to
obtain even more explicit results in terms of certain
special functions. This can be expected, since there
are no Fermi points in the massive regime and since
the functions that enter naturally into the description
of correlation functions are periodic or quasi-periodic
with period or quasi-period $\p$. In fact, in some
cases the multiple-integral formulae for the form
factors of the XXZ chain in the antiferromagnetic massive
regime at zero temperature and magnetic field, that
were obtained within the vertex-operator approach \cite{JiMi95,%
Lashkevich02,LuTe03}, could be evaluated in closed form.

More recently we considered form factors of the usual
transfer matrix in the antiferromagnetic massive
regime from a Bethe Ansatz perspective \cite{DGKS14app}.
In this case the ground state magnetization is zero,
and the excitations are characterized by Bethe root
patterns that involve non-real roots, organized in
so-called wide pairs and two-strings satisfying 
a set of `higher-level' Bethe Ansatz equations
\cite{Woynarovich82c,ViWo84,BVV83}. These
remain non-trivial even in the thermodynamic limit.
The appearance of the higher-level Bethe Ansatz equations
as well as the singular behaviour in the thermodynamic
limit of norms of Bethe states involving two-strings
made the analysis rather delicate. In the end we
obtained novel formulae for the form factors in
the thermodynamic limit for which we found numerical
agreement with rather differently looking formulae
obtained within the vertex operator approach
\cite{Lashkevich02,LuTe03}.

As we shall see below, in the limit $T \rightarrow 0_+$
the Bethe roots which determine the spectrum of the
quantum transfer matrix at vanishing magnetic field
satisfy a set of higher-level Bethe Ansatz equations
of the same form as in case of the usual transfer matrix.
If the magnetic field is non-zero however, we observe
a dramatic reorganization of the Bethe root patterns
at low temperatures. Like in the massless case it turns
out to be possible to interpret them entirely in terms of
particle-hole excitations. At small finite temperatures the
positions of the particle and hole parameters in the
complex plane are still determined by a set of higher-level
Bethe Ansatz equations. But as the temperature goes to
zero they become free parameters on two branches of a
curve in the complex plane. The eigenvalue ratios, which
determine the correlation lengths, are explicit functions
of these parameters. This is the main result of this work.
It will lead to yet another form factor series for the
ground state two-point functions of the XXZ chain in
the thermodynamic limit. We shall report it in a separate
publication.

The paper is organized as follows. In the remaining part
of this introduction we recall the Hamiltonian and its
ground state phase diagram. We also recall the expansion
of thermal correlation functions in terms of eigenstates
and eigenvalues of the quantum transfer matrix. The main
subject of this work, which is the calculation of the spectrum
of correlation lengths of the XXZ chain in the massive
antiferromagnetic regime at finite magnetic field for
low temperatures, is then explored in section~\ref{sec:main}
and followed by conclusions in section~\ref{sec:conclusions}.
Appendix~\ref{app:functions} contains a summary of the
properties of the basic functions that determine the
low-temperature behaviour of the correlation lengths,
namely, the dressed momentum, dressed energy and dressed
phase which, in the massive antiferromagnetic regime, can
be explicitly expressed in terms of elliptic functions
and $q$-Gamma functions. Appendix~\ref{app:numerics}
contains important complementary material on the numerical
study of the Bethe Ansatz equations and of the eigenvalues
of the quantum transfer matrix at finite Trotter number.

\subsection{Hamiltonian and ground state phase diagram}
The Hamiltonian of the spin-$\2$ XXZ chain in a homogeneous
magnetic field parallel to the direction of its anisotropy
axis may be written as
\begin{equation} \label{ham}
     H = J \sum_{j = - L + 1}^L \Bigl( \s_{j-1}^x \s_j^x + \s_{j-1}^y \s_j^y
                               + \D \bigl( \s_{j-1}^z \s_j^z - 1 \bigr) \Bigr)
         - \frac{h}{2} \sum_{j = - L + 1}^L \s_j^z \epc
\end{equation}
where the $\s_j^\a$ are Pauli matrices $\s^\a$ acting on the $j$th factor of
the tensor-product space of states ${\cal H} = {\mathbb C}^{\otimes 2L}$
of $2L$ spins $\2$. We choose the number of spins in the chain to be even,
since this implies that the ground state of the Hamiltonian (\ref{ham})
is unique for $h = 0$, $\D > - 1$ and every $L \in 2 {\mathbb N}$. Below we
will consider the system in the thermodynamic limit $L \rightarrow \infty$
and at finite temperature $T$. The Hamiltonian (\ref{ham}) depends on
three parameters, the strength of the exchange interaction, $J > 0$,
which sets the energy scale, the strength of the magnetic field,
$h \ge 0$, and the anisotropy parameter~$\D$.

The ground state phase diagram in the $\D$-$h$ parameter plane was
obtained by Yang and Yang \cite{YaYa66d} (see Figure~\ref{fig:phasediagram}).
For $\D > 1$ the zero-temperature magnetization vanishes below a
lower critical field $h_\ell$ (right red line in Figure~%
\ref{fig:phasediagram}). This region of the $\D$-$h$ parameter
plane is called the antiferromagnetic massive regime. It contains
the Ising chain ($h = 0$, $\D = + \infty$) as a particular point.
The physical properties of the model in the antiferromagnetic
massive regime are believed to be approximately accessible by
perturbation theory around the Ising limit ($\D \rightarrow + \infty$).
The ground state in this regime becomes two-fold degenerate in
the thermodynamic limit, and the lowest excited states are
separated from the degenerate ground states by a finite energy
gap (or `mass gap'). The two degenerate ground states in the
thermodynamic limit may be thought of as `finite-$\D$ deformations'
of the two N\'eel states which are the ground states of the Ising chain.

\subsection{Correlation functions and correlation lengths by Bethe Ansatz}
In this work we study the model in the antiferromagnetic massive
regime non-perturbatively, using its integrable structure encoded in
the $R$-matrix
\begin{equation} \label{rmatrix}
     \begin{array}{cc}
     R(x,y) = \begin{pmatrix}
                  1 & 0 & 0 & 0 \\
		  0 & b(x,y) & c(x,y) & 0 \\
		  0 & c(x,y) & b(x,y) & 0 \\
		  0 & 0 & 0 & 1
		 \end{pmatrix} \epc &
     \begin{array}{c}
     b(x, y) = \frac{\sin(y - x)}{\sin(y - x + \i \g)} \\[2ex]
     c(x, y) = \frac{\sin(\i \g)}{\sin(y - x + \i \g)}
    \end{array}
    \end{array} \epc
\end{equation}
of the six-vertex model. As is well known \cite{Baxter72b} the Hamiltonian
(\ref{ham}) at $h = 0$ is proportional to the logarithmic derivative of
the transfer matrix of the homogeneous six-vertex model. Its anisotropy
parameter $\D$ is then a function of the deformation parameter $\g$
of the $R$-matrix, $\D = \ch (\g)$. In the antiferromagnetic massive
regime $\g$ must be real. In the following we restrict ourselves to
$\g > 0$. Later on we shall also use the common notation $q = \re^{- \g}$.

For the calculation of temperature correlation functions and their
correlation lengths we need the statistical operator $\re^{- H/T}$
rather than the Hamiltonian. This operator can be related to the
monodromy matrix of an inhomogeneous auxiliary six-vertex model
defined for every lattice site $j \in \{- L + 1, \dots, L\}$ by
\begin{equation}
     T_j (x ) =
        q^{\k \s_j^z}
        R_{j \overline{N}} \bigl(x, \tst{\frac{\i \be}{N}} \bigr)
	R_{\overline{N-1} j}^{t_1}
	   \bigl(- \tst{\frac{\i \be}{N}}, x \bigr) \dots
        R_{j \overline{2}} \bigl(x, \tst{\frac{\i \be}{N}} \bigr)
	R_{\bar 1 j}^{t_1} \bigl(- \tst{\frac{\i \be}{N}}, x \bigr)
	\epp
\end{equation}
Here $N \in 2 {\mathbb N}$ is called the Trotter number, and the indices
$\bar j = \bar 1, \dots, \overline N$ refer to $N$ auxiliary sites
in `Trotter direction'. Furthermore
\begin{equation}
     \be = - \frac{2 J \sh(\g)}{T} \epc \qqd \k = - \frac{h}{2 \g T}
\end{equation}
are rescaled inverse temperature and magnetic field.

Define
\begin{equation}
     \r_{N, L} = \Tr_{\bar 1 \dots \overline N} \{T_{- L + 1} (0) \dots T_L (0)\} \epp
\end{equation}
Then it is easy to see \cite{GKS04a} that
\begin{equation}
     \re^{- H/T} = \lim_{N \rightarrow \infty} \r_{N, L} \epp
\end{equation}
We call $\r_{N, L}$ a finite Trotter number approximant of the statistical
operator. Using $\r_{N, L}$ we obtain simple expressions for the finite
Trotter number approximants of correlation functions. Namely, for any
product of local operators ${\cal O}^{(j)} \in \End {\mathbb C}^2$,
$j = 1, \dots, m$, acting on $m$ consecutive sites
\begin{multline} \label{opexpapr}
     \bigl\< \CO_1^{(1)} \dots \CO_m^{(m)} \bigr\>_N
        = \lim_{L \rightarrow \infty}
	  \frac{\Tr_{- L + 1 \dots L} \bigl\{ \r_{N, L} \CO_1^{(1)} \dots
	             \CO_m^{(m)}\bigr\}}
               {\Tr_{- L + 1 \dots L} \{\r_{N, L}\}} \\[1.5ex]
        = \frac{\<\Ps_0| \Tr \{\CO^{(1)} T(0)\} \dots \Tr \{\CO^{(m)} T(0)\}|\Ps_0\>}
               {\<\Ps_0|\Ps_0\> \La_0^m (0)} \epc
\end{multline}
where $\La_0 (0)$ is the unique eigenvalue of largest modulus
of the quantum transfer matrix $t(\la) = \Tr T(\la)$ at $\la = 0$,
and where $|\Ps_0\>$ is the corresponding eigenvector (see
\cite{GKS04a} for more details). The other eigenstates will be
denoted $|\Ps_n\>$, the associated eigenvalues $\La_n (\la)$.

Replacing $m$ by $m + 1$ in (\ref{opexpapr}) and considering e.g.\
$\CO^{(1)} = \s^-$, $\CO^{(m+1)} = \s^+$ and the intermediate operators
to be unity, we obtain a finite temperature asymptotic expansion for
the transverse correlation functions,
\begin{equation} \label{spm}
     \<\s_1^- \s_{m+1}^+\>_N = \sum_n A_n^{-+} \r_n^m \epc
\end{equation}
if we insert a complete set of states. Here we have used
the notation
\begin{equation} \label{defxia}
     \r_n = \re^{- 1/\x_n} = \frac{\La_n (0)}{\La_0 (0)} \epc \qd
     A_n^{-+} = \frac{\<\Ps_0|T^1_2 (0)|\Ps_n\>}{\La_n (0) \<\Ps_0|\Ps_0\>} \,
                \frac{\<\Ps_n|T^2_1 (0)|\Ps_0\>}{\La_0 (0) \<\Ps_n|\Ps_n\>} \epp
\end{equation}
Similar expansions hold for the longitudinal two-point functions
and for their generating function \cite{DGK13a}. The sum on the
right hand side of (\ref{spm}) is finite as long as the Trotter number
is finite. In the Trotter limit $N \rightarrow \infty$ it turns
into a series that provides an easy access to the large-distance
asymptotic behaviour of the thermal correlation function
$\<\s_1^- \s_{m+1}^+\>$, since $|\r_n| < 1$ for $n \in {\mathbb Z}_+$.
The $\r_n$ are ratios of higher eigenvalues of the quantum transfer
matrix to the dominant eigenvalue $\La_0$. The specific choice of
the operators in~(\ref{opexpapr}) ($\s^-$, $\s^+$ in our example
(\ref{spm})) determines which amplitudes will be non-zero and, hence,
which eigenvalue ratios will appear. The numbers $\xi_n$ are called
the correlation lengths. They are generally non-real and describe
the rate of exponential decay with distance of a given term in
the series (\ref{spm}) as well as its oscillatory behaviour. The
coefficients $A_n^{-+}$ are called correlation amplitudes. They are
products of two factors which we called thermal form factors in
\cite{DGK13a}.

In this work we analyse the low-temperature behaviour of the
eigenvalue ratios $\r_n$ in the massive regime, \emph{i.e.}\ we
concentrate on the spectrum of the quantum transfer matrix.
An exploration of the low-temperature properties of the correlation
amplitudes for the transverse and longitudinal correlation functions
as well as of the behaviour of the corresponding series representations
of two-point functions is deferred to a separate publication.

For any finite Trotter number the eigenvalues of the quantum
transfer matrix are determined by the algebraic Bethe Ansatz
(see e.g.\ \cite{GKS04a}\footnote{Note that we are using a slightly
different parameterization of the `Boltzmann weights' $b$ and
$c$ here which is more suitable for $\D > 1$ (see eqn.\
(\ref{rmatrix})).}):
\begin{equation} \label{baev}
     \La (x) = \re^\frac{h}{2T}
        \biggl( \frac{\sin (x + \i \be/N)}{\sin(x + \i \be/N + \i \g)} \biggr)^\frac N 2
	\biggl[ \prod_{j=1}^M \frac{\sin(x - x_j^r + \i \g/2)}{\sin(x - x_j^r - \i \g/2)}
	        \biggr]
	\bigl( 1 + \fa (x - \i \g/2) \bigr) \epc
\end{equation}
where the auxiliary function $\fa$ is defined by
\begin{multline} \label{baaux}
     \fa(x) = \fa \bigl(x \big| \{x_k^r\}_{k=1}^M \bigr)\\ = \re^{- \frac h T}
        \biggl( \frac{\sin (x + \i \g/2 - \i \be/N) \sin(x + 3 \i \g/2 + \i \be/N)}
	             {\sin (x + \i \g/2 + \i \be/N) \sin(x - \i \g/2 - \i \be/N)}
		      \biggr)^\frac N 2
	\prod_{k=1}^M \frac{\sin(x - x_k^r - \i \g)}{\sin(x - x_k^r + \i \g)}.
\end{multline}
and where the Bethe roots $x_j^r$ are subject to the Bethe Ansatz equations
\begin{equation} \label{baes}
     \fa \bigl(x_j^r \big| \{x_k^r\}_{k=1}^M \bigr) = - 1 \epc \qd j = 1, \dots, M \epp
\end{equation}
Throughout this work we shall assume that the Bethe roots $x_j^r$
are pairwise distinct. This is not a severe limitation in that,
should some of the roots coincide, it would be enough to slightly
deform the auxiliary function $\fa$ (and thus the Bethe Ansatz
equations) by introducing additional `deformation parameters', then
perform the analysis of the equations and send the deformation
parameters to zero in the end.

At this point we can formulate our goal on a technical level:
we want to analyse equations (\ref{baev})-(\ref{baes}) in the
Trotter limit for small $T$, for $h$ below the lower critical
field, and for fixed value of the `spin'
\begin{equation}
     s = N/2 - M \epp
\end{equation}
Note that $s = 0$ for the longitudinal two-point functions, while
$s = 1$ for the transverse two-point functions (\ref{spm}).
Spin $s = - 1$ belongs to the transverse two-point function
$\<\s_1^+ \s_{m+1}^-\>$, while larger values of $|s|$ occur
in the study of multi-point correlation functions or
if we study higher form factors with several local operators
acting on neighbouring sites. Instead of the Bethe Ansatz
equations (\ref{baes}) and the defining equation (\ref{baaux})
we shall introduce an equivalent characterization of the
auxiliary function $\fa$ by means of a nonlinear integral
equation in the next section. Using the nonlinear integral
equation we can easily perform the Trotter limit and also get
access to the low-temperature limit in which the nonlinear
integral equation linearizes.
\section{Low-temperature spectrum of correlation lengths}
\label{sec:main}
\subsection{Nonlinear integral equation for the auxiliary function}
\label{subsec:nlie}
The nonlinear integral equation will connect $\ln \fa$ and $\ln (1 + \fa)$.
As we shall see, in the antiferromagnetic massive regime it turns
out to be important to keep control over the overall phase of
the logarithms. For taking the logarithm of equation (\ref{baaux})
we introduce for every $\de > 0$ the functions
\begin{subequations}
\begin{align}
     & K(x|\de) = \frac{1}{2 \p \i} \bigl( \ctg(x - \i \de) - \ctg(x + \i \de) \bigr)
                  \epc \\[1ex]
     & \th (x|\de) = 2 \p \i \int_{\G_x} \rd y \: K(y|\de) \epc
\end{align}
\end{subequations}
where $\G_x$ is a piecewise straight contour starting at $- \p/2$, running
parallel to the imaginary axis to $- \p/2 + \i \, \Im x$ and then parallel
to the real axis from $- \p/2 + \i \, \Im x$ to $x$. Accordingly, the
function $\th (x|\de)$ is defined in the cut complex plane with
cuts along the line segments $(- \infty \pm \i \de, - \p \pm \i \de]
\cup [\pm \i \de, \pm \i \de + \infty)$.
If $\de = \g$ we shall write $K(x) = K(x|\g)$ and $\th(x) = \th(x|\g)$
for short.

Hereafter we will use the following properties of these functions:
\begin{subequations}
\label{qppkth}
\begin{align}
     & K(x|\de) = \frac{1}{2 \p} \, \frac{\sh(2 \de)}{\sh^2 (\de) + \sin^2 (x)} > 0 \epc
                  \qd \text{for $x \in {\mathbb R}$,} \\[1ex]
     & K(x + \p|\de) = K(x|\de) \epc \qd K(- x|\de) = K(x|\de) \epc \\[1ex]
     \label{thetaqp} \displaybreak[0]
     & \th (x + \p|\de) = \th(x|\de) +
                          \begin{cases} 
			     2 \p \i & |\Im x| < \de \\
                             0 & |\Im x| > \de,
                          \end{cases}\\[1ex]
     \label{thetaminus}
     & \th (- x|\de) = - \th(x|\de) +
                         \begin{cases} 
			    2 \p \i & |\Im x| < \de \\
                            0 & |\Im x| > \de.
                         \end{cases}
\end{align}
\end{subequations}
It follows from (\ref{thetaqp}) and (\ref{thetaminus}) that $\th(\p/2|\de)
= 2 \p \i$ and $\th(0|\de) = \p \i$. Setting $x = u + \i v$, $u, v \in
{\mathbb R}$ we further find the asymptotic behaviour
\begin{equation} \label{asytheta}
     \lim_{v \rightarrow \pm \infty} \th(u + \i v) = \mp 2 \g \epp
\end{equation}
Note that $\th (x|\de)$ is a determination of the logarithm 
of $\sin(x - \i \de)/\sin(x + \i \de)$ which coincides with
its principle branch $\Ln \bigl( \sin(x - \i \de)/\sin(x + \i \de) \bigr)$
in the strip $- \p < \Re x < \nolinebreak 0$. Here and in the following
we define the latter in such a way that $- \p < \Im \Ln (x) \le \p$.

Using $\th$ in (\ref{baaux}) we can define the function $\ln \fa$ as
\begin{equation} \label{deflna}
     \ln \fa (x) = - \frac{\e_0^{(N)} (x)} T
                   - \frac N 2 \th(x + \i \g/2 + \i \be/N)
                   + \sum_{j=1}^M \th(x - x_k^r) \epc
\end{equation}
where
\begin{equation} \label{defenulln}
     \e_0^{(N)} (x) =
        h - \frac {N T} 2
	\bigl[ \th(x + \i \be/N|\g/2) - \th(x - \i \be/N|\g/2) \bigr] \epp
\end{equation}

Starting from this equation we may derive an integral equation for $\fa$.
The form of this integral equation will depend on our initial choice
of the integration contour. We choose a rectangular, positively oriented
contour ${\cal C}$ starting and ending at $- \p/2$ and defined in such a
way that its upper edge ${\cal C}_+$ joins $\p/2$ with $- \p/2$ while
its lower edge ${\cal C}_-$ joins $- \p/2 - \i \g^-$ with $\p/2 - \i \g^-$. We
further denote its left edge by ${\cal C}_\ell$ and its right edge by
${\cal C}_r$. By definition $\g^- = \g - 0_+$.
\begin{figure}

\setlength{\unitlength}{1mm}

\begin{center}

\begin{picture}(120,85)

\put(14,43){${\cal C}_\ell$}
\put(74,51){${\cal C}_r$}
\put(67,29){${\cal C}_-$}
\put(34,63){${\cal C}_+$}

\put(14,62){$- \p/2$}
\put(77,62){$\p/2$}

\put(10,59){\vector(1,0){90}}
\put(50,5){\vector(0,1){85}}

\thicklines

\put(20,35){\framebox(60,24)}

\put(49,83){\line(1,0){2}}
\put(52,82){$\i \g$}
\put(42,31){$- \i \g^-$}
\put(49,11){\line(1,0){2}}
\put(40,10){$- 2 \i \g$}

\put(19.9,59){\line(0,1){1}}
\put(80.1,59){\line(0,1){1}}

\put(60,34){$>$}
\put(30,58){$<$}

\end{picture}

\end{center}

\caption{\label{fig:massive_contour} 
The integration contour ${\cal C}$.}
\end{figure}
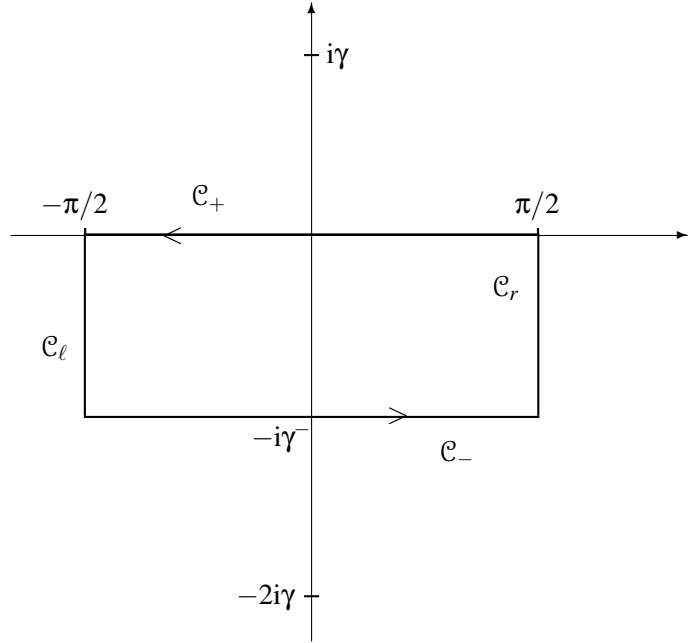

Relative to the contour $\cal C$ we introduce the following terminology
for the roots of the equation
\begin{equation} \label{nameroots}
     1 + \fa (x) = 0 \epp
\end{equation}
A root $x$ of (\ref{nameroots}) is called a Bethe root if $x \in
\{x_k^r\}_{k=1}^M$. Bethe roots outside $\cal C$ are called particle
roots or particles. A particle $x$ is called close if $x - \i \g$ is
inside $\cal C$, far otherwise. We denote the number of close particles
$n_c$ and the number of far particles $n_f$. The close and far
particles themselves will be denoted $x_j^c$ and $x_k^f$. Roots of
(\ref{nameroots}) inside $\cal C$ other than Bethe roots are called
holes. They are denoted $x_j^h$ while their number is $n_h$ by
definition.

It follows from (\ref{thetaqp}) and (\ref{deflna}) that, for real
$x$ and $|\be/N| < \g/2$,
\begin{equation} \label{qplna}
     \ln \fa (x + \p) = \ln \fa (x) - 2 \p \i (s + n_f) \epp
\end{equation}
Similarly, using (\ref{asytheta}) in (\ref{deflna}) we conclude that
\begin{equation} \label{asylna}
     \lim_{\Im x \rightarrow \pm \infty} \ln \fa (x)
        = - \frac h T \pm 2 \g s \epp
\end{equation}

The functions $\fa$ and $1 + \fa$ are meromorphic and have the
same poles inside the strip $- \p/2 < \Im x \le \p/2$: two
$N/2$-fold poles at $\pm \i (\g/2 + \be/N)$ and $M$ simple
poles at $x_k^r - \i \g$ (note that we assume that all roots of
(\ref{nameroots}) are simple). It follows that the only
singularities of $\6_x \ln (1 + \fa (x)) = \fa' (x)/(1 + \fa (x))$
inside $\cal C$ are the simple poles:
\renewcommand{\arraystretch}{1.6}
\begin{center}
\begin{tabular}{lr}
\toprule
location & residue \\
\midrule
$x \in \{x_k^r\}_{k=1}^M \setminus
   \bigl\{ \{x_k^c \}_{k=1}^{n_c} \cup \{x_k^f \}_{k=1}^{n_f} \bigr\}$ & $1$ \\
$x \in \{x_k^h\}_{k=1}^{n_h}$ & $1$ \\
$x \in \{x_k^c - \i \g \}_{k=1}^{n_c}$ & $- 1$ \\
$x = - \i (\g/2 + \be/N) $ & $- N/2$ \\
\bottomrule
\end{tabular}
\end{center}
Moreover, if $x, y \in {\cal C}$ or inside, then $\th(x - y)$ is
holomorphic as a function of $y$. It follows that, for $- \g < \Im (x) < 0$,
\begin{multline} \label{nlide}
     \ln \fa (x) = - \frac{\e_0^{(N)} (x)} T
                   - \sum_{j=1}^{n_h} \th(x - x_j^h)
                   + \sum_{j=1}^{n_c} \bigl( \th(x - x_j^c) + \th(x - x_j^c + \i \g) \bigr)
                   + \sum_{j=1}^{n_f} \th(x - x_j^f) \\
		   + \int_{\cal C} \frac{\rd y}{2 \p \i} \th(x - y) \6_y \ln (1 + \fa (y)) \epp
\end{multline}

This equation can be transformed into a nonlinear integral equation for
the auxiliary function by partial integration. Again some care is
necessary with the definition of the logarithms. First of all, there
is no ambiguity in the definition of the function $\6_x \ln (1 + \fa (x))$
which is simply defined as $\fa' (x)/(1 + \fa (x))$, similarly $\6_x \ln \fa (x)
= \fa' (x)/\fa (x)$ by definition. For any point $x$ on $\cal C$ and
$f = \fa, 1 + \fa, 1 + \fa^{-1}$ we now define
\begin{equation} \label{deflnc}
     \ln_{\cal C} f (x) = \int_{{\cal C}_x} \rd y \: \6_y \ln f (y) \epp
\end{equation}
Here ${\cal C}_x$ is the simple contour which starts at $- \p/2 - \i 0$
and runs along ${\cal C}$ up to the point~$x$. The function $\ln_{\cal C} f$
is holomorphic along ${\cal C}$ by construction\footnote{We assume that none
of the zeros or poles of $f$ are on ${\cal C}$. This can always be achieved
by slightly deforming the contour if necessary.} and can be used in
partial integration. The monodromy of $\ln_{\cal C} (1 + \fa) (x)$ along
${\cal C}$ is generally nontrivial. Using the above tabular we find that
\begin{equation} \label{defd}
     d = \int_{\cal C} \frac{\rd y}{2 \p \i} \6_y \ln (1 + \fa (y))
       = \frac{\ln_{\cal C} (1 + \fa) (- \p/2)}{2 \p \i} 
       = n_h - 2 n_c - n_f - s
\end{equation}
which may be generally non-zero. Performing now the partial integration
in (\ref{nlide}) we arrive at the following
\begin{lemma}
The auxiliary function $\fa$ defined in (\ref{baaux}) satisfies the
nonlinear integral equation
\begin{multline} \label{nlie}
     \ln \fa (x) = - \frac{\e_0^{(N)} (x)} T
                   - \sum_{j=1}^{n_h} \th(x - x_j^h)
                   + \sum_{j=1}^{n_c} \bigl( \th(x - x_j^c) + \th(x - x_j^c + \i \g) \bigr)
                   + \sum_{j=1}^{n_f} \th(x - x_j^f) \\
		   + d \th(x + \p/2)
		   + \int_{\cal C} \rd y \: K(x - y) \ln_{\cal C} (1 + \fa) (y) \epp
\end{multline}
This equation determines $\fa$ directly inside the strip $- \g < \Im x < 0$
and, by analytic continuation, in the entire complex plane. In particular,
for $x \in {\cal C}_\pm$ the integral term should be understood as an
appropriate boundary value of a Cauchy-like operator.

The particles and holes have to be determined such that they satisfy
the subsidiary conditions
\begin{equation} \label{subs}
     1 + \fa \bigl(x_j^{h, p, f}\bigr) = 0
\end{equation}
with $x_j^{h, p, f}$ in their respective domains of definition.
\end{lemma}
In equation (\ref{nlie}) the Trotter limit is easily performed
by substituting $\e_0^{(N)} (x)$ by
\begin{equation} \label{defenull}
     \e_0 (x) = \lim_{N \rightarrow \infty} \e_0^{(N)} (x)
              = h - \frac{4 J (\D^2 - 1)}{\D - \cos (2x)} \epp
\end{equation}
Equation (\ref{nlie}) is a good starting point for studying
the system numerically or, as we shall see below, analytically
in the low-temperature limit. Before moving to the latter
subject in the next subsection we wish to add three remarks.
\begin{Remark}
Equations (\ref{defd}), (\ref{nlie}) are compatible with the
quasi-periodicity property (\ref{qplna}) and with the asymptotic
behaviour (\ref{asylna}) of $\ln \fa$. In fact, using (\ref{qppkth})
in (\ref{nlie}) we obtain for real $x$
\begin{equation}
     \ln \fa (x + \p) = \ln \fa (x) +  2 \p \i (- n_h + 2 n_c + d)
\end{equation}
which turns into (\ref{qplna}) after inserting (\ref{defd}).

To see the compatibility with the asymptotic behaviour we note
the following. Due to the $\p$-periodicity of the auxiliary function,
Bethe roots are only defined modulo $\p$. If there is a Bethe root
with real part $- \p/2$ it must be identified with the same root 
shifted by $\p$, having real part $\p/2$. Then the contour must
be deformed such as to exclude one of the two points, the one
with real part $- \p/2$, say. This can be achieved by infinitesimally
shifting the contour to the right which then also can be slightly
narrowed. Then the right hand side of (\ref{nlie}) determines
$\ln \fa (x)$ for $\Re x = - \p/2 + \e$, $\e > 0$ outside $\cal C$
by analytic continuation, and we can calculate the limit along
the line $\Re x = - \p/2 + \e$ using (\ref{asytheta}) (which is
possible if there is no Bethe root with real part $- \p/2$ anyway),
\begin{equation}
     \lim_{v \rightarrow \infty} \ln \fa (- \p/2 + \e + \i v) =
        - \frac h T - 2 \g (- n_h + 2 n_c + n_f + d) \epp
\end{equation}
Inserting (\ref{defd}) we see that this is compatible with (\ref{asylna}).
\end{Remark}
\begin{Remark}
So far we have shown that (\ref{baaux}) and (\ref{baes}) imply
(\ref{defd}) and (\ref{nlie}). It is not difficult to see that the
converse is also true. Thus, the two pairs of equations
(\ref{baaux}), (\ref{baes}) and (\ref{defd}), (\ref{nlie})
are equivalent.
\end{Remark}
\begin{Remark}
Equations (\ref{defd}), (\ref{nlie}) depend on our definition of
the contour and on the specific class of solutions we are considering.
Let us restrict ourselves for a moment to solutions with $n_f = 0$,
and let us deform the contour ${\cal C} \rightarrow {\cal C}_s$
in such a way that the points $x_j^c - \i \g$ are outside ${\cal C}_s$.
Suppose further that $d = 0$ along ${\cal C}_s$. Then (\ref{nlie})
turns into
\begin{equation} \label{nliemassless}
     \ln \fa (x) = - \frac{\e_0^{(N)} (x)} T
                   - \sum_{j=1}^{n_h} \th(x - x_j^h)
                   + \sum_{j=1}^{n_c} \th(x - x_j^c)
		   + \int_{{\cal C}_s} \rd y \: K(x - y) \ln_{{\cal C}_s} (1 + \fa) (y) \epp
\end{equation}
This equation has the same form as the one we used in the analysis
of the massless antiferromagnetic regime for $|\D| < 1$ \cite{DGK13a,%
DGK14a} (in that case we had an additional term $- \i \p s$ which is
absorbed into the different definition of the function $\th$ here). In
the massless regime for $\D > 1$ (when the magnetic field is between
lower and upper critical field \cite{YaYa66d}) equation (\ref{nliemassless})
is an appropriate starting point for the analysis of the low-temperature
behaviour of correlation lengths. In this case the relevant `excitations'
(leading to correlation lengths that diverge for $T \rightarrow 0$) are
characterized by particles and holes $x_j^c$, $x_k^h$ very close to
two Fermi points $\pm Q$, $Q \in (0,\p/2)$. The condition $d = 0$ can be
satisfied by performing a small deformation of the contour in the
vicinity of the Fermi points and by lifting up the lower part of the
contour. This means that the contour ${\cal C}_s$ has to be
self-consistently determined in the course of the calculation (see
\cite{DGK13a,DGK14a}).
\end{Remark}
\subsection{The auxiliary function in the low-temperature limit}
For the analysis of the low-temperature behaviour of (\ref{nlie})
in the antiferromagnetic massive regime we split the integral
over $\cal C$ into contributions from its rectilinear parts
${\cal C}_\ell, \dots, {\cal C}_+$ (see Figure~\ref{fig:massive_contour}),
\begin{multline} \label{splitcontour}
     \int_{\cal C} \rd y \: K(x - y) \ln_{\cal C} (1 + \fa) (y) =
        \int_{{\cal C}_+} \rd y \: K(x - y) \ln_{\cal C} \fa (y) \\
	+ \bigl( \th(x + \p/2 + \i \g^-) - \th(x + \p/2) \bigr)
	  \int_{{\cal C}_-} \frac{\rd y}{2 \p \i} \6_y \ln (1 + \fa (y)) \\
	+ \int_{{\cal C}_+} \rd y \: K(x - y) \ln_{\cal C} (1 + \fa^{-1}) (y)
	+ \int_{{\cal C}_-} \rd y \: K(x - y) \ln_{\cal C} (1 + \fa) (y) \epp
\end{multline}
Here we have used the definition of $\ln_{\cal C}$ introduced in 
(\ref{deflnc}). The integrals over the left and right partial
contours contribute the second line of (\ref{splitcontour}). In
order to see this define a contour ${\cal C}_{y, y + \p}$ running
along $\cal C$ from a point $y$ on ${\cal C}_\ell$ to $y + \p$ on
${\cal C}_r$. Then
\begin{multline}
     \int_{{\cal C}_\ell + {\cal C}_r} \rd y \: K(x - y) \ln_{\cal C} (1 + \fa (y)) \\
        = \int_{{\cal C}_\ell} \rd y \: K(x - y)
	  \bigl( \ln_{\cal C} (1 + \fa (y)) - \ln_{\cal C} (1 + \fa (y + \p)) \bigr) \\
        = - \int_{{\cal C}_\ell} \rd y \: K(x - y)
	  \int_{{\cal C}_{y, y + \p}} \rd z \: \6_z \ln (1 + \fa (z)) 
        = - \int_{{\cal C}_\ell} \rd y \: K(x - y)
	  \int_{{\cal C}_-} \rd z \: \6_z \ln (1 + \fa (z)) \\
	= \bigl( \th(x + \p/2 + \i \g^-) - \th(x + \p/2) \bigr)
	  \int_{{\cal C}_-} \frac{\rd y}{2 \p \i} \6_y \ln (1 + \fa (y)) \epp
\end{multline}
Note that, due to the $\p$-periodicity of $\fa$, the last integral on
the right hand side must be an integer. To understand the splitting of
the integral over ${\cal C}_+$ in (\ref{splitcontour}) use $\6_x
\ln(1 + \fa (x)) = \6_x \ln \fa (x) + \6_x \ln \bigl( 1 + \fa^{-1} (x) \bigr)$.

From the leading behaviour of the driving term in (\ref{nlie}) we expect
that $\fa^{-1} (x) = {\cal O} (T^\infty)$ for $x \in {\cal C}_+$ and
$\fa (x) = {\cal O} (T^\infty)$ for $x \in {\cal C}_-$. Our strategy will
be to assume such behaviour and to determine $\fa$ self-consistently,
\emph{i.e.}\ we shall show a posteriori that the solutions we obtain based
on such an assumption have indeed the assumed properties. Then we shall
argue that the set of self-consistent solutions is complete.

If $\fa (x) = {\cal O} (T^\infty)$ for $x \in {\cal C}_-$, then
$\Ln (1 + \fa (x))$ is holomorphic on ${\cal C}_-$ and
\begin{equation}
     \int_{{\cal C}_-} \rd y \: \6_y \ln (1 + \fa (y)) =
        \int_{{\cal C}_-} \rd y \: \6_y \Ln (1 + \fa (y)) = 0 \epc
\end{equation}
because of the $\p$-quasi-periodicity of $\fa$. It follows that
\begin{multline} \label{splitprince}
     \int_{\cal C} \rd y \: K(x - y) \ln_{\cal C} (1 + \fa (y)) =
        \int_{{\cal C}_+} \rd y \: K(x - y) \ln \fa (y) \\
	+ \int_{{\cal C}_+} \rd y \: K(x - y) \Ln \bigl(1 + \fa^{-1} (y)\bigr)
	+ \int_{{\cal C}_-} \rd y \: K(x - y) \Ln (1 + \fa (y)) - 2 \p \i d \\
	+ \ln \fa (- \p/2) + \Ln \bigl(1 + \fa^{-1} (- \p/2)\bigr)
	- \Ln \bigl(1 + \fa (- \p/2 - \i \g^-)\bigr) \\
	+ \int_{{\cal C}_\ell} \rd y \: \6_y \ln (1 + \fa (y)) \epp
\end{multline}
Note that the last two lines are equal to a pure phase,
\begin{multline} \label{phasek}
     \ln \fa (- \p/2) + \Ln \bigl(1 + \fa^{-1} (- \p/2)\bigr)
	- \Ln \bigl(1 + \fa (- \p/2 - \i \g^-)\bigr) \\
	+ \int_{{\cal C}_\ell} \rd y \: \6_y \ln (1 + \fa (y)) = - 2 \p \i k \epc
\end{multline}
$k \in {\mathbb Z}$.

At this point it is convenient to switch the notation by introducing
the function
\begin{equation} \label{defu}
     u(x) = - T \ln \fa (x) \epp
\end{equation}
Further, setting\footnote{%
When $x \in {\cal C}_+$, the integral over ${\cal C}_-$ has to be
understood as a boundary value of a Cauchy operator. In order to
deal with a more regular expression one could slightly shift up
the lower contour ${\cal C}_-$ by a small finite $\de > 0$.
This might produce additional contributions if solutions of
$1 + \fa (x) = 0$ were located between ${\cal C}_-$ and ${\cal C}_-
+ \i \de$. In order to rule out the existence of roots between
${\cal C}_-$ and ${\cal C}_- + \i \de$ we would then have to
distinguish one more case in the subsequent analysis of the
higher-level Bethe Ansatz equations. In the end the result would
be the same. In order to lighten the discussion we will simply
assume that no roots of $1 + \fa (x) = 0$ exist in tight finite
strips around ${\cal C}_\pm$, and, thus, that $r[u] (x)$ is smooth
for $x \in {\cal C}_\pm$.}
\begin{equation} \label{defr}
     r[u] (x) =
	\int_{{\cal C}_+} \rd y \: K(x - y) \Ln \bigl(1 + \re^{\frac{u(y)}T}\bigr)
	+ \int_{{\cal C}_-} \rd y \: K(x - y) \Ln \bigl(1 + \re^{- \frac{u(y)}T}\bigr)
\end{equation}
and inserting (\ref{defu}) and (\ref{defr})  together with
(\ref{splitprince}) and (\ref{phasek}) into (\ref{nlie}) we obtain
the following form of the nonlinear integral equation in the
Trotter limit:
\begin{equation} \label{nlielowt}
     u (x) = \e_0 (x) + T \Th (x)
             - \int_{- \p/2}^{\p/2} \rd y \: K(x - y) u (y)
	     - T \cdot r[u] (x) \epc
\end{equation}
where
\begin{multline}
    \Th (x) = 2 \p \i k + \sum_{j=1}^{n_h} \th(x - x_j^h)
	      - \sum_{j=1}^{n_c} \bigl( \th(x - x_j^c) + \th(x - x_j^c + \i \g) \bigr) \\
	      - \sum_{j=1}^{n_f} \th(x - x_j^f)
	      - d \th(x - \p/2) \epp
\end{multline}
By construction, equation (\ref{nlielowt}) is valid as long as
$|\fa (x)| = \re^{- \Re u(x)/T} < 1$ on ${\cal C}_-$ and
$|\fa^{-1} (x)| = \re^{\Re u(x)/T} < 1$ on ${\cal C}_+$. But
then $|r[u]|$ is bounded, and $T \Th (x)$ as well as $T \cdot r[u] (x)$
can be neglected for small $T$ if none of the singularities of
$\Th$ is close to ${\cal C}_+$ or~${\cal C}_-$.

Assuming the latter as part of the conditions which have to be
self-consistently satisfied, we obtain, to leading order in $T$,
a linear integral equation of the same form as the integral
equation for the dressed energy $\e$,
\begin{equation} \label{lineps}
     \e (x) = \e_0 (x) - \int_{- \p/2}^{\p/2} \rd y \: K(x - y) \e (y) \epp
\end{equation}
This equation can be solved explicitly in terms of elliptic functions
(see Appendix~\ref{app:energy}). From the explicit solution we can
see that $\e(x) < 0$ on ${\cal C}_+$ and $\e (x) > 0$ on ${\cal C}_-$
as long as
\begin{equation}
     0 < h < h_\ell =
        \frac 1 \p 8 J K \sh \Bigl(\frac{\p K'}{K} \Bigr) \dn(K|k) \epp
\end{equation}
Here $h_\ell > 0$ is the lower critical field that determines the
boundary of the antiferromagnetic massive phase \cite{YaYa66d}
(see Figure~\ref{fig:phasediagram}, and note that we have introduced
the parameterization $\g = \p K'/K$ of the anisotropy parameter,
where $K(k)$ and $K' (k) = K(\sqrt{1 - k^2})$ are complete
elliptic integrals and $k$ is the elliptic module; $\dn$ is
a Jacobi elliptic function). Thus, if the singularities of the
driving term in (\ref{nlielowt}) are away from ${\cal C}_\pm$,
the remainder $r[u]$, which contains the nonlinear dependence
of the integrands on $u$, is of order $T^\infty$ and can be
self-consistently neglected in order to obtain $u$ on ${\cal C}_\pm$
and in the domain containing ${\cal C}_\pm$ where $r[u]$
is analytic.

In order to obtain the ${\cal O} (T)$ contributions to $u$
we introduce the dressed charge $Z$ and the dressed phase
$\ph(*,z)$ which we define on $[-\p/2,\p/2]$ as solutions
of the linear integral equations
\begin{subequations}
\label{linzphi}
\begin{align}
     & Z(x) = 1 - \int_{- \p/2}^{\p/2} \rd y \: K(x - y) Z (y) \epc \\
     & \ph (x,z) = 
        \th (x - z) - \2 \one_{|\Im z| < \g} \th (x - \p/2)
        - \int_{- \p/2}^{\p/2} \rd y \: K(x - y) \ph (y,z) \epp
	\label{dressphase}
\end{align}
\end{subequations}
In (\ref{dressphase}) we have introduced the notation
\[
     \one_{\rm condition} = \begin{cases} 1 & \text{if condition is satisfied} \\
                                          0 & \text{else.}
                            \end{cases}
\]
The values of the functions $Z$ and $\ph(*,z)$ in the entire complex
plane are obtained by analytic continuation of the solutions of
(\ref{linzphi}), which can be constructed explicitly. In particular,
$Z(x) = 1/2$. For a detailed description of $\ph(*,z)$ see Appendix~%
\ref{app:phase}. Note that different continuations of $\ph(*,z)$
are possible depending on the system of cuts that is chosen.
Just as for the logarithm, any two such continuations differ
locally at most by a constant in $2 \p \i {\mathbb Z}$.

Using (\ref{lineps}), (\ref{linzphi}) in (\ref{nlielowt}) we find that
\begin{equation} \label{uuone}
     u(x) = u_1 (x) + {\cal O} (T^\infty) \qd \text{for $- \g < \Im x \le 0$} \epc
\end{equation}
where
\begin{multline} \label{uone}
     u_1 (x) = \e (x)
               + T \Bigl\{ \i \p k + \sum_{j=1}^{n_h} \ph(x, x_j^h)
                   - \sum_{j=1}^{n_c} \bigl( \ph(x, x_j^c) + \ph(x, x_j^c - \i \g) \bigr) \\
                   - \sum_{j=1}^{n_f} \ph(x, x_j^f)
		   - 2 (d - n_h/2 + n_c) \ph(x, \p/2) \Bigr\} \epp
\end{multline}
The last term in this equation vanishes due to a linear relation between
the numbers of particles and holes and the spin. Namely, if
$|\fa (x)| = \re^{- \Re u(x)/T} < 1$ on ${\cal C}_-$ and $|\fa^{-1} (x)| =
\re^{\Re u(x)/T} < 1$ on ${\cal C}_+$, so in particular in the
low-$T$ limit, we have
\begin{multline}
     d = \int_{\cal C} \frac{\rd y}{2 \p \i} \6_y \ln (1 + \fa (y))
       = \int_{{\cal C}_+ + {\cal C}_-} \frac{\rd y}{2 \p \i} \6_y \ln (1 + \fa (y)) \\
       = \frac{1}{2 \p \i} \bigl\{
           \ln \fa (- \p/2) - \ln \fa (\p/2)
         + \Ln \bigl(1 + \fa^{-1} (- \p/2)\bigr)
         - \Ln \bigl(1 + \fa^{-1} (\p/2)\bigr) \\
         + \Ln \bigl(1 + \fa (\p/2 - \i \g^-)\bigr)
         - \Ln \bigl(1 + \fa (- \p/2 - \i \g^-)\bigr) \bigr\} \\
       = \frac{1}{2 \p \i} \bigl\{ \ln \fa (- \p/2) - \ln \fa (\p/2) \bigr\} \epp
\end{multline}
Here we have used the quasi-periodicity of $\fa$ and the fact
that $\fa^{-1} = {\cal O} (T^\infty)$ on ${\cal C}_+$ and
$\fa = {\cal O} (T^\infty)$ on ${\cal C}_-$. Inserting
(\ref{uuone}) and (\ref{uone}) into the above equation and
using the $\p$-periodicity of $\e$ as well as the $\p$-quasi-periodicity
of $\ph(*,z)$ we obtain
\begin{equation}
     d = n_h/2 - n_c \epp
\end{equation}

Thus we have derived the following
\begin{lemma} \label{lem:ustrip}
In the strip $- \g < \Im x \le 0$ the nonlinear integral equations
(\ref{nlielowt}) have self-consistent low-temperature solutions of
the form
\begin{equation}
     u(x) = u_1 (x) + {\cal O} (T^\infty) \epc
\end{equation}
where
\begin{equation} \label{u1}
     u_1 (x) = \e (x)
               + T \Bigl\{ \i \p k + \sum_{j=1}^{n_h} \ph(x, x_j^h)
                           - \sum_{j=1}^{n_c} \bigl( \ph(x, x_j^c)
			                           + \ph(x, x_j^c - \i \g) \bigr)
                           - \sum_{j=1}^{n_f} \ph(x, x_j^f) \Bigr\}
\end{equation}
and where the numbers of particles and holes are related to the spin
by the condition
\begin{equation}
     n_h - 2 n_c - 2 n_f = 2 s \epp
\end{equation}
\end{lemma}
In order do be able to discuss the subsidiary conditions (\ref{subs})
and to derive higher-level Bethe Ansatz equations we need to know
the function $u$ in the full complex plane. We shall obtain it
by analytic continuation from the nonlinear integral equation
(\ref{nlielowt}). Since analytic continuation and low-$T$ limit do
not commute, we will have to be careful with contributions stemming
from the remainder $r[u]$. The analytic continuation of the integrals
in (\ref{nlielowt}) is determined by the following elementary
\begin{lemma} \label{lem:contint}
Let $g_\pm$ by analytic in some strip ${\cal S}_\pm$ around
${\cal C}_\pm$ and let
\begin{equation}
     f_\pm (x) = \int_{{\cal C}_\pm} \rd y \: K(x - y) g_\pm (y) \epp
\end{equation}
Then $f_+$ is analytic for $- \g < \Im x < \g$ while $f_-$ is
analytic for $- 2 \g < \Im x < 0+$. Denote the analytic continuations
of $f_\pm$ by the same letters. Then
\begin{subequations}
\begin{align}
     & f_+ (x) = \int_{{\cal C}_+} \rd y \: K(x - y) g_+ (y) -
       \begin{cases}
          g_+ (x - \i \g) & \Im x > \g,\ x \in {\cal S}_+ + \i \g \\
	  0 & |\Im x| < \g \\
	  g_+ (x + \i \g) & \Im x < - \g,\ x \in {\cal S}_+ - \i \g \epc
       \end{cases} \\[1ex]
     & f_- (x) = \int_{{\cal C}_-} \rd y \: K(x - y) g_- (y) +
       \begin{cases}
          g_- (x - \i \g) & \Im x > 0, \ x \in {\cal S}_- + \i \g \\
	  0 & - 2 \g < \Im x \le 0 \\
	  g_- (x + \i \g) & \Im x < - 2 \g, \ x \in {\cal S}_- - \i \g \epp
       \end{cases}
\end{align}
\end{subequations}
\end{lemma}
Applying this lemma to equation (\ref{nlielowt}) we obtain
\begin{multline} \label{ueverywhere}
     u (x) = \e_0 (x) + T \Th (x)
             - \int_{- \p/2}^{\p/2} \rd y \: K(x - y) u (y)
	     - T r[u] (x) \\[1ex]
	     + \begin{cases}
	       2 \p \i k' T & \Im x > \g \\
	       - T \Ln \bigl(1 + \re^{- \frac{u(x - \i \g)} T} \bigr) & 0 < \Im x < \g \\
	       0 & - \g < \Im x < 0 \\
	       - u (x + \i \g) + T \Ln \bigl(1 + \re^{\frac{u(x + \i \g)}T} \bigr) &
	       - 2 \g < \Im x < - \g \\
	       2 \p \i k'' T & \Im x < - 2 \g \epc
	       \end{cases}
\end{multline}
where $k', k'' \in{\mathbb Z}$. The function $r[u]$ in (\ref{ueverywhere})
is given by (\ref{defr}) for all $x$ in the respective strips. Thus,
$ r[u] (x) = {\cal O} (T^\infty)$ in (\ref{ueverywhere}). Using
Lemma~\ref{lem:ustrip} in (\ref{ueverywhere}) we obtain
\begin{multline} \label{uu1every}
     u (x) = \e_0 (x) + T \Th (x)
             - \int_{- \p/2}^{\p/2} \rd y \: K(x - y) u_1 (y)
	     + {\cal O} (T^\infty) \\[1ex]
	     + \begin{cases}
	       2 \p \i k' T & \Im x > \g \\
	       - T \Ln \bigl(1 + \re^{- \frac{u_1(x - \i \g)} T}
	          (1 + {\cal O} (T^\infty)) \bigr) & 0 < \Im x < \g \\
	       0 & - \g < \Im x < 0 \\
	       - u_1 (x + \i \g) + T \Ln \bigl(1 + \re^{\frac{u_1(x + \i \g)}T}
	          (1 + {\cal O} (T^\infty)) \bigr) &
	       - 2 \g < \Im x < - \g \\
	       2 \p \i k'' T & \Im x < - 2 \g \epp
	       \end{cases}
\end{multline}
Notice that we have replaced $u$ by $u_1$ everywhere on the right
hand side. In each case the ${\cal O} (T^\infty)$ remainder is a
smooth function. In particular, it has no singularities.
We further simplify equation (\ref{uu1every}). Applying
Lemma~\ref{lem:contint} to the linear integral equation satisfied
by $u_1$ (equation (\ref{nlielowt}) without the remainder term)
we obtain
\begin{equation} \label{u1cont}
     \e_0 (x) + T \Th (x) - \int_{- \p/2}^{\p/2} \rd y \: K(x - y) u_1 (y)
	       = \begin{cases}
	            u_1 (x) + u_1 (x - \i \g) & \Im x > \g \\
	            u_1 (x) & |\Im x| < \g \\
	            u_1 (x) + u_1 (x + \i \g) & \Im x < - \g \epp
	         \end{cases}
\end{equation}
Inserting this identity into (\ref{uu1every}) and using (\ref{defu})
we end up with
\begin{lemma} \label{lem:alowentire}
Low-temperature form of the auxiliary function in the complex
plane.
\begin{equation} \label{alowentire}
     \fa (x) = \begin{cases}
                  \re^{- \frac 1 T (u_1 (x) + u_1 (x - \i \g))}
		  & \Im x > \g \\[1ex]
                  \re^{- \frac 1 T u_1 (x)} +
                  \re^{- \frac 1 T (u_1 (x) + u_1 (x - \i \g))}
		  & 0 < \Im x < \g \\[1ex]
                  \re^{- \frac 1 T u_1 (x)}
		  & - \g < \Im x < 0 \\[1ex]
		  \Bigl[
                  \re^{\frac 1 T u_1 (x)} +
                  \re^{\frac 1 T (u_1 (x) + u_1 (x + \i \g))}
		  \Bigr]^{-1}
		  & - 2 \g < \Im x < - \g \\[1ex]
                  \re^{- \frac 1 T (u_1 (x) + u_1 (x + \i \g))}
		  & \Im x < - 2 \g 
               \end{cases}
\end{equation}
up to multiplicative corrections of the form $1 + {\cal O} (T^\infty)$
(in front of each exponent, cf.\ (\ref{uu1every})).
\end{lemma}
Interestingly, there are only two independent functions occurring
on the right hand side. These can be written explicitly in
terms of special functions by means of Lemma~\ref{lem:ustrip}
and the formula collected in Appendix~\ref{app:functions}.
For this purpose we split the far roots into two sets
$\{x_j^f\}_{j=1}^{n_f} = \{x_j^+\}_{j=1}^{n_+} \cup \{x_j^-\}_{j=1}^{n_-}$,
where the $x_j^+$ have imaginary parts greater than $\g$ while
the $x_j^-$ have imaginary parts less than $- \g$. Then the
functions on the right hand side of (\ref{alowentire}) can
be expressed in terms of
\begin{multline}\label{aplus}
     \fa^{(+)} (x) =
     \re^{- \frac 1 T (u_1 (x) + u_1 (x - \i \g))} =
        \re^{ - \frac h T}
	\biggl[ \prod_{j=1}^{n_h}
	        \frac{\sin(x - x_j^h)}{\sin(x - x_j^h - \i \g)} \biggr]
	\biggl[ \prod_{j=1}^{n_c}
	        \frac{\sin(x - x_j^c - \i \g)}{\sin(x - x_j^c + \i \g)} \biggr] \\
        \times
	\biggl[ \prod_{j=1}^{n_+}
	        \frac{\sin(x - x_j^+ - \i \g)}{\sin(x - x_j^+ + \i \g)} \biggr]
	\biggl[ \prod_{j=1}^{n_-}
	        \frac{\sin(x - x_j^- - 2 \i \g)}{\sin(x - x_j^-)} \biggr]
\end{multline}
and
\begin{multline}\label{a0}
     \fa^{(0)} (x) =
     \re^{- \frac 1 T u_1 (x)} =
        (-1)^k \re^{ - \frac{\e (x)} T - \sum_{j=1}^{n_h} \ph (x,x_j^h)}
	\biggl[ \prod_{j=1}^{n_c}
	        \frac{\sin(x - x_j^c)}{\sin(x - x_j^c + \i \g)} \biggr] \\
        \times
	\biggl[ \prod_{j=1}^{n_+}
	        \frac{\sin(x - x_j^+)}{\sin(x - x_j^+ + \i \g)} \biggr]
	\biggl[ \prod_{j=1}^{n_-}
	        \frac{\sin(x - x_j^- - \i \g)}{\sin(x - x_j^-)} \biggr] \epp
\end{multline}
For later convenience we also define $\fa^{(-)} (x) =
\fa^{(+)} (x + \i \g)$. These functions together with
Lemma~\ref{lem:alowentire} will be used to discuss the
subsidiary conditions (\ref{subs}) and to derive a set
of `higher-level Bethe Ansatz equations' in the next
subsection.
\subsection{Root patterns and higher-level Bethe Ansatz
equations for non-zero magnetic field}
In this subsection we discuss the subsidiary conditions
(\ref{subs}) for $h > 0$ and $T \rightarrow 0_+$. We consider
the different types of roots in their respective domains of
definition.

Far roots $x_j^+$ are located at $\Im x > \g$.
In the low-temperature limit they are zeros of $1 + \fa^{(+)}$
according to Lemma~\ref{lem:alowentire}. Because of the
prefactor $\re^{-h/T}$, the function $\fa^{(+)}$ goes to
zero pointwise as $T \rightarrow 0_+$. Let $x_\ell^+
\in \{x_j^+\}_{j=1}^{n_+}$ such that
\begin{equation} \label{xplusmax}
     \Im x_\ell^+ \ge \Im x_j^+ \epc \qd j = 1, \dots, n_+ \epp
\end{equation}
For $x_\ell^+$ to be a root of $1 + \fa^{(+)}$ it must be
close to a pole of $\fa^{(+)} (x)$. The only possible poles
of $\fa^{(+)} (x)$ in the region $\Im x > \g$ are at
$x_j^+ - \i \g$. They cannot be close to $x_\ell^+$ due
to~(\ref{xplusmax}). Hence, $x_\ell^+$ cannot exist, and $n_+ = 0$.

Close roots $x_j^c$ are located in the strip $0 < \Im x < \g$.
From the formulae in Appendix~\ref{app:functions} we may infer that
\begin{equation} \label{zerospolesephi}
     \re^{- \ph(x, x^h)} = \re^{- \i (\p/2 + x - x^h)}
        \prod_{k=1}^\infty
	\frac{\bigl(1 - \re^{-2 [(2k - 1)\g - \i (x - x^h)]}\bigr)
	      \bigl(1 - \re^{-2 [2k\g + \i (x - x^h)]}\bigr)}
	     {\bigl(1 - \re^{-2 [(2k - 1)\g + \i (x - x^h)]}\bigr)
	      \bigl(1 - \re^{-2 [2k\g - \i (x - x^h)]}\bigr)} \epp
\end{equation}
Hence, this function has a zero at $x - x^h = - \i \g$ and a
pole at $x - x^h = \i \g$. These are its only poles and zeros in
the strip $- 2 \g < \Im x < \g$. It follows that
\begin{equation}
     A(x) = \re^{- \sum_{j=1}^{n_h} \ph(x, x_j^h)}
            \prod_{j=1}^{n_h}
	    \frac{\sin(x - x_j^h - \i \g)}{\sin(x - x_j^h + \i \g)}
\end{equation}
is analytic and non-zero in this strip. Thus, the subsidiary
condition for close roots at low temperature, $\fa^{(0)} (x)+ \fa^{(+)} (x)
= - 1$, takes the form
\begin{multline} \label{subsclose}
     (-1)^k A(x) \re^{- \frac{\e(x)} T}
           \biggl[ \prod_{j=1}^{n_h}
	   \frac{\sin(x - x_j^h + \i \g)}{\sin(x - x_j^h - \i \g)}
	   \biggr] \biggl[ \prod_{j=1}^{n_c}
	   \frac{\sin(x - x_j^c)}{\sin(x - x_j^c + \i \g)}
	   \biggr] \biggl[ \prod_{j=1}^{n_-}
	   \frac{\sin(x - x_j^- - \i \g)}{\sin(x - x_j^-)} \biggr] \\[1ex]
         + \re^{- \frac h T}
           \biggl[ \prod_{j=1}^{n_h}
	   \frac{\sin(x - x_j^h)}{\sin(x - x_j^h - \i \g)}
	   \biggr] \biggl[ \prod_{j=1}^{n_c}
	   \frac{\sin(x - x_j^c - \i \g)}{\sin(x - x_j^c + \i \g)}
	   \biggr] \biggl[ \prod_{j=1}^{n_-}
	   \frac{\sin(x - x_j^- - 2 \i \g)}{\sin(x - x_j^-)} \biggr] = - 1 \epc
\end{multline}
where we have already inserted $n_+ = 0$.

Let us assume for the moment that none of the factors
$\sin(x - x_j^c)$ in the numerator of the first term 
on the right hand side of equation (\ref{subsclose})
is canceled by a factor $\sin(x - x_j^h - \i \g)$ in
the denominator (no `exact particle-hole strings'). Then,
for $x \in \{x_j^c\}_{j=1}^{n_c}$, the first term on the
left  hand side vanishes, while the second term goes to
zero pointwise as $T \rightarrow 0_+$. Hence, if there are
no exact strings, close roots can only exist close to
the points $x_j^h + \i \g$ (recall (see footnote to
(\ref{defr})) that we assume that the $x_j^c$ stay away
from the integration contour, which means that
$\sin(x_m^c - x_j^c + \i \g)$ cannot be small). In other
words, for each close root $x_j^c$ there is a
$k \in \{1, \dots, n_h\}$ such that
\begin{equation} \label{phstring}
     x_j^c = x_k^h + \i \g + \i \de_k \epc
\end{equation}
where $\de_k \rightarrow 0$ for $T \rightarrow 0_+$.
Here and in the following we shall rest on the further 
technical assumption that no two particles or holes are
exponentially close to each other. Based on this assumption
we will rule out the possibility of exact strings below.

Far roots $x_j^-$ are located at $\Im x < - \g$. Inside the
strip $- 2 \g < \Im x < - \g$ they satisfy the low-temperature
subsidiary condition
\begin{multline}
     1/\fa(x) = 1/\fa^{(0)} (x) + 1/\fa^{(-)} (x) \\
        = \frac{(-1)^k}{A(x)} \re^{\frac{\e(x)} T}
           \biggl[ \prod_{j=1}^{n_h}
	   \frac{\sin(x - x_j^h - \i \g)}{\sin(x - x_j^h + \i \g)}
	   \biggr] \biggl[ \prod_{j=1}^{n_c}
	   \frac{\sin(x - x_j^c + \i \g)}{\sin(x - x_j^c)}
	   \biggr] \biggl[ \prod_{j=1}^{n_-}
	   \frac{\sin(x - x_j^-)}{\sin(x - x_j^- - \i \g)} \biggr] \\[1ex]
	  + \re^{\frac h T}
	  \biggl[ \prod_{j=1}^{n_h}
	          \frac{\sin(x - x_j^h)}{\sin(x - x_j^h + \i \g)} \biggr]
          \biggl[ \prod_{j=1}^{n_c}
	          \frac{\sin(x - x_j^c + 2 \i \g)}{\sin(x - x_j^c)} \biggr]
          \biggl[ \prod_{j=1}^{n_-}
	          \frac{\sin(x - x_j^- + \i \g)}{\sin(x - x_j^- - \i \g)} \biggr]
       = - 1
\end{multline}
according to Lemma~\ref{lem:alowentire}. Assuming that none
of the factors $\sin(x - x_j^-)$ in the numerator of the first
term on the right hand side is canceled (no exact strings)
we see that this term is zero at $x_j^-$, $j = 1, \dots, n_-$,
and conclude in a similar way as above that far roots $x_j^-$
are determined by the subsidiary condition
\begin{equation} \label{subsminus}
     \re^{ - \frac h T}
     \biggl[ \prod_{j=1}^{n_h}
             \frac{\sin(x - x_j^h + \i \g)}{\sin(x - x_j^h)} \biggr]
     \biggl[ \prod_{j=1}^{n_c}
             \frac{\sin(x - x_j^c)}{\sin(x - x_j^c + 2 \i \g)} \biggr]
     \biggl[ \prod_{j=1}^{n_-}
             \frac{\sin(x - x_j^- - \i \g)}{\sin(x - x_j^- + \i \g)} \biggr]
	     = - 1 \epc
\end{equation}
which is the same as for far roots $x_j^-$ with $\Im x_j^- <
- 2 \g$. Again the expression on the left hand side vanishes
pointwise for $T \rightarrow 0_+$. Thus, far roots $x_j^-$
can only exist close to the poles at $x_k^- - \i \g$ or at
$x_m^c - 2 \i \g$. Now choose $x_\ell^-$ such that
$\Im x_\ell^- \ge \Im x_j^-$ for $j \in \{1, \dots, n_-\}$.
Then $x_\ell^-$ cannot be located close to $x_j^- - \i \g$.
Hence, $x_\ell^-$ must be close to $x_m^c - 2 \i \g$ for some
$m \in \{1, \dots, n_c\}$. It follows that the factor
$\sin(x_m^c - x_\ell^- - 2 \i \g)$ in the numerator of $\fa^{(+)}
(x_m^c)$ is small and must be balanced by a small factor
$\sin(x_m^c - x_n^h - \i \g)$ for some $n \in \{1, \dots, n_h\}$
(recall again that we assume that the $x_j^c$ stay away from the
integration contour, which means that $\sin(x_m^c - x_j^c + \i \g)$
cannot be small). Thus, $x_\ell^-$, $x_m^c$ and $x_n^h$ form
a three-string, implying that the factor $\sin(x_\ell^- - x_n^h + \i \g)$
in $\fa^{(-)} (x_\ell^-)$ is small as well. Then, for
$x_\ell^-$, $x_m^c$ and $x_n^h$ to satisfy the subsidiary
conditions (assuming that no two holes or roots are exponentially
close to each other), the conditions
\begin{subequations}
\begin{align}
     & \frac{x_\ell^- - x_m^c + 2 \i \g}{x_\ell^- - x_n^h + \i \g} =
        1 + \frac{x_n^h - x_m^c + \i \g}{x_\ell^- - x_n^h + \i \g} =
	{\cal O} \bigl( \re^{- h/T} \bigr) \epc \\[1ex]
     & \frac{x_m^c - x_\ell^- - 2 \i \g}{x_m^c - x_n^h - \i \g} =
        1 - \frac{x_\ell^- - x_n^h + \i \g}{x_n^h - x_m^c + \i \g} =
	{\cal O} \bigl( \re^{h/T} \bigr)
\end{align}
\end{subequations}
must hold, which cannot both be true. Thus, $x_\ell^-$ cannot exist,
and $n_- = 0$.

Let us now turn to the possible locations of holes. By definition
holes $x_j^h$ are located in the strip $- \g < \Im x < 0$, where
they have to satisfy the subsidiary condition $1 + \fa^{(0)} (x)
= 0$ at low temperatures. More explicitly, if there are no
exact strings, this condition reads
\begin{equation} \label{subsholes}
     (-1)^k A(x) \re^{- \frac{\e(x)} T}
           \biggl[ \prod_{j=1}^{n_h}
	   \frac{\sin(x - x_j^h + \i \g)}{\sin(x - x_j^h - \i \g)}
	   \biggr] \biggl[ \prod_{j=1}^{n_c}
	   \frac{\sin(x - x_j^c)}{\sin(x - x_j^c + \i \g)} \biggr]
	   = - 1 \epc
\end{equation}
where we have already taken into account that $n_- = n_+ = 0$.
Thus, holes can exist in a vicinity of the curve $\Re \e (x) = 0$ 
or close to $x_j^c - \i \g$ if such point is in a region
where $\Re \e > 0$ (since $\re^{- \e(x)/ T}$ is small in that
case). Note that the latter is consistent with (\ref{phstring}).

Let us now discuss the issue of exact strings or possible
cancellations between different types of roots. We would
like to rule out the possibilities that $\fa^{(0)} (x_j^c)
\ne 0$ and that $1/\fa^{(0)} (x_j^-) \ne 0$. We may have
$\fa^{(0)} (x_j^c) \ne 0$ only if a factor
$\sin(x - x_j^c)$ in $\fa^{(0)} (x)$ is canceled by
a factor $\sin(x - x_k^h - \i \g)$ for some $k \in
\{1, \dots, n_h\}$, \emph{i.e.}\ if $x_j^c = x_k^h + \i \g$.
In this case it follows that
\begin{equation}
     \frac{\sin(x - x_k^h + \i \g)}{\sin(x - x_k^h - \i \g)}
     \frac{\sin(x - x_j^c)}{\sin(x - x_j^c + \i \g)} =
     \frac{\sin(x - x_k^h + \i \g)}{\sin(x - x_k^h)} \epc
\end{equation}
but such a pole cannot exist, since $\fa^{(0)} (x_k^h) = - 1$.
It must be canceled by $\sin(x - x_\ell^- - \i \g)$, say. Then
$x_\ell^- = x_k^h - \i \g$, and
\begin{equation} \label{dropout}
     \frac{\sin(x - x_k^h + \i \g)}{\sin(x - x_k^h - \i \g)}
     \frac{\sin(x - x_j^c)}{\sin(x - x_j^c + \i \g)}
     \frac{\sin(x - x_\ell^- - \i \g)}{\sin(x - x_\ell^-)} = 1 \epp
\end{equation}
All the factors on the left hand side of (\ref{dropout})
drop out from $\fa^{(0)}$, and $x_j^c$, $x_k^h$ and
$x_\ell^-$ form an exact three-string, $x_k^h = x_j^c - \i \g
= x_\ell^- + \i \g$.

We shall show that the existence of such three-strings
would imply that two roots would come exponentially close
to each other for small $T$: After inserting (\ref{dropout})
into (\ref{subsclose}) we see that, for $\Re \e (x_j^c) < 0$,
equation (\ref{subsclose}) can only be satisfied if there
is an $x_n^c$, such that $x_n^c - x_j^c = {\cal O} \bigl(
\re^{\Re \e(x_j^c)/T} \bigr)$ (one could imagine making the
second term in (\ref{subsclose}) exponentially large by
approaching a hole such that it would compensate the first
term, but this would also increase the size of the first
term, rendering such a compensation impossible). For
$\Re \e (x_j^c) > 0$ we can satisfy (\ref{subsclose})
only if there is an $x_n^h$ such that $x_n^h + \i \g - x_j^c =
{\cal O} \bigl( \max \{ \re^{- \Re \e(x_j^c)/T}, \re^{- h/T} \} \bigr)$,
but then $x_n^h - x_k^h$ would be of the same exponentially small
order. Finally, if $\Re \e (x_j^c) \sim 0$ then $\Re \e (x_k^h) =
\Re \e (x_j^c - \i \g) = h - \Re \e (x_j^c) \sim h$, and
it follows from $\fa^{(0)} (x_k^h) = - 1$ that there
must be an $x_n^c$ such that $x_k^h - x_n^c + \i \g =
{\cal O} (\re^{- h/T})$, and therefore also $x_n^c - x_j^c =
{\cal O} (\re^{- h/T})$. Thus, if we could exclude the
existence of roots coming exponentially close to each other
for small $T$, we could exclude the existence of the above type
of three-strings. Similar arguments hold for the remaining
case $1/\fa^{(0)} (x_j^-) \ne 0$.

For the remaining part of this work we shall simply assume
that no two roots can come exponentially close to each other,
and, as stated earlier, that no roots or holes exist close to
the real axis or close to the line $\Im x = - \g$. In other
words we shall consider the class of eigenstates of
the quantum transfer matrix characterized by particle and
hole patterns having these two properties. As we have seen
above, the patterns in this class cannot contain exact strings
or far particle roots.

After excluding the existence of far particles at low temperature
and also establishing that each close particle is combined with
a hole into a two-string, we remain with the equations
\begin{equation} \label{prehlbae1}
     \re^{- \frac h T}
     \biggl[ \prod_{k=1}^{n_h}
             \frac{\sin(x_j^c - x_k^h)}
	          {\sin(x_j^c - x_k^h - \i \g)} \biggr]
     \biggl[ \prod_{k=1}^{n_c}
             \frac{\sin(x_j^c - x_k^c - \i \g)}
	          {\sin(x_j^c - x_k^c + \i \g)} \biggr] = - 1 \epc
\end{equation}
$j = 1, \dots, n_c$,
\begin{equation} \label{prehlbae2}
     (-1)^k \re^{ - \frac{\e (x_j^h)} T
                  - \sum_{k=1}^{n_h} \ph (x_j^h,x_k^h)}
	\biggl[ \prod_{k=1}^{n_c}
	        \frac{\sin(x_j^h - x_k^c)}
		     {\sin(x_j^h - x_k^c + \i \g)} \biggr] = - 1 \epc
\end{equation}
$j = 1, \dots, n_h$, where $n_h = 2 n_c + 2 s$ and
\begin{equation} \label{phstring2}
     x_j^c = x_j^h + \i \g + \i \de_j \epc \qd j = 1, \dots, n_c \epp
\end{equation}
Here we took the liberty to relabel the holes to our
convenience.

We shall see in a minute that the $\de_j$ are exponentially
small. Hence, we can consistently remove the holes
$x_j^h$, $j = 1, \dots, n_c$, from our equations. For this
purpose we introduce the notation
\begin{align} \label{relabel}
     y_j & = x_j^c \epc & j = 1, \dots, n_c \epc \notag \\
     x_j & = x_{j + n_c}^h \epc & j = 1, \dots, n_c + 2s \epp
\end{align}
Inserting (\ref{phstring2}) and (\ref{relabel}) into
(\ref{prehlbae1}) we obtain
\begin{equation} \label{deltaexp}
     \de_j \sim \sh(\g) \re^{- \frac h T}
	\biggl[ \prod_{\substack{k = 1 \\ k \ne j}}^{n_c}
	        \frac{\sin(y_j - y_k - \i \g)}
		     {\sin(y_j - y_k)} \biggr]
	\biggl[ \prod_{k=1}^{n_c + 2s}
	        \frac{\sin(y_j - x_k)}
		     {\sin(y_j - x_k - \i \g)} \biggr] \epp
\end{equation}
This is consistently ${\cal O} (\re^{- h/T})$ if no two $y_j$
or $x_j$ are exponentially close to each other.

Inserting (\ref{phstring2}) and (\ref{relabel}) into
(\ref{prehlbae2}) and using (\ref{deltaexp}) and the functional
equation (\ref{funphi}) we obtain
\begin{subequations}
\begin{align}
     & \re^{ - \frac{\e (x_j)} T
             + \sum_{k=1}^{n_c} \ph (x_j,y_k)
             - \sum_{k=1}^{n_c + 2s} \ph (x_j,x_k)} = (-1)^{k+1}
	     + {\cal O} (T^\infty) \epc \\[1ex]
     & \re^{ - \frac{\e (y_\ell)} T
             + \sum_{k=1}^{n_c} \ph (y_\ell,y_k)
             - \sum_{k=1}^{n_c + 2s} \ph (y_\ell,x_k)} = (-1)^{k+1}
	     + {\cal O} (T^\infty) \epc
\end{align}
\end{subequations}
where $j = 1, \dots, n_c + 2s$ and $\ell = 1, \dots, n_c$. Upon
taking the logarithm we end up with
\begin{lemma}
The higher-level Bethe Ansatz equations. Up to corrections of
the order $T^\infty$ the independent holes $x_j$, $j = 1, \dots,
n_c + 2s$ and the particles in particle-hole strings $y_\ell$,
$\ell = 1, \dots, n_c$ are determined by the higher-level
Bethe Ansatz equations
\begin{subequations}
\label{hlbaes}
\begin{align}
     & \frac{\e(x_j)} T = \p \i n + \sum_{k=1}^{n_c} \ph (x_j,y_k)
                             - \sum_{k=1}^{n_c + 2s} \ph (x_j,x_k) \epc \\
     & \frac{\e(y_\ell)} T = \p \i m + \sum_{k=1}^{n_c} \ph (y_\ell,y_k)
                             - \sum_{k=1}^{n_c + 2s} \ph (y_\ell,x_k) \epc
\end{align}
\end{subequations}
where $n, m$ are even if $k$ is odd, while $n, m$ are odd if $k$ is even,
and where $- \g < \Im x_j < 0$, $0 < \Im y_\ell < \g$ by definition.
\end{lemma}
\begin{figure}
\begin{center}
\includegraphics[width=.90\textwidth]{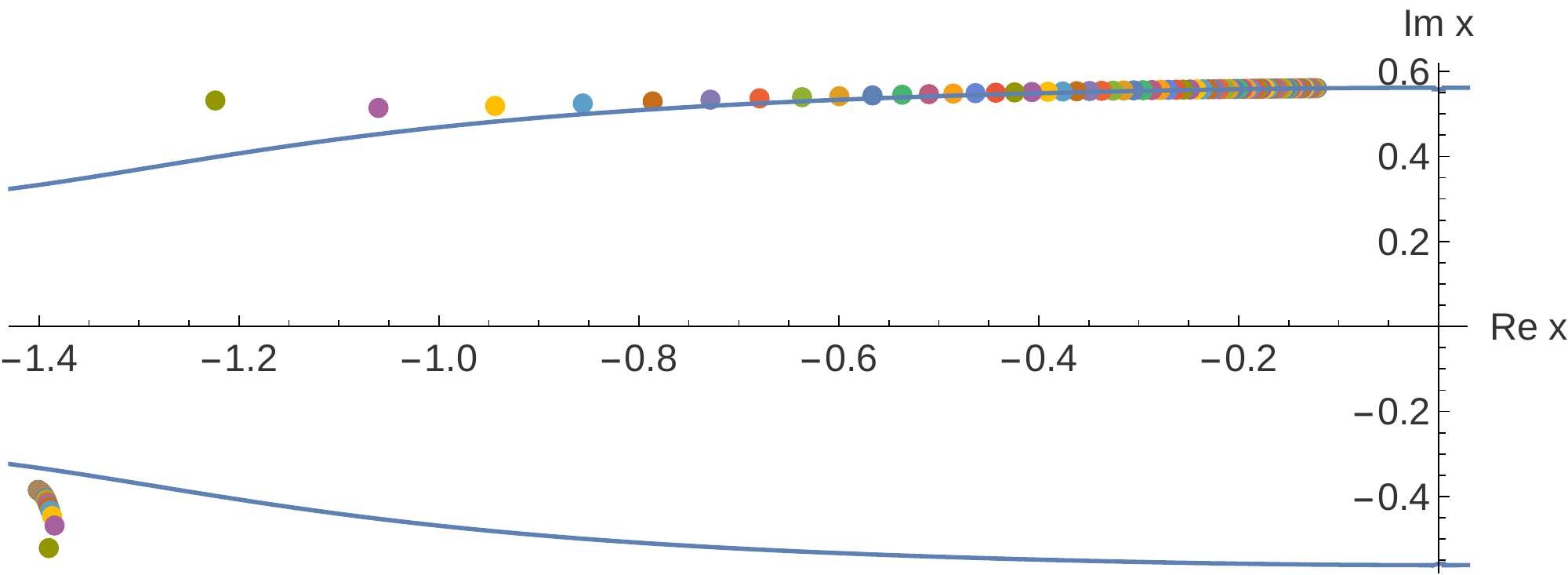}
\end{center}
\caption{\label{fig:1p1hexcitation} 
Single particle-hole pair excitations ($s = 0$, $n_c = 1$) according to
equations (\ref{hlbaes}). $T/J = 0.1$, $h/h_\ell = 2/3$, $\D = 1.7$, $h_\ell/J
= 0.76$. Shown are particle-hole pairs for $n = 1$ fixed and
$m$ running from $-1$ to $- 70$. The interaction with the particles
slightly influences the hole position. The blue lines are the curves
$\Re \e (x) = 0$.
}
\end{figure}
It is not difficult to solve the higher-level Bethe Ansatz
equations (\ref{hlbaes}) numerically. A simple example is shown in 
Figure~\ref{fig:1p1hexcitation}. The example shows $70$ single
particle-hole pairs ($s = 0$, $n_c = 1$). The quantum number $n$
of the hole is always $1$, the quantum number $m$ of the
particle varies from $-1$ to $-70$. The particle and the hole
in each pair are depicted by dots of the same colour.

For the calculation of correlation lengths below we need to
calculate integrals over $\cal C$ that involve the auxiliary function
$\fa$. Hence, we need $\fa$ on $\cal C$. Using the low-temperature
picture obtained above, we obtain 
\begin{equation} \label{asyaux}
     \fa (x) = (-1)^k \re^{- \frac{\e (x)} T
             + \sum_{k=1}^{n_c} \ph (x,y_k)
             - \sum_{k=1}^{n_c + 2s} \ph (x,x_k)}
	       \bigl(1 + {\cal O} (T^\infty) \bigr)
\end{equation}
for $x \in {\cal C}$.

\begin{figure}
\begin{center}
\includegraphics[width=.70\textwidth]{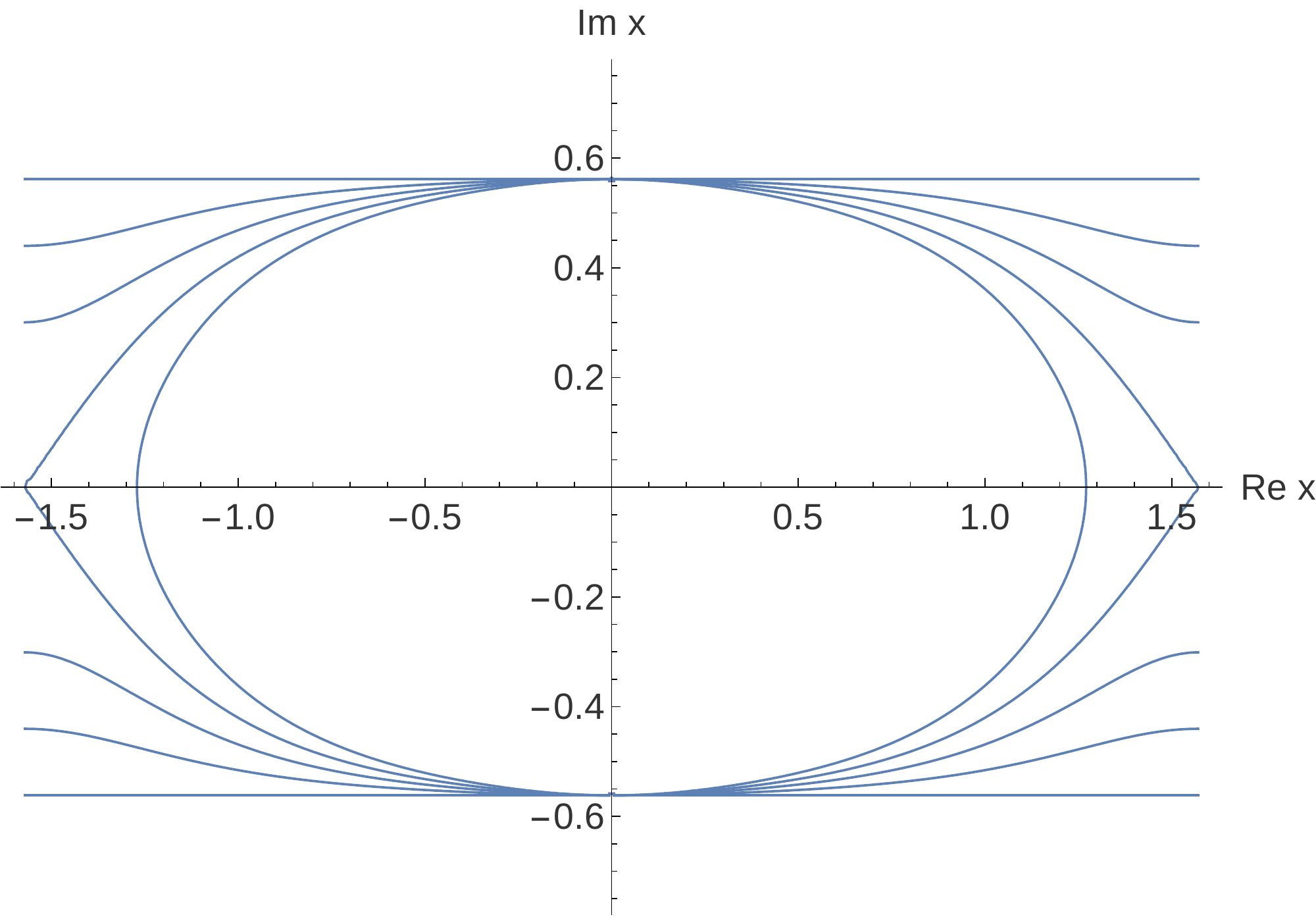}
\end{center}
\caption{\label{fig:regimes} The curves $\Re \e (x) = 0$ for various
values of the magnetic field. Here $\D = 1.7$, $h_\ell/J = 0.76$.
The values of the magnetic field decrease proceeding from the inner
to the outer curve: $h/h_\ell = 1.34, 1, 2/3, 1/3, 0$. The curves are
closed for $h_\ell < h < h_u$. At the lower critical field $h =
h_\ell$ they develop two cusps, and a gap opens for $0 < h < h_\ell$.}
\end{figure}

We believe that at sufficiently low temperatures and finite
magnetic field our self-consistent solutions to the nonlinear
integral equations, described by the higher-level Bethe equations
(\ref{hlbaes}) and by the asymptotic auxiliary function
(\ref{asyaux}), are complete. This means, in particular, that
two particles or holes are not exponentially close to each
other and not exponentially close to the real axis or to the
line $\Im x = - \g$ for $T \rightarrow 0_+$, for all solutions
leading to finite eigenvalue ratios in this limit. This belief
is supported by numerical calculations for finite Trotter number
(see Appendix~\ref{app:numerics}) and by the free Fermion
picture that arises in the limit $T \rightarrow 0$. It leads
us to the following
\begin{conjecture*}
For low temperatures at finite magnetic field every eigenstate of
the quantum transfer matrix is parameterized by one of the solutions
of the higher-level Bethe equations (\ref{hlbaes}). This implies that
all eigenstates can be interpreted in terms of particle-hole
excitations.
\end{conjecture*}
In the limit $T \rightarrow 0_+$ at finite $s$ and $n_c$ the higher-%
level Bethe Ansatz equations (\ref{hlbaes}) decouple, $\i \p n T$
and $\i \p m T$ turn into independent continuous variables, and the
particles and holes become free parameters on the curves
\begin{subequations}
\begin{align}
     & \Re \e (y) = 0 \epc \qd 0 < \Im  y < \g \epc \\
     & \Re \e (x) = 0 \epc \qd - \g < \Im x < 0 \epp
\end{align}
\end{subequations}
These curves are shown in Figure~\ref{fig:regimes}. Clearly, the
massive regime is distinguished from the massless regime by the
opening of a `band gap' at the critical field $h_\ell$.

\begin{figure}
\begin{center}
\includegraphics[width=.70\textwidth]{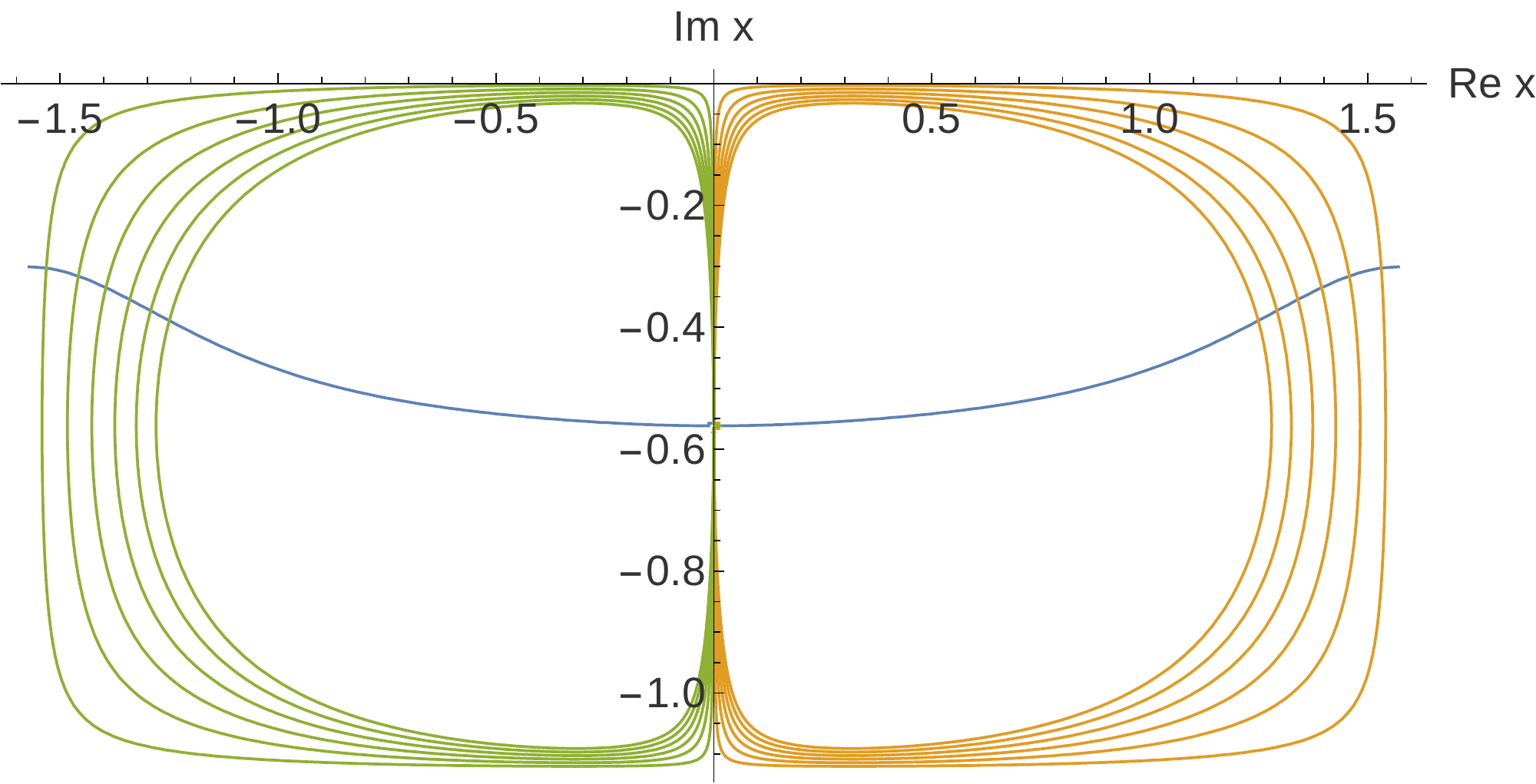}
\end{center}
\caption{\label{fig:dominantbr} 
Bethe roots of the dominant state according to (\ref{brdominant})
depicted as the intersections of the curves $\Re \eps (x) = 0$
and $\Im \eps (x) = n \p T$ for $T/J = 0.01$, $n = \pm 1,
\pm 3, \dots, \pm 11$, $h/h_\ell = 2/3$, $\D = 1.7$, $h_\ell/J
= 0.76$. For growing $|n|$ the sequence of roots becomes more
and more dense with an accumulation point at $- \i \g/2$ (intersection
point of the curve $\Re \eps (x) = 0$ with the imaginary axis).
}
\end{figure}

The case $s = n_h = 0$ is special. In this case there are no
higher-level Bethe Ansatz equations, and the auxiliary function
is $\fa (x) = \pm \re^{- \eps (x)/T}$. The Bethe roots of the
corresponding states are determined by $\fa (x) = \mp 1$ for
$- \g < \Im x < 0$, or
\begin{equation} \label{brdominant}
     \eps (x) = \i \p n T \epc \qd - \g < \Im  x < 0 \epc
\end{equation}
where the $n$ are odd integers if $\fa (x) = \re^{- \eps (x)/T}$
and even integers in the other case. Thus, the Bethe Ansatz
equations decouple for $s = n_c = 0$ (up to corrections of
order $T^\infty$). We shall identify the corresponding states
as the dominant state and a state which is degenerate up
to corrections of the order $T^\infty$. An example of a Bethe
root pattern of the dominant state is depicted in
Figure~\ref{fig:dominantbr}.

With our low-temperature approximation (\ref{asyaux}) we can also
check the consistency of the phase $k$ defined in (\ref{phasek}).
Let us consider the simple example $s = n_c = 0$ and $k$ odd.
Recall that we agreed to restrict the Bethe roots $x^r$ to
$- \p/2 < \Re x^r \le \p/2$. Thus, the contours ${\cal C}_\ell$
and ${\cal C}_r$ should be slightly shifted (say, by a quantity of
order $T^2$) on the line $\Re \e(x) = 0$. This way one excludes
the zero of $1 + \fa$ which, for $k$ odd, has its real part
very close (up to a correction of order $T^\infty$) to $- \p/2$.
At the same time one includes the $\p$-shifted zero close to~%
${\cal C}_r$. With this convention for ${\cal C}_\ell$ we
can now calculate the phase $k$. First of all,
\begin{multline} \label{pvphase}
     \int_{{\cal C}_\ell} \rd y \: \6_y \ln (1 + \fa (y)) =
       \PV \int_{-\p/2}^{-\p/2 - \i \g^-} \rd y \: \6_y \ln (1 + \fa (y)) - \i \p \\ =
       \Ln (1 + \fa (-\p/2 - \i \g^-)) - \Ln (- 1 - \fa (-\p/2)) - \i \p \epp
\end{multline}
Here we took into account (\ref{asyaux}), which tells us that
(up to a multiplicative correction of order $1 + T^\infty$)
$\fa$ is real on ${\cal C}_\ell$ with large negative values
at $- \p/2$ and small positive values at $- \p/2 - \i \g^-$.
It further follows that the arguments of the logarithms on the
right hand side of (\ref{pvphase}) have large positive real parts
for small $T$. Inserting (\ref{pvphase}) into (\ref{phasek}) 
and using that $\ln \fa(- \p/2) = - \e(- \p/2)/T - \i \p k$
by (\ref{u1}), we obtain that $k = 1$ (which is consistent
with $k$ odd). Similar arguments apply for $s = n_c = 0$ and $k$
even, in which case we find that $k = 0$.

\subsection{Root patterns and higher-level Bethe Ansatz
equations for zero magnetic field}
We would like to briefly comment on the case of vanishing
magnetic field, $h = 0$. We shall maintain our assumption
that there are no exact strings. If $h = 0$ the function
$\fa^{(+)}$ becomes
\begin{multline}
     \fa^{(+)} (x) = 
           \biggl[ \prod_{j=1}^{n_h}
	   \frac{\sin(x - x_j^h)}{\sin(x - x_j^h - \i \g)} \biggr] \\ \times
	   \biggl[ \prod_{j=1}^{n_c}
	   \frac{\sin(x - x_j^c - \i \g)}{\sin(x - x_j^c + \i \g)} \biggr]
	   \biggl[ \prod_{j=1}^{n_+}
	   \frac{\sin(x - x_j^+ - \i \g)}{\sin(x - x_j^+ + \i \g)} \biggr]
	   \biggl[ \prod_{j=1}^{n_-}
	   \frac{\sin(x - x_j^- - 2 \i \g)}{\sin(x - x_j^-)} \biggr] \epp
\end{multline}
If there are no exact strings, then $\fa^{(0)} (x_j^c) = 0$, and
the close roots and the far roots $x_j^+$ are determined by
\begin{equation} \label{hnullplusclose}
     \fa^{(+)} (x_j^c) = - 1 \epc \qd \fa^{(+)} (x_j^+) = - 1 \epp
\end{equation}
This means that we cannot exclude the existence of far roots
$x_j^+$ in this case, neither can we conclude as before that close
roots form particle-hole strings. Similarly, far roots $x_j^-$ may
exist as well and satisfy
\begin{equation} \label{hnullminus}
     \fa^{(-)} (x_j^-) = - 1 \epc
\end{equation}
where $\fa^{(-)} (x) = \fa^{(+)} (x + \i \g)$.

Holes must satisfy the equation
\begin{equation} \label{hnullhole}
     \fa^{(0)} (x_j^h) = - 1 \epc
\end{equation}
where
\begin{multline}
     \fa^{(0)} (x) = (-1)^k A(x) \re^{- \frac{\e(x)} T}
           \biggl[ \prod_{j=1}^{n_h}
	   \frac{\sin(x - x_j^h + \i \g)}{\sin(x - x_j^h - \i \g)}
	   \biggr] \\ \times \biggl[ \prod_{j=1}^{n_c}
	   \frac{\sin(x - x_j^c)}{\sin(x - x_j^c + \i \g)}
	   \biggr] \biggl[ \prod_{j=1}^{n_-}
	   \frac{\sin(x - x_j^- - \i \g)}{\sin(x - x_j^-)} \biggr]
	   \biggl[ \prod_{j=1}^{n_+}
	   \frac{\sin(x - x_j^+)}{\sin(x - x_j^+ + \i \g)} \biggr] \epp
\end{multline}
Clearly holes may exist close to the line $\Im x = - \g/2$,
where $\Re \e (x) = 0$. Let us discuss the situation away
from this line. Suppose that $\Re \e (x_j^h) < 0$. Then, for
(\ref{hnullhole}) to hold, there must be an $x_k^-$ such that
$x_j^h - x_k^- - \i \g \rightarrow 0$ as $T \rightarrow 0_+$.
This implies that $\fa^{(-)} (x_k^-) \rightarrow 0$ which
cannot be. If, on the other hand, $\Re \e (x_j^h) > 0$, then
(\ref{hnullhole}) can only hold if there is an $x_k^c$ such
that $x_j^h - x_k^c + \i \g \rightarrow 0$ as $T \rightarrow 0_+$.
But this implies that $1/\fa^{(+)} (x_k^c) \rightarrow 0$
which is impossible.

Thus, holes can exist only close to the line $\Im x = - \g/2$.
As $T \rightarrow 0_+$ they densely fill the line segment
$- \i \g/2 + [- \p/2, \p/2]$. 
This means that the holes become
free parameters as $T \rightarrow 0_+$, whereas close roots
and far roots remain constrained by the higher-level Bethe
Ansatz equations (\ref{hnullplusclose}), (\ref{hnullminus}).
This is in stark contrast to the case of finite magnetic
field where both, particle and hole parameters, become free
for $T \rightarrow 0_+$.

Note that the higher-level Bethe Ansatz equations become
uniform if we perform the following change of variables,
\begin{equation}
     \{\chi_j\}_{j=1}^{n_\chi} = \{x_j^c - \i \g/2 \}_{j=1}^{n_c}
	\cup \{x_j^+ - \i \g/2 \}_{j=1}^{n_+}
	\cup \{x_j^- + \i \g/2 \}_{j=1}^{n_-} \epp
\end{equation}
With these variables the higher-level Bethe Ansatz equations
(\ref{hnullplusclose}), (\ref{hnullminus}) turn into
\begin{equation} \label{hlbaehnull}
     - 1 = \biggl[ \prod_{k=1}^{n_h}
	   \frac{\sin(\chi_j - x_k^h + \i \g/2)}
	        {\sin(\chi_j - x_k^h - \i \g/2)} \biggr]
           \biggl[ \prod_{j=1}^{n_\chi}
	   \frac{\sin(\chi_j - \chi_k - \i \g)}
	        {\sin(\chi_j - \chi_k + \i \g)} \biggr] \epc
\end{equation}
$j = 1, \dots, n_\chi$, where
\begin{equation}
     n_\chi = n_c + n_f = \frac{n_h}{2} - s \epp
\end{equation}
Equations (\ref{hlbaehnull}) are of the same form as the
higher-level Bethe Ansatz equations for the ordinary
transfer matrix in the antiferromagnetic massive
regime \cite{BVV83,ViWo84} (see also our recent discussion
in \cite{DGKS14app}).

\subsection{Correlation lengths at low temperatures and finite
magnetic field}
We now turn back to finite magnetic field $h > 0$.
Starting from equation (\ref{baev}) which expresses the quantum
transfer matrix eigenvalues in terms of Bethe roots and
employing a similar reasoning as in the derivation of the
nonlinear integral equations in Section~\ref{subsec:nlie}
we obtain the representation
\begin{multline} \label{evirep}
     \La (x) = \biggl( \frac{\cos(\i \g/2 + x)}{\cos(\i \g/2 - x)} \biggr)^d
	\biggl[ \prod_{j=1}^{n_h}
	        \frac{\sin(x - x_j^h - \i \g/2)}
		     {\sin(x - x_j^h + \i \g/2)} \biggr]
	\biggl[ \prod_{j=1}^{n_c}
	        \frac{\sin(x - x_j^c + 3 \i \g/2)}
		     {\sin(x - x_j^c - \i \g/2)} \biggr] \\ \times
        \exp \biggl\{ \frac{h}{2 T}
	            - \int_{\cal C} \rd y \: K(x - y|\g/2)
		         \ln_{\cal C} (1 + \fa(y)) \biggr\}
\end{multline}
valid for $- \g/2 < \Im x < \g/2$.
This is the general expression for $\La$, still valid for any
temperature and magnetic field $h \ge 0$, in the case that
there are no far roots. If far roots are present, an additional
factor of
\begin{equation}
     \prod_{j=1}^{n_f}
        \frac{\sin(x - x_j^f + \i \g/2)}
             {\sin(x - x_j^f - \i \g/2)}
\end{equation}
appears on the right hand side of (\ref{evirep}).

In the low-$T$ limit at $h > 0$ there are no far roots,
and the close roots form strings with holes. Using
(\ref{phstring2}) and (\ref{relabel}) as well as the
facts that $\fa^{-1} (x) = {\cal O} (T^\infty)$
for $x \in {\cal C}_+$ and $\fa (x) = {\cal O} (T^\infty)$
for $x \in {\cal C}_-$ in (\ref{evirep}) we obtain
\begin{multline} \label{evireplowt}
     \La (x) = \biggl( \frac{\cos(\i \g/2 + x)}{\cos(\i \g/2 - x)} \biggr)^d
	\biggl[ \prod_{j=1}^{n_c + 2s}
	        \frac{\sin(x - x_j - \i \g/2)}
		     {\sin(x - x_j + \i \g/2)} \biggr]
	\biggl[ \prod_{j=1}^{n_c}
	        \frac{\sin(x - y_j + \i \g/2)}
		     {\sin(x - y_j - \i \g/2)} \biggr] \\ \times
        \exp \biggl\{ \frac{h}{2 T}
	             + \int_{- \p/2}^{\p/2} \rd y \: K(x - y|\g/2)
		          \ln \fa(y) \biggr\}
			  \bigl(1 + {\cal O} (T^\infty) \bigr) \epp
\end{multline}
We shall denote the eigenvalue with $n_c = s = k = 0$ by $\La_0 (x)$.
This eigenvalue has the representation
\begin{equation} \label{domina}
     \La_0 (x) =
         \exp \biggl\{
           \frac{h}{2 T}
	 - \frac 1 T \int_{- \p/2}^{\p/2} \rd y \: K(x - y|\g/2) \e (y) \biggr\}
	   \bigl(1 + {\cal O} (T^\infty) \bigr) \epp
\end{equation}
We shall argue below that this eigenvalue is the dominant eigenvalue.
We are interested in the ratio $\r (0|\a)$ of the eigenvalue of
a twisted excited state (with $h \rightarrow h' = h - 2 \a \g T$,
$\a$ twist parameter) and the untwisted dominant state. Using
(\ref{evireplowt}) we find that, up to corrections of the order
${\cal O} (T^\infty)$,
\begin{multline} \label{ratiobareform}
     \r (0|\a) = (-1)^k
	\biggl[ \prod_{j=1}^{n_c + 2s}
	        \frac{\sin(x_j + \i \g/2)}
		     {\sin(x_j - \i \g/2)} \biggr]
	\biggl[ \prod_{j=1}^{n_c}
	        \frac{\sin(y_j - \i \g/2)}
		     {\sin(y_j + \i \g/2)} \biggr] \\ \times
        \exp \biggl\{ \int_{- \p/2}^{\p/2} \rd x \: K(x|\g/2)
	              \biggl[
		         \sum_{k=1}^{n_c} \ph (x,y_k)
			 - \sum_{k=1}^{n_c + 2s} \ph (x,x_k)
		      \biggr] \biggr\} \epp
\end{multline}
Note that the dependence on the twist parameter $\a$ has been 
entirely absorbed by the roots $x_j$, $y_j$.

Equation (\ref{ratiobareform}) can be further simplified. For
this purpose we recall (see e.g.\ \cite{DGKS14app}) that the
momentum $p$ satisfies the linear integral equation
\begin{equation} \label{linp}
     p(x) = \frac{p_0 (x)}{2 \p} - \frac{\th(x - \p/2)}{4 \p \i}
            - \int_{- \p/2}^{\p/2} \rd y \: K(x - y) p(y) \epc
\end{equation}
where $p_0 (x) = - \i \th(x|\g/2)$ (for an explicit form of the
solution of the integral equation see Appendix~\ref{app:momentum}).
Combining (\ref{linp}) with (\ref{dressphase}) we obtain
\begin{equation} \label{linpphi}
     p(x) + \frac{\ph(x,\p/2)}{2 \p \i} = \frac{p_0 (x)}{2 \p} 
            - \int_{- \p/2}^{\p/2} \rd y \: K(x - y)
	      \biggl( p(y) + \frac{\ph(y,\p/2)}{2 \p \i} \biggr) \epp 
\end{equation}
Then, applying partial integration and the dressed function trick
to the integral in (\ref{ratiobareform}), we find that
\begin{equation}
     \int_{- \p/2}^{\p/2} \rd x \: K(x|\g/2) \ph(x, z) =
        \i \p + 2 \p \i p(z) - \i p_0 (z) \epp
\end{equation}

Using the latter equation in (\ref{ratiobareform}) we arrive at the
main result of this work which is a formula for the eigenvalue ratios
$\r (0|\a)$ in the antiferromagnetic massive regime at finite
magnetic field in terms of the particle and hole roots $y_j$,
$x_k$ determined by the higher-level Bethe Ansatz equations
(\ref{hlbaes}):
\begin{align} \label{evarat} \notag
     \r (0|\a) & = (-1)^k \exp \biggl\{
                    2 \p \i \Bigl[ \sum_{j=1}^{n_c} p(y_j)
		                 - \sum_{j=1}^{n_c + 2s} p(x_j) \Bigr] \biggr\} \\[1ex]
               & = (-1)^{k}
	\biggl[ \prod_{j=1}^{n_c}
	        \frac{\dh_1(y_j - \i \g/2|q^2)}
		     {\dh_4(y_j - \i \g/2|q^2)} \biggr]
	\biggl[ \prod_{j=1}^{n_c + 2s}
	        \frac{\dh_4(x_j - \i \g/2|q^2)}
		     {\dh_1(x_j - \i \g/2|q^2)} \biggr] \epc
\end{align}
this being valid up to multiplicative corrections of the order
$\bigl(1 + {\cal O} (T^\infty)\bigr)$. Remarkably, the second
representation of the eigenvalue ratios in (\ref{evarat}) is
completely explicit in terms of the particle and hole parameters.

\begin{figure}
\begin{center}
\includegraphics[width=.85\textwidth]{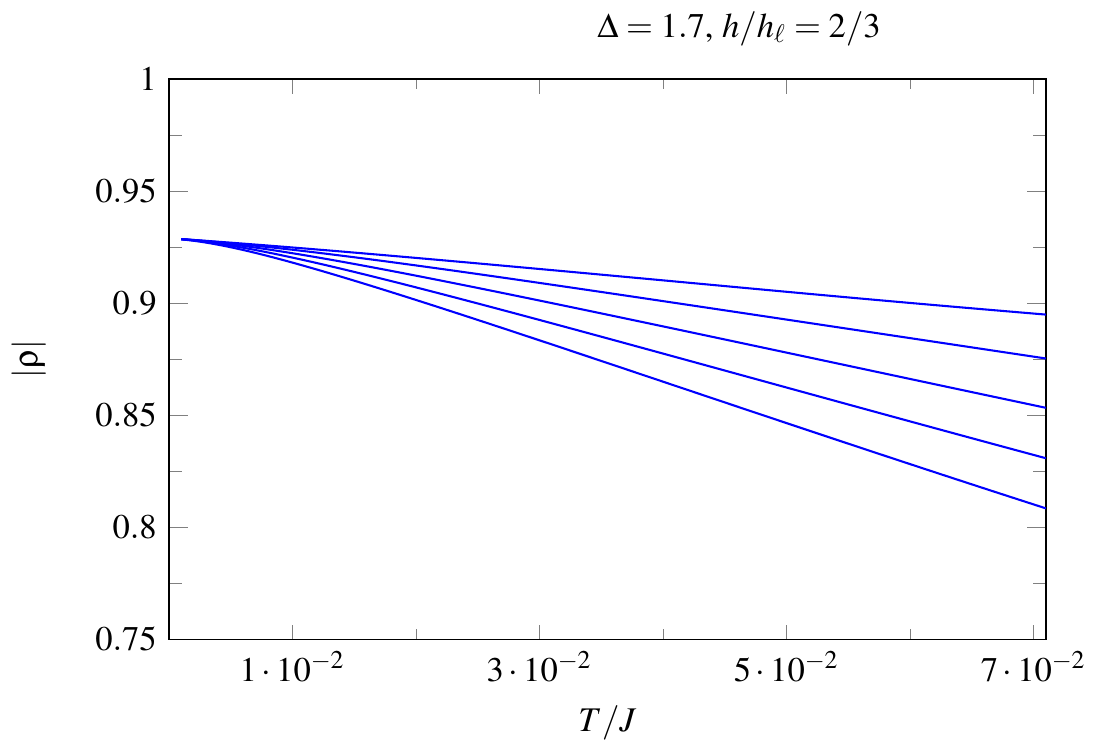}
\end{center}
\caption{\label{fig:cls} 
The behaviour of $|\r|$ as a function of temperature
for one hole and one particle. The particle and
hole roots are obtained from the higher-level Bethe Ansatz
equations (\ref{hlbaes}) for $n = 1$, $m = -1, -3, -5 , -7, -9$
(from top to bottom) and inserted into (\ref{evarat}). The
parameters are chosen as $h/h_\ell = 2/3$, $\D = 1.7$,
$h_\ell/J = 0.76$, $\a = 0$.
}
\end{figure}%
Setting $s = n_c = 0$ we obtain $\r (0|\a) = \pm 1$. For
all other states not both, $s$ and $n_c$, are zero. Hence,
using (\ref{compareimpi}) in (\ref{evarat}) we conclude that
$|\r (0|\a)| < 1$ for all other states. Then, referring to
our conjecture of the previous subsection, one of the
states with $s = n_c = 0$ must be the dominant state (or
the dominant state is two-fold degenerate). It is possible
to see, by solving the nonlinear integral equation
(\ref{nlie}) numerically, that the dominant state corresponds
to $k = 0$ or, equivalently, to $\fa (x) = \re^{- \eps (x)/T}$
and $\La_0 (x)$, equation (\ref{domina}), at low temperatures.
The gap between the dominant state and the almost degenerate
state is of the order $T^\infty$.
\subsection{Discussion}
The behaviour of the correlation lengths as functions of
temperature can be extracted from equations (\ref{hlbaes}) and
(\ref{evarat}). An example is shown in Figure~\ref{fig:cls}.
In this example $s = 0$ and $n_c = 1$ as in Figure~%
\ref{fig:1p1hexcitation}. States with these quantum numbers
appear in the form factor expansion of the longitudinal
two-point functions $\<\s_1^z \s_n^z\>$. The figure shows
$|\r|$ where $\r = \r(0,0)$. The corresponding correlation
lengths are $\x = - 1/\ln |\r|$. The excitations leading to
the eigenvalue ratios shown in the figure have
quantum numbers $n=1$ and $m = - 1, -3, -5, -7, -9$
from top to bottom. The top curve corresponds to the
leading correlation length. As expected $|\r|$ and hence
the corresponding correlation length $\x$ decreases with
increasing temperature. For $T \rightarrow 0_+$ infinitely
many correlation lengths degenerate and the spectrum becomes
dense. Thus, a summation (or integration) over these
infinitely many contributions is necessary for the calculation
of the two-point functions at zero temperature. This will
result in correlation lengths of the zero-temperature
correlation functions which are different from the values of
$\x$ for $T \rightarrow 0_+$. In other words, our above formulae
determine the physical correlation lengths only for small
but finite temperatures.

Nevertheless it is interesting to calculate the value of
$|\r|$ for $T \rightarrow 0_+$. It can be obtained
from the energy-momentum relation (\ref{enmomrel}). Denoting
the particle position by $y$ and the hole position by $x$
we have, according to (\ref{evarat}), $|\r| = |\re^{2 \p \i p (y)}|
|\re^{- 2 \p \i p (x)}|$, where $x$ and $y$ are determined
by the higher-level Bethe Ansatz equations (\ref{hlbaes}).
For $T \rightarrow 0_+$ they simplify to $\e (x) = \e (y) = 0$.
For this reason we can determine $|\r|$ from (\ref{enmomrel})
by setting the left hand side equal to zero and solving for
$\re^{2 \p \i p}$. Equations (\ref{compareimpi}) allow us to
detect which solution of the quadratic equation belongs to the
particle and which one to the hole. After an elementary
calculation we end up with
\begin{equation}
     |\r| = \Biggl[
            \sqrt{\frac{1}{k^2} - \biggl( \frac{1}{k^2} - 1 \biggr)
                  \biggl( \frac{h}{h_\ell} \biggr)^2} -
            \sqrt{\biggl( \frac{1}{k^2} - 1 \biggr)
                  \biggl( 1 - \biggl( \frac{h}{h_\ell} \biggr)^2 \biggr)}
            \Biggr]^2
\end{equation}
which reduces to a well-known result \cite{JKM73} for $h = 0$.
The monotonic behaviour as a function of $h$ with maximum $|\r| = 1$
at $h = h_\ell$ corresponds to the linear decrease of the mass gap
(see (\ref{enmomrel})) as a function of the magnetic field. At
$h = h_\ell$ the mass gap closes, and the correlation length
diverges in accordance with our intuition.

\section{Conclusions}
\label{sec:conclusions}
We have considered the spectrum of the quantum transfer
matrix of the XXZ chain in the antiferromagnetic massive
regime and have worked out in detail the case of low
temperatures at finite magnetic field $0 < h < h_{\ell}$.

The nonlinear integral equation for a suitable auxiliary function
was cast into a form such that the low-temperature limit could be
immediately taken. A careful analysis of the resultant simplified
equations in the entire complex rapidity plane showed that roots
away from the distribution center (far roots) cannot exist. More
importantly, our analysis suggests that at low temperatures all
excitations can be classified as particle-hole excitations. The
ratios of the eigenvalues to the dominant eigenvalue of the quantum
transfer matrix turned out to be explicit functions of the particle
and hole parameters which satisfy a set of higher-level Bethe Ansatz
equations at small finite~$T$. These parameters become free on
two curves in the complex plane as $T \rightarrow 0_+$. We
conjecture that for finite magnetic field at low temperatures all
eigenvalue ratios are of this form.

Our conjecture implies that only particle-hole excitations have to
be taken into account in the calculation of correlation functions
by means of a form factor expansion. This is rather reminiscent of
previous studies of the massless regime \cite{DGK13a,DGK14a}. Still,
there are two differences. The configurations of particle-hole pairs
differ in that, in the massive case considered here, another hole is
rigidly attached in a string-like manner to every particle. Yet, this
may be seen as an artefact related to our choice of the integration
contour. More importantly, the summation of all particle-hole excitations
may have a different meaning in the massive and in the massless case.
In the massless case it is based on a `critical form factor summation
formula'. In the context of the analysis of the large-distance
asymptotics of correlation functions this formula first appeared in
\cite{KMS11b}, where it was applied to the interacting Bose gas at
finite temperature. A proof and an analysis of the ground state two-point
functions of the XXZ chain in the critical regime were supplied in
\cite{KKMST11b}, and an analysis of the finite temperature case
followed in \cite{DGK13a,DGK14a}. In those works it was argued that
a restriction of the summation to the gapless particle-hole excitations
would give the low-energy long-wave length contribution to the
two-point functions and hence their large-distance asymptotics.
By way of contrast, we expect that a summation over all particle-hole
contributions in the massive antiferromagnetic regime will give
the full two-point functions at any distance in a similar way as
in the zero temperature case \cite{JiMi95,DGKS14app}.
 
We believe that we are now in the position to study two-point
correlation functions at low temperatures in the antiferromagnetic
massive regime. We hope to report first progress in near future.
\\[1ex]
\noindent {\bf Acknowledgment.}
The authors are grateful to A. Kl\"umper for helpful discussions and 
for his interest in this work. MD and FG acknowledge financial support by
the Volkswagen Foundation and by the Deutsche Forschungsgemeinschaft
under grant number Go 825/7-1. KKK is supported by the CNRS. His work has
been partly financed by a Burgundy region PARI 2013-2014 FABER grant
`Structures et asymptotiques d'int\'egrales multiples'. KKK also enjoys
support from the ANR `DIADEMS' SIMI 1 2010-BLAN-0120-02. JS is
supported by a JSPS Grant-in-Aid for Scientific Research (C) No.\ 15K05208.

\clearpage

\setcounter{footnote}{0}

{\appendix
\Appendix{Basic functions}%
\label{app:functions}
In this appendix we gather explicit representations of the basic
functions appearing in the low-temperature analysis of the
antiferromagnetic massive regime. These are the momentum, the
dressed energy and the dressed phase. Fourier series are the
starting point for the derivation of the various representations
and properties of these functions. The Fourier coefficients can
be directly obtained from the respective linear integral
equations by means of the convolution theorem. Derivations
of most of the formulae can be found in the appendices of
\cite{DGKS14app}.
\subsection{Momentum}
\label{app:momentum}
Fourier series representation:
\begin{equation}
     p(x) = \4 + \frac x {2 \p}
            + \sum_{n=1}^\infty \frac 1 {n \p} \;
	                        \frac{\sin(2 n x)}{q^n + q^{-n}} \epp
\end{equation}
Representation in terms of Jacobi-theta functions:\footnote{We are
using the conventions of Whittaker and Watson \cite{WhWa63} for
Jacobi-theta functions and elliptic functions.}
\begin{equation} \label{ptheta4}
      p(x) = \4 + \frac x {2 \p}
             + \frac 1 {2\p\i} \ln \biggl(
	       \frac{\dh_4 (x + \i \g/2|q^2)}{\dh_4 (x - \i \g/2|q^2)}
	       \biggr) \epp
\end{equation}
Behaviour above and below the real axis:
\begin{equation}
\label{compareimpi}
\begin{split}
     & \Re \bigl[ 2 \p \i p(x) \bigr] < 0 \qd
       \text{for}\ 0 < \Im x < \g \epc \\[.5ex]
     & \Re \bigl[ 2 \p \i p(x) \bigr] > 0 \qd
       \text{for}\ - \g < \Im x < 0 \epp
\end{split}
\end{equation}
From (\ref{ptheta4}) we obtain the formula
\begin{equation} \label{cospissn}
     \cos \bigl( 2 \p p(x)\bigr) =
        - \sn \biggl( \frac{2 K x}{\p} \bigg| k \biggr) \epc
\end{equation}
where now $k$ is the elliptic modulus, $K = K(k)$ is the complete elliptic
integral of the first kind and $K' = K(\sqrt{1 - k^2})$. The elliptic
modulus parameterizes $\g$ as $\g = \p K'/K$. The function $\sn$ is the
Jacobi-elliptic $\sn$-function.
\subsection{Dressed energy}
\label{app:energy}
The dressed energy is the solution of the linear integral equation
(\ref{lineps}). It has the following series representations.

\noindent Fourier series representation:
\begin{equation}
     \e(x) = \frac h 2 - 4 J \sh(\g)
             \sum_{n \in {\mathbb Z}} \frac{\re^{2 \i n x}}{q^n + q^{-n}} \epp
\end{equation}
Poisson-resummed series:
\begin{equation} \label{epspoisson}
     \e (x) = \frac h 2 - \frac{2 \p J \sh(\g)}{\g}
              \sum_{n \in {\mathbb Z}}
	      \frac{1}{\ch \bigl(\frac \p \g (x - n \p)\bigr)} \epp
\end{equation}

Equation (\ref{epspoisson}) defines the dressed energy as a meromorphic
function in the complex plane. We see from this formula that $\e$ is
double periodic. It can be expressed it in terms of the Jacobi-dn function.
\begin{equation} \label{epsdn}
     \e(x) = \frac h 2 - \frac{4 J K \sh (\g)}{\p}
             \dn \biggl( \frac{2 K x}{\p} \bigg| k \biggr)
\end{equation}
From (\ref{epspoisson}) we can also readily read off the functional equation
\begin{equation} \label{funeps}
     \e(x) + \e(x + \i \g) = h
\end{equation}
and the periodicity
\begin{equation}
     \e(x + \p) = \e(x) \epp
\end{equation}

Let us recall that the lower critical field $h_\ell$ is defined
by the condition $\e (\p/2)~=~0$, implying that
\begin{equation} \label{defhell}
     h_\ell = \frac{8 J K \sh (\p K'/K)}{\p} \dn (K|k) \epp
\end{equation}
One can show that $\e$ is even and monotonously increasing for
$0 < x < \p/2$. Hence, $0 < h < h_\ell$ implies that $\e (x) < 0$
for all $x \in [- \p/2, \p/2]$, and it follows that $\Re \bigl( - T
\ln (\fa) \bigr)$ is positive on ${\cal C}_+$ if $T$ is small enough.
Then (\ref{epspoisson}) implies that $\e(x + \i \g) = h - \e (x) > 0$,
whence $\Re \bigl( - T \ln (\fa) \bigr)$ is negative on ${\cal C}_-$
if only $T$ is small enough. These were the conditions from which
we started our low-temperature analysis. They are self-consistently
satisfied if $0 \le h < h_\ell$, i.e.\ if we are in the 
antiferromagnetic massive regime.

Reinserting (\ref{defhell}) into (\ref{epsdn}) and using (\ref{cospissn})
we obtain the energy-momentum relation
\begin{equation} \label{enmomrel}
     \e (p) = \frac h 2
        - \frac{h_\ell} 2 \sqrt{\frac{1 - k^2 \cos^2 (2 \p p)}{1 - k^2}} \epp
\end{equation}
\subsection{Dressed phase}
\label{app:phase}
Fourier series representation of $\ph(*,z)$ for $|\Im z| < \g$:
\begin{equation}
     \ph(x,z) = \i (\p/2 + x - z)
                + \sum_{n=1}^\infty \frac{2 \i} n \;
		  \frac{\sin\bigl(2 n (x - z)\bigr)}{1 + q^{- 2n}} \epp
\end{equation}
This series can be resummed and expressed in terms of $q$-$\G$ functions:
\begin{equation}
     \ph(x_1, x_2) = \i \Bigl( \frac \p 2 + x_{12} \Bigr)
        + \ln \Biggl\{ \frac{\G_{q^4} \bigl(1 + \frac{\i x_{12}}{2\g}\bigr)
	                     \G_{q^4} \bigl(\2 - \frac{\i x_{12}}{2\g}\bigr)}
		            {\G_{q^4} \bigl(1 - \frac{\i x_{12}}{2\g}\bigr)
			     \G_{q^4} \bigl(\2 + \frac{\i x_{12}}{2\g}\bigr)}
			     \Biggr\} \epc
\end{equation}
where $x_{12} = x_1 - x_2$ and $|\Im x_2| < \g$. Here $\G_q$ is
defined by the infinite product
\begin{equation}
     \G_q (x) = (1 - q)^{1 - x}
                \prod_{n=1}^\infty \frac{1 - q^n}{1 - q^{n + x -1}} \epp
\end{equation}
Let us also recall the definition of a $q$-number
\begin{equation} \label{phigammaq}
     [x]_q = \frac{1 - q^x}{1 - q} \epp
\end{equation}
Using $q$-numbers the fundamental recursion relation of the
$q$-$\G$ functions becomes
\begin{equation}
     \G_q (x + 1) = [x]_q \G_q (x) \epc \qd \G_q (1) = 1 \epp
\end{equation}
It implies that the dressed phase obeys the functional equation
\begin{equation} \label{funphi}
     \re^{\ph(x_1, x_2) + \ph(x_1 + \i \g, x_2)}
        = \frac{\sin(x_1 - x_2)}{\sin(x_1 - x_2 + \i \g)} \epp
\end{equation}
We further have the quasi-periodicity
\begin{equation}
     \ph(x + \p, z) = \ph(x, z) + \i \p \epp
\end{equation}

For $|\Im z| > \g$ we have the explicit representation
\begin{equation}
     \re^{\ph (x, z)} =
        \begin{cases}
	   \displaystyle \frac{\sin(x - z)}{\sin(x - z + \i \g)} &
	   \text{if $\Im z > \g$} \\[4ex]
	   \displaystyle \frac{\sin(x - z - \i \g)}{\sin(x - z)} &
	   \text{if $\Im z < - \g$} \epp
        \end{cases}
\end{equation}

%
%
\Appendix{Numerical study}%
\label{app:numerics}
In this appendix we supply numerical results for finite Trotter
numbers in order to support claims made in the main text. 
 
It is not hard to solve numerically the Bethe Ansatz equations for
low-lying excited states of the ordinary transfer matrix of the six-vertex
model which determine the spectrum of the XXZ Hamiltonian and its
correlation functions at $T=0$. In this case the magnetic field $h$
does not change the location of the roots: it merely changes the
energy eigenvalues through the Zeeman term in the Hamiltonian. The
situation is completely different in the finite temperature case:
the Bethe Ansatz equations depend on both $h$ and $T$. The root
distributions then exhibit diverse behavior with changes in $h$
or in $T$, depending on the states under consideration. There occur,
moreover, crossings of energy levels. These  prevent us from drawing
a simple conclusion about the general root patterns pertaining
to the relevant eigenstates. 

We thus do not trace Bethe roots for specially chosen (low-lying)
states, but study all possible configurations of the Bethe roots 
for a fixed Trotter number $N$. The actual procedure combines the
numerical diagonalization of the quantum transfer matrix with symbolic
manipulations on Baxter's TQ relation \cite{Baxter72} as proposed for
the first time in \cite{ADM92}. This, of course, restricts the possible
values of $N$ to small numbers: typically we shall set $N=8$ or 10,
while the logic of the quantum transfer matrix formalism would rather 
require $N$ infinity. At least the parameter $|\beta/N|$ must be chosen
small. We shall impose the condition $|\beta/N| < 1/2$ which implies 
that very low values of $T$ cannot be reached. Nevertheless, the
full list of Bethe-root distributions reveals a simple characterization
of the relevant eigenstates for a non-vanishing magnetic field.
 
\subsection{Reduction of strings}
The existence of longer strings implies the existence of far roots.
Below we argue that the dominant contributions to the spectrum of
correlation lengths do not include longer strings in the low temperature
limit at non-zero magnetic fields.
 
We fix the value of $h$. For sufficiently high temperatures longer
strings \emph{do} exist. With decrease in temperature, roots change
their locations continuously except at discrete values, $T_m
= \frac{h}{2m \gamma}$,  $m \in \mathbb{N}$. By approaching $T_m$
from above, some of the roots go to infinity ($\Im x_j \rightarrow +\infty$).
We call them diverging roots. To be precise, led by numerical
investigations, we arrive at the following
%
\begin{conjecture}\label{conj:divergingroot}
Consider eigenstates in the sector with spin $s$, or equivalently 
$M=\frac{N}{2}-s$ Bethe roots, at $T \sim T_m$. If $s+1 \le m \le
\frac{N}{2}$, there are ${N \choose N/2-m}$ eigenstates (out of totally   
${N \choose M}$ states in the sector) with $m-s$ diverging roots.
\end{conjecture}
Case studies for $N =4, \dots, 10$\footnote{$1 \le M \le 4$ for $N=10$.}  
with various choices of parameters confirm that there is no exception
to the above rule. There is, however, a subtle point. Due to the condition
$|\beta/N| < 1/2$, $h$ can not be too small in order to reach the singular
values of $T$. Some of the trials leading to the above conjecture 
were thus performed outside of the range $h < h_{\ell}$. As far as we
have observed, the massive phase and the massless phase share the same
divergent behavior. Thus, we temporarily neglect the condition $h < h_{\ell}$.
After crossing a singular value $T_m$ of $T$, the diverging roots come back
to finite locations. We observe that their positions are different from
those at $T$ slightly above $T_m$. This abrupt change of locations at
singular values of $T$ is the main mechanism for the reduction of strings.
We demonstrate this with the example of a 4-string solution, $N=8, s=0,
\gamma=1$. In order to achieve low $T$ we take relatively large magnetic
field, $h=10$. The snapshots of Bethe roots with decrease in $T$ are
depicted in Figure~\ref{FourStrings}. Clearly, the string becomes shorter
every time $T$ crosses a singular value.
\begin{figure}[!h]

\begin{center}
\begin{tabular}{c}
 \begin{minipage}{0.33\hsize}
  \begin{center}
   \includegraphics[width=40mm]{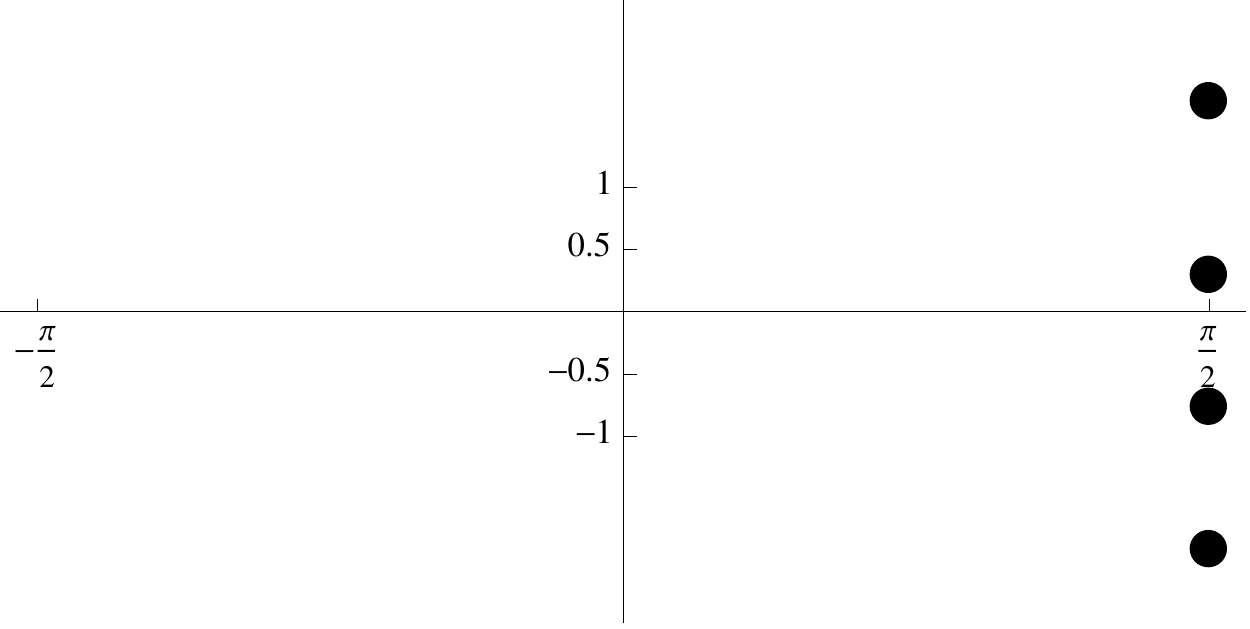}
  \end{center}
 \end{minipage}
 \begin{minipage}{0.33\hsize}
 \begin{center}
  \includegraphics[width=40mm]{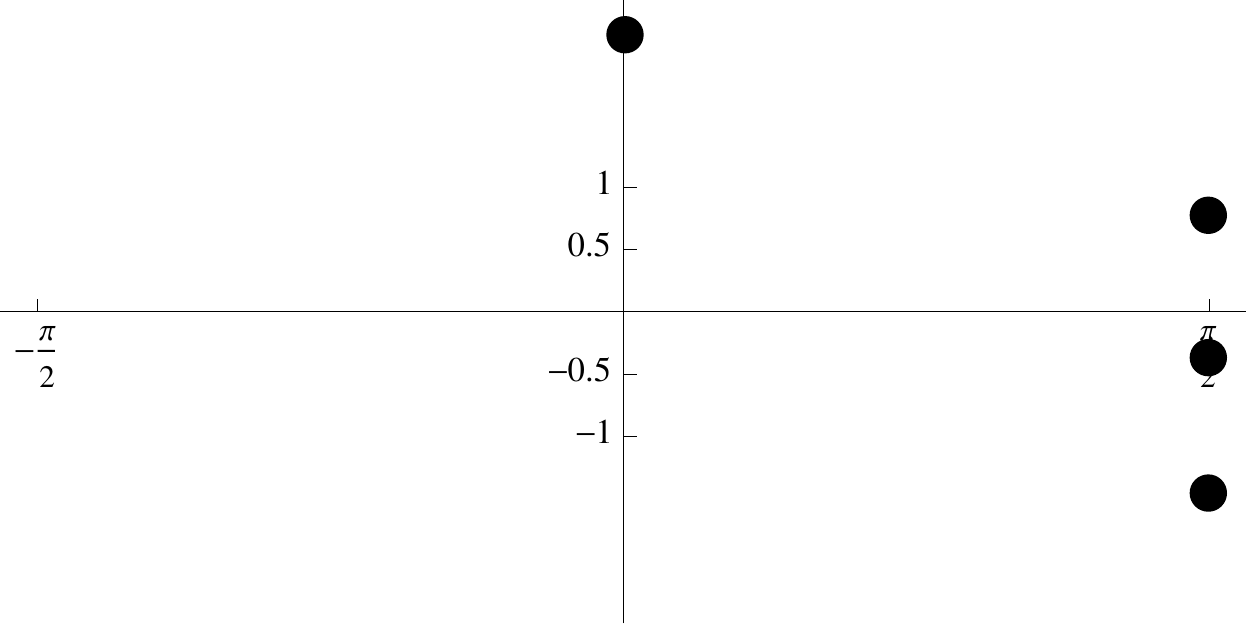}
 \end{center}
 \end{minipage}
 \begin{minipage}{0.33\hsize}
 \begin{center}
  \includegraphics[width=40mm]{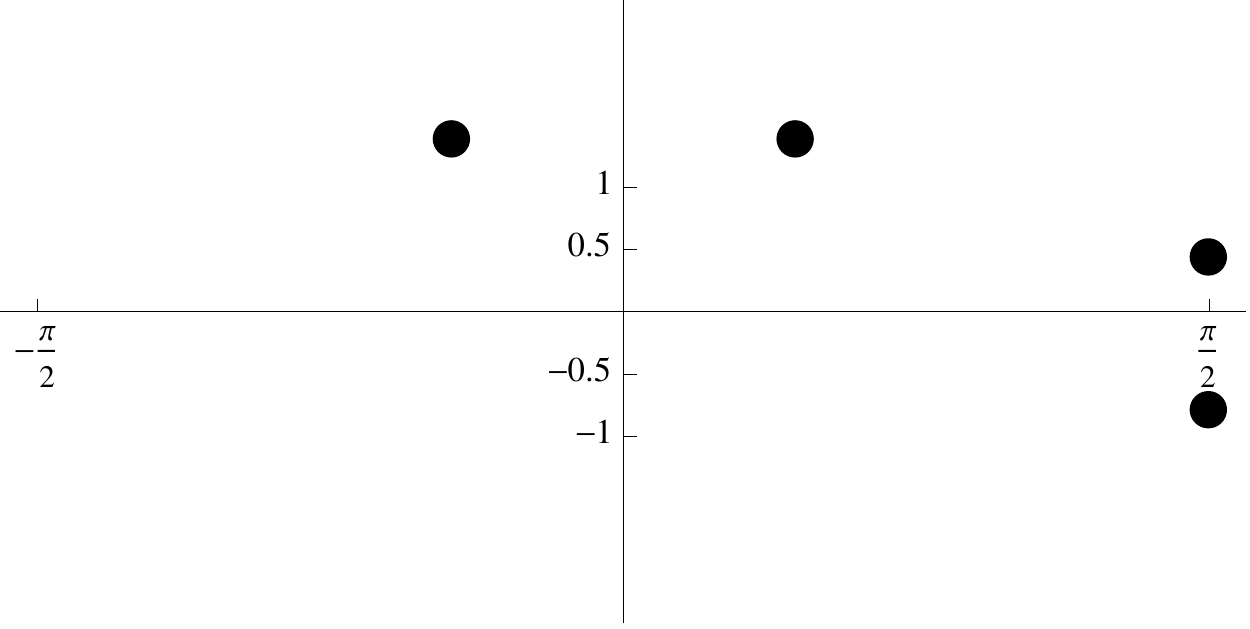}
 \end{center}
 \end{minipage}
  \end{tabular}
 \end{center}
 \begin{center}
\begin{tabular}{c}
 \begin{minipage}{0.33\hsize}
 \begin{center}
  \includegraphics[width=40mm]{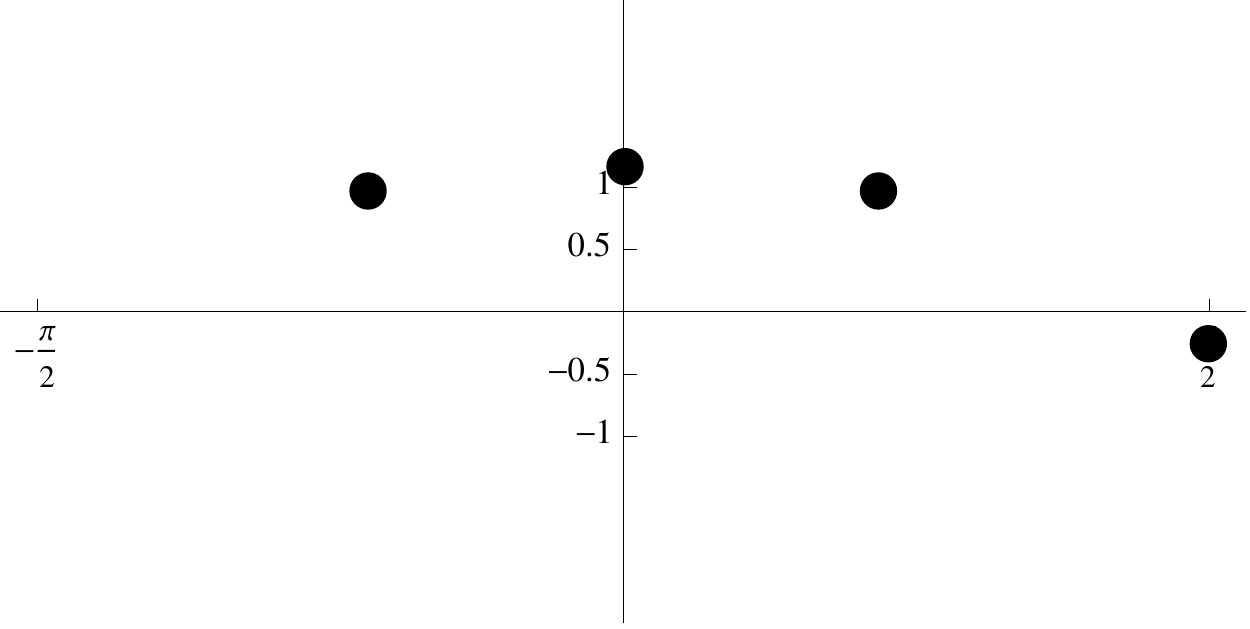}
 \end{center}
 \end{minipage}
 \begin{minipage}{0.33\hsize}
 \begin{center}
  \includegraphics[width=40mm]{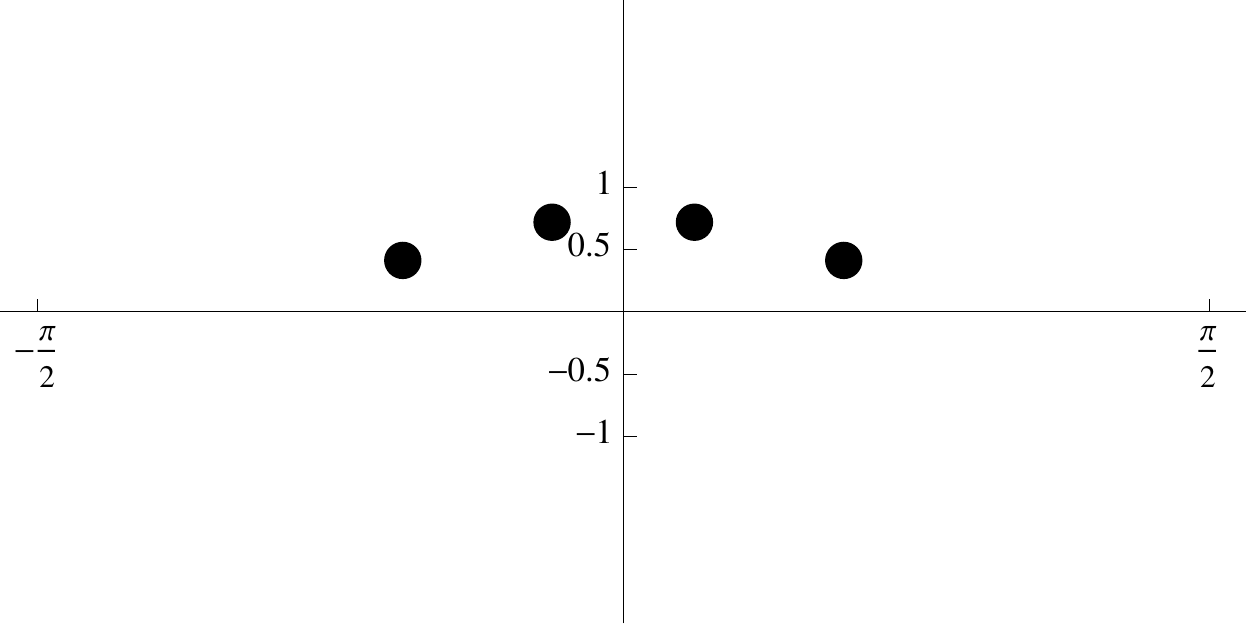}
 \end{center}
 \end{minipage}
 \end{tabular}
\caption{The fate of Bethe roots composing a 4 string at sufficiently high
$T$. We fix $h=10$ and $\gamma=1$. $T=10,4,2$ (upper row), $T=1.5,1.1$
(lower row) from the left to right.}
   \label{FourStrings}
 \end{center}
\end{figure}

The states with diverging roots involve strings at sufficiently high
temperatures. We have no explanation why the converse should be true,
but it seems empirically the case. Let $\Lambda_j$ be the $j$th
eigenvalue of the quantum transfer matrix. We  arrange the
$|\Lambda_j|$  in  decreasing order and refer to them as
`energy levels'. We then count the number of states with
diverging roots in every 10 consecutive `energy levels'
at $T$ slightly above $T_m$. An example for $N=8, \gamma=3,
h=16$ and $s=0$ is depicted in Figure~\ref{fig:divergingroots}.
The sector contains 70 states. Thus, we divide them in 7 portions.  
The horizontal axis represents the `energy level'. The height
of the leftmost bin represents the number of states with
diverging roots among the first 10 `energy levels', and so on.
The figure manifestly shows that, as temperature goes down,
states with diverging roots become higher and higher excited
states. This suggests that, after sufficiently many crossings
of singular values $T_m$, even if there still remain states
containing strings, these should be highly excited. We thus
conclude that they do not significantly contribute to correlations
at finite temperatures.

\begin{figure}[!h]
\begin{center}
\includegraphics[width=.35\textwidth]{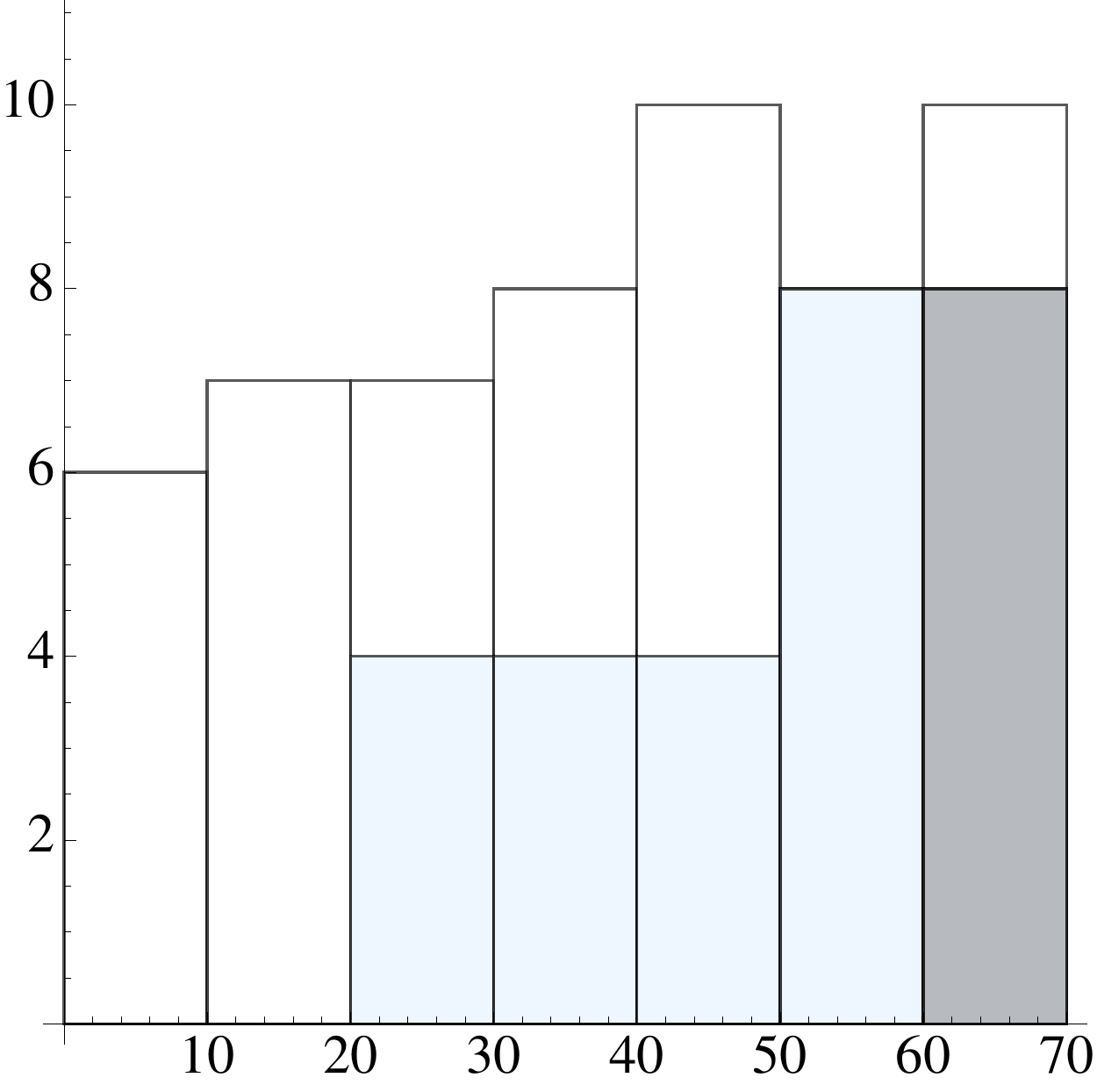}
\end{center}
\caption{The white, light blue and grey  bins represent the distributions of
diverging-root states slightly above $T_1 = h/2\gamma, T_2 = h/4\gamma$
and $T_3= h/6\gamma$, respectively. The heights ($= 8$) of the rightmost
bins are identical for $T_2$ and $T_3$.
}
\label{fig:divergingroots}
\end{figure}

\subsection{Dominant contributions}
 
We proceed further and claim that free holes and particle-hole
pairs provide a complete description of the dominant excitations.
Let us illustrate this with an example, $N = \nolinebreak 8$, $h = 3.7$,
$\gamma = 2$ and $s=0$. Consider the case $T=2$. We present the three 
Bethe-root configurations which correspond to the first three 
`energy levels' in Figure~\ref{fig:top5T2}. We omit configurations
with complex conjugate eigenvalues, which are obtained by a simple
reflection. The figure illustrates the high-temperature regime,
$T>h/2\gamma$. One observes 2-string states but no particle-hole
bound pairs. Figure~\ref{fig:worst5T2} shows the three 
configurations  corresponding to the last three 
`energy levels'. They are all characterized by 4-strings, and
the holes are more densely distributed on the line $\Im x =-\gamma/2$.

\begin{figure}[!h]
\begin{center}
\begin{tabular}{c}
   \begin{minipage}{0.33\hsize}
        \begin{center}
            \includegraphics[width=40mm]{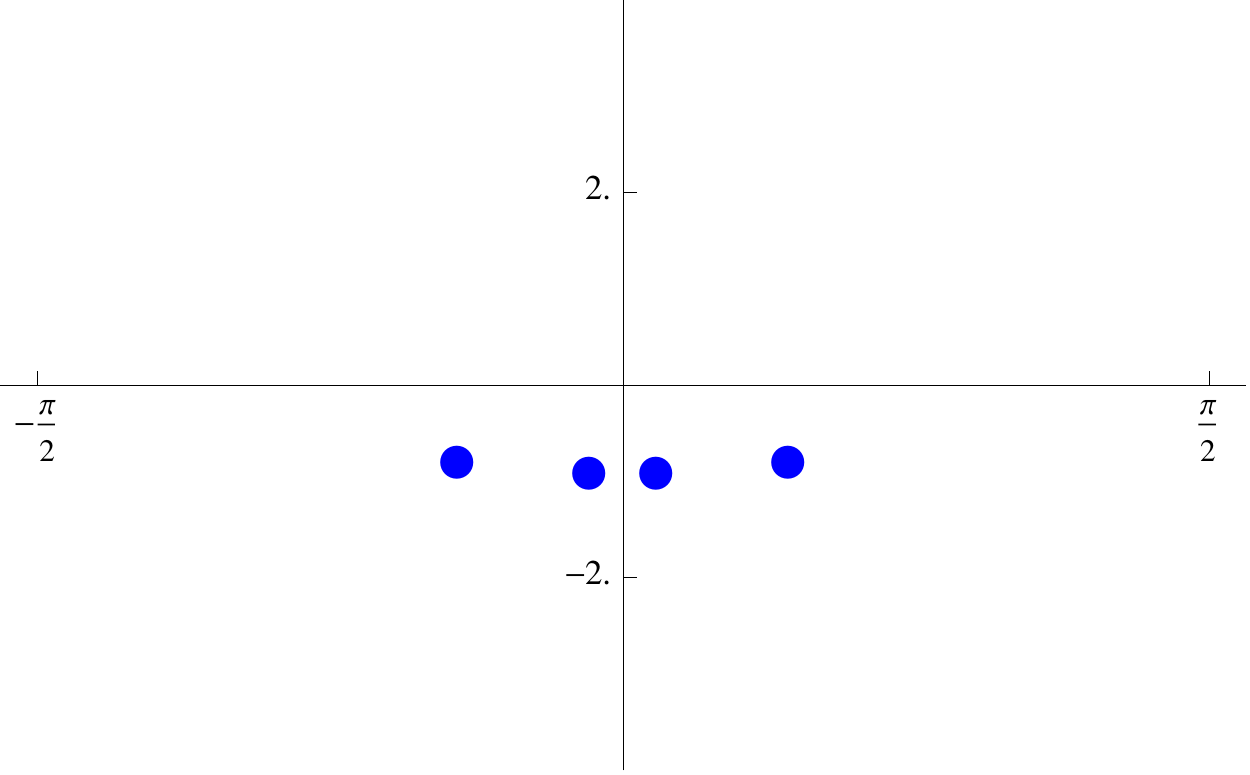}
        \end{center}
    \end{minipage}
    \begin{minipage}{0.33\hsize}
           \begin{center}
                  \includegraphics[width=40mm]{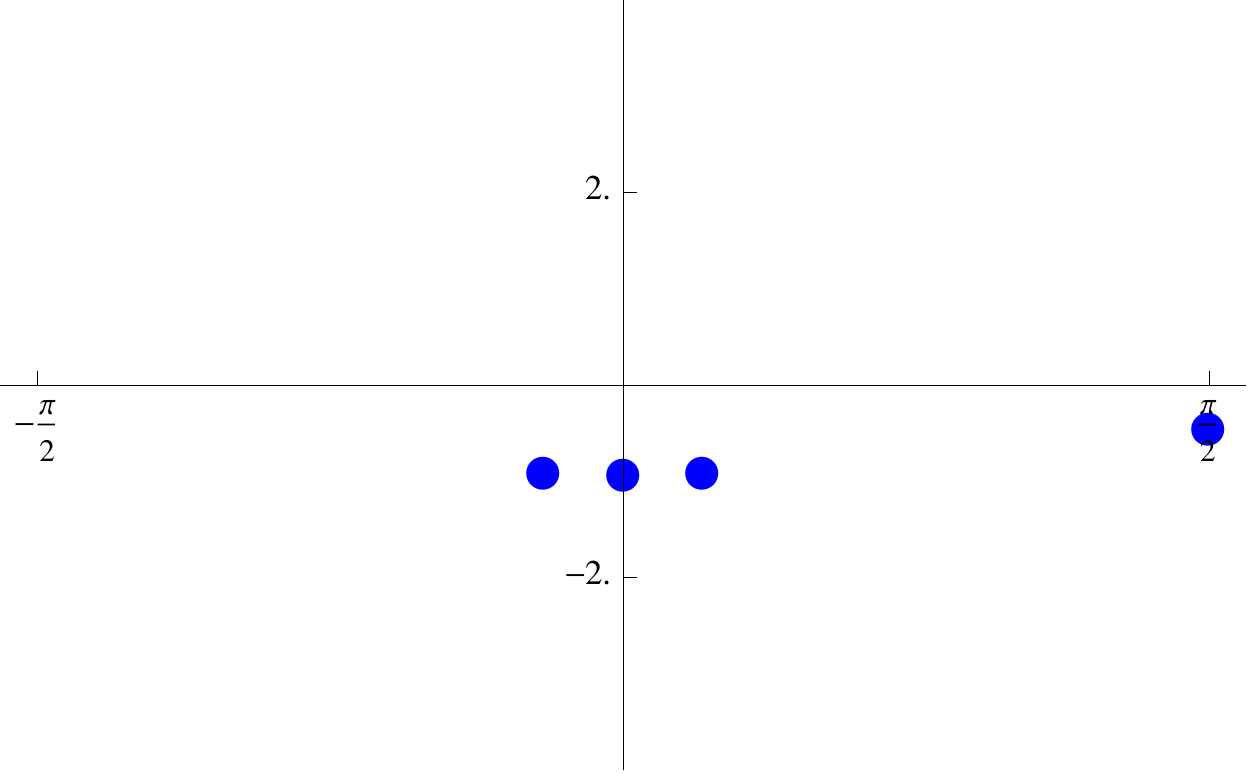}
            \end{center}
    \end{minipage}
     \begin{minipage}{0.33\hsize}
            \begin{center}
                   \includegraphics[width=40mm]{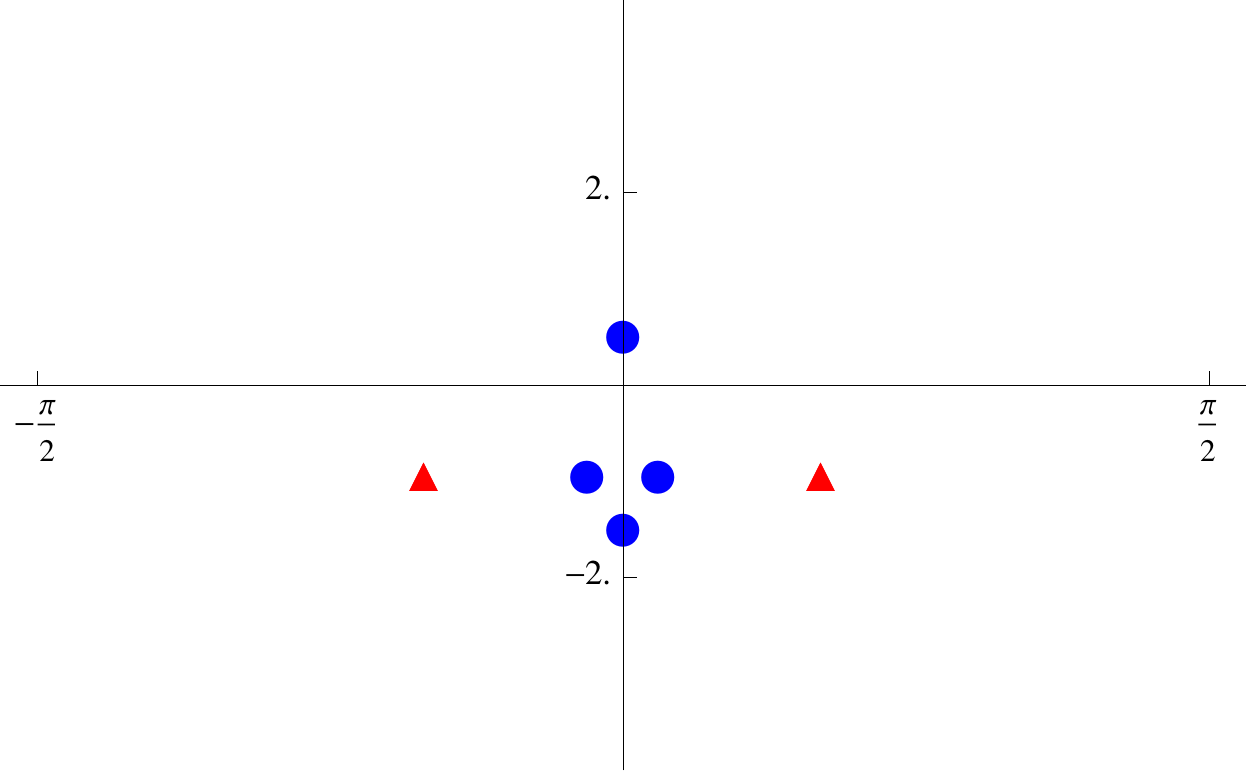}
           \end{center}
      \end{minipage}
\end{tabular}
\caption{`Top 3' configurations at $T=2$. Blue circles represent Bethe
roots while red triangles are holes.}
   \label{fig:top5T2}
 \end{center}
\end{figure}

\begin{figure}[!h]
  \begin{center}
  \begin{tabular}{c}
        \begin{minipage}{0.33\hsize}
           \begin{center}
                \includegraphics[width=40mm]{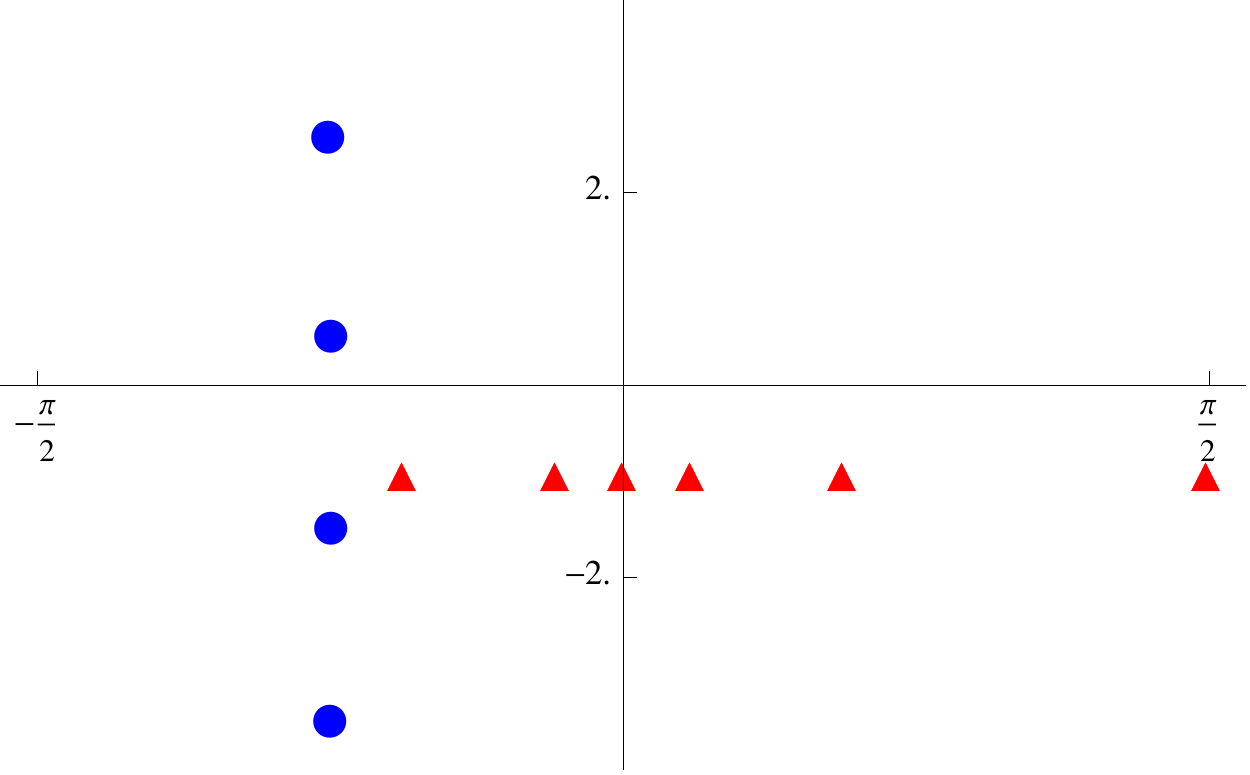}
           \end{center}
         \end{minipage}
          \begin{minipage}{0.33\hsize}
              \begin{center}
                    \includegraphics[width=40mm]{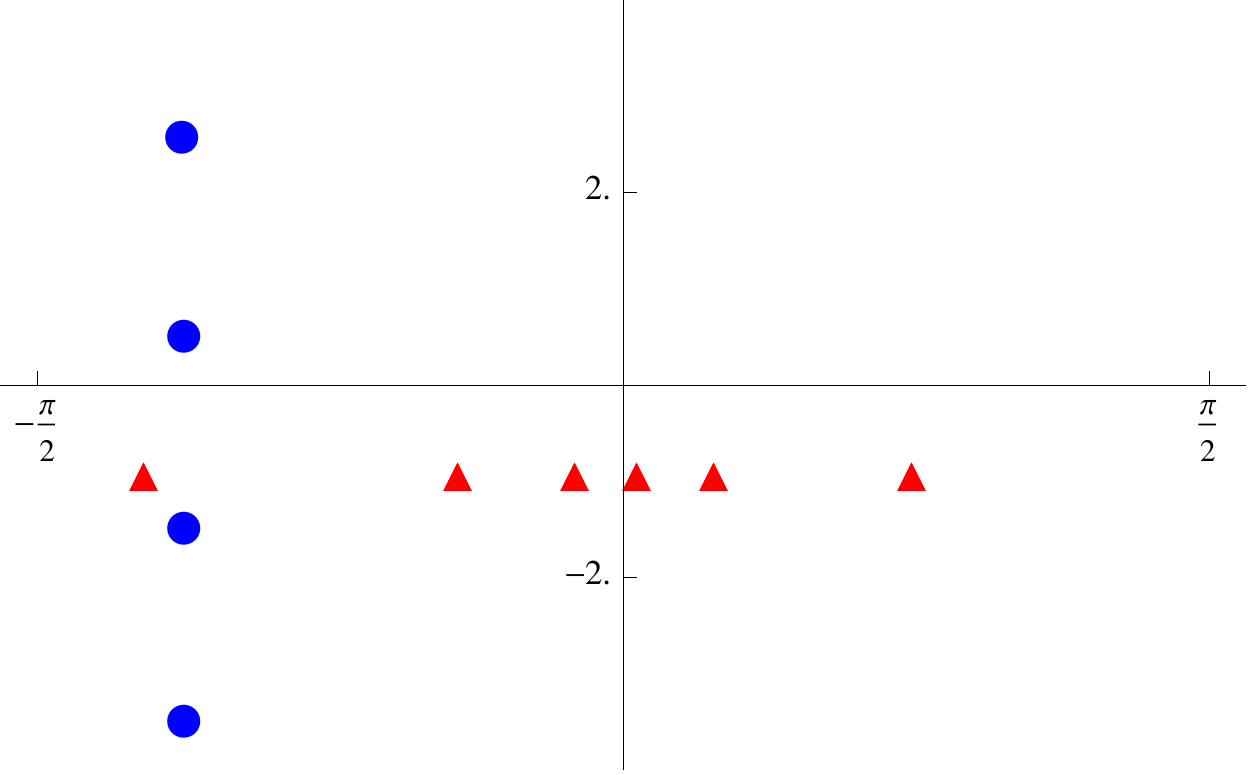}
              \end{center}
          \end{minipage}
          \begin{minipage}{0.33\hsize}
               \begin{center}
                     \includegraphics[width=40mm]{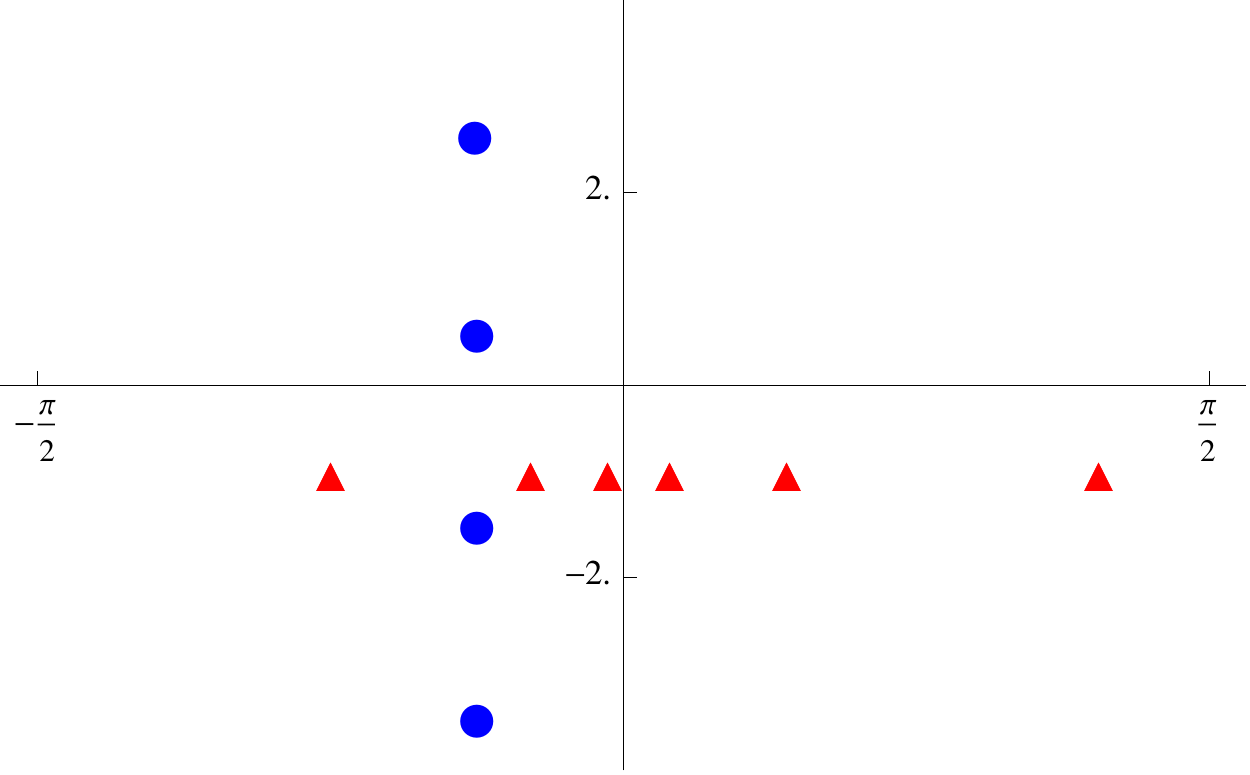}
              \end{center}
          \end{minipage}
     \end{tabular}
    \caption{`Worst 3' configurations at $T=2$.  }

   \label{fig:worst5T2}
 \end{center}
\end{figure}

Next, we consider the case $T=0.46<h/4\gamma$, which
is reached after passing two critical temperatures.
The three 
Bethe root configurations corresponding to the three
eigenvalues of largest modulus are given in Figure~\ref{fig:top5T046}.
We do not observe strings anymore. Instead, we observe a particle-hole
pair in each state (except for the dominant one).

\begin{figure}[!h]
\begin{center}
\begin{tabular}{c}
     \begin{minipage}{0.33\hsize}
            \begin{center}
                  \includegraphics[width=40mm]{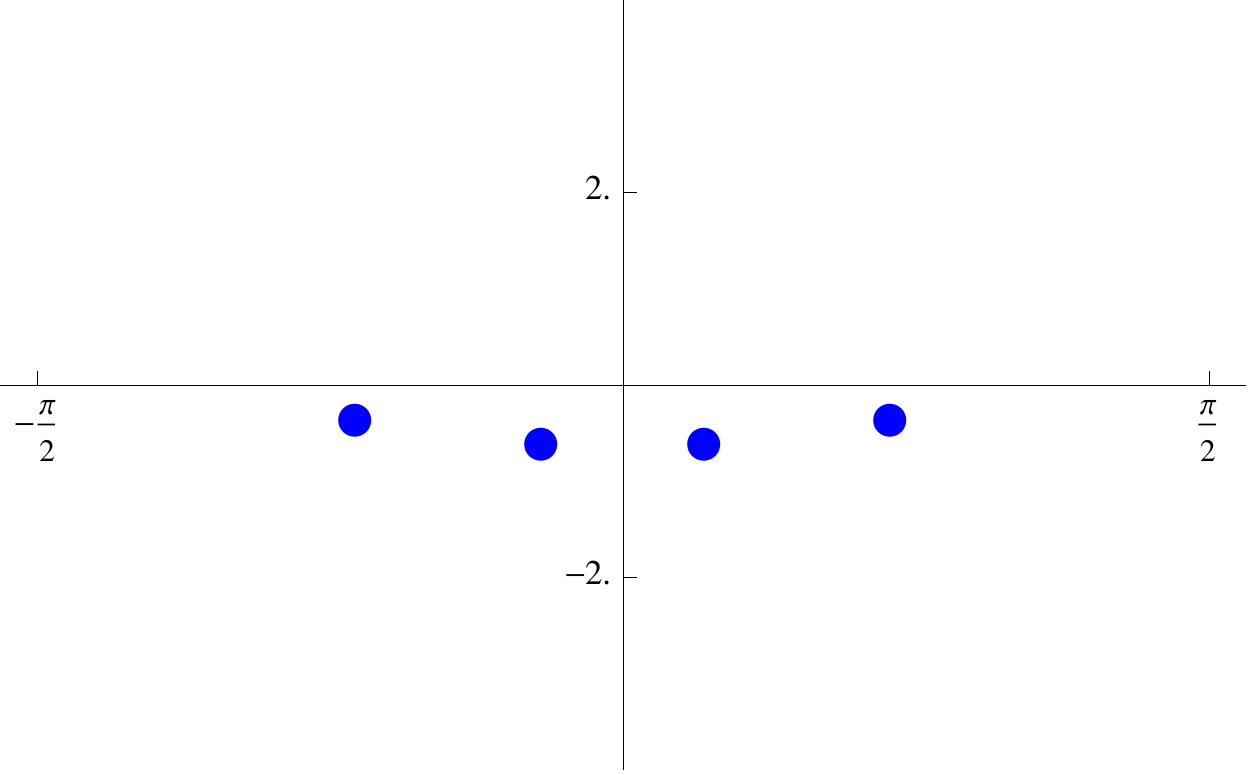}
            \end{center}
      \end{minipage}
      \begin{minipage}{0.33\hsize}
            \begin{center}
                 \includegraphics[width=40mm]{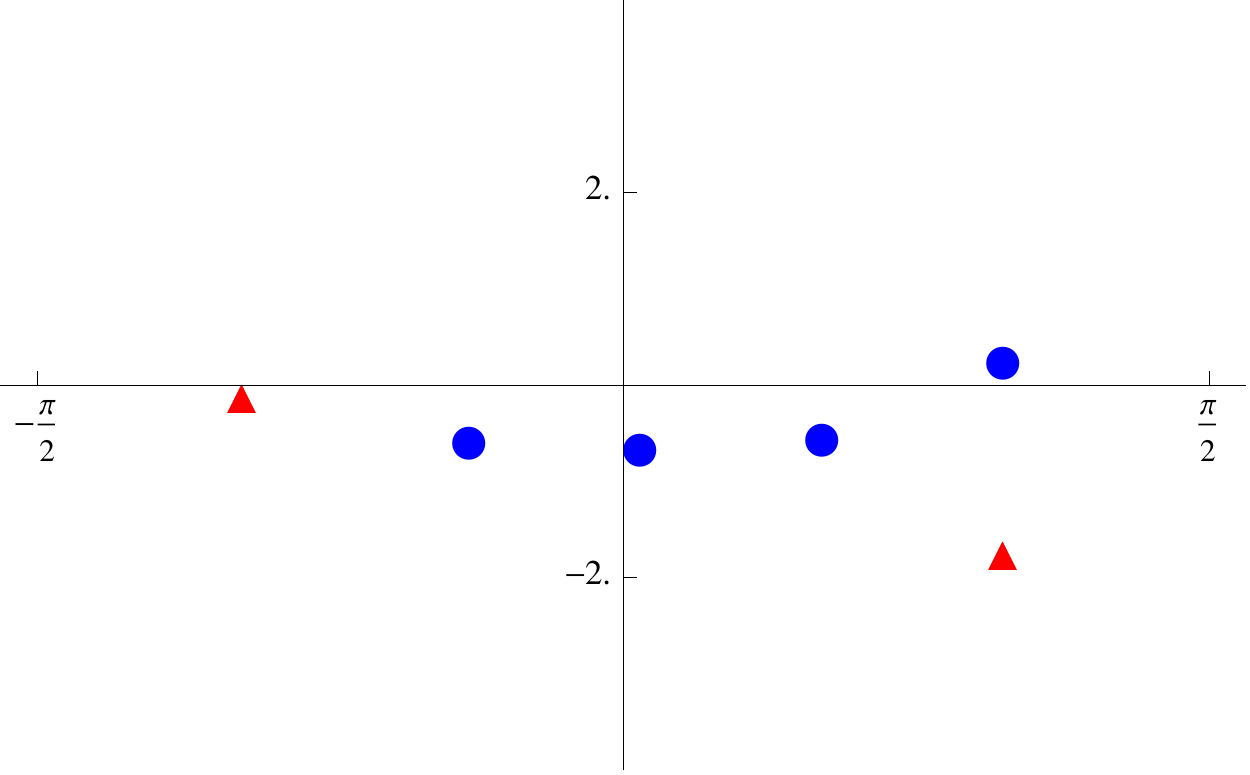}
            \end{center}
       \end{minipage}
       \begin{minipage}{0.33\hsize}
            \begin{center}
                 \includegraphics[width=40mm]{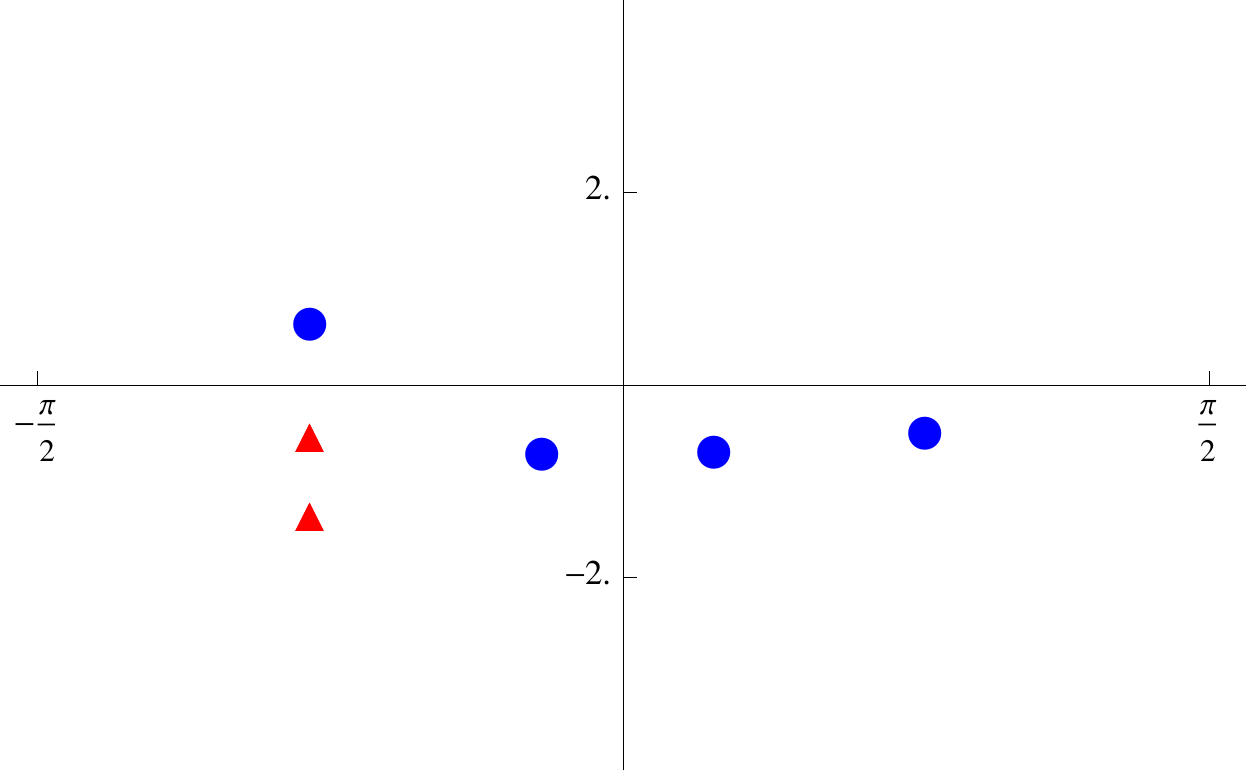}
             \end{center}
       \end{minipage}
 \end{tabular}
  \caption{`Top 3' configurations at $T=0.46$.  }
   \label{fig:top5T046}
 \end{center}
\end{figure}
\begin{figure}[!h]
\begin{center}
   \begin{tabular}{c}
      \begin{minipage}{0.33\hsize}
        \begin{center}
            \includegraphics[width=40mm]{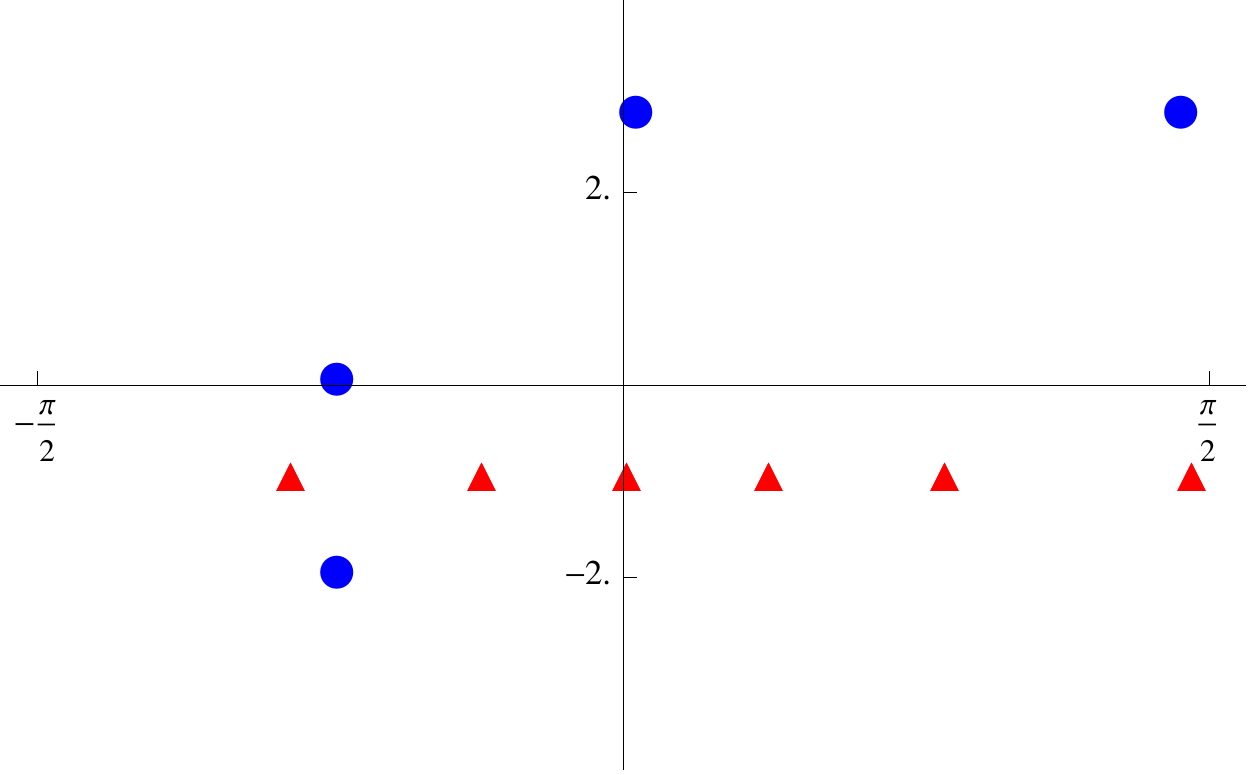}
        \end{center}
     \end{minipage}
     \begin{minipage}{0.33\hsize}
          \begin{center}
               \includegraphics[width=40mm]{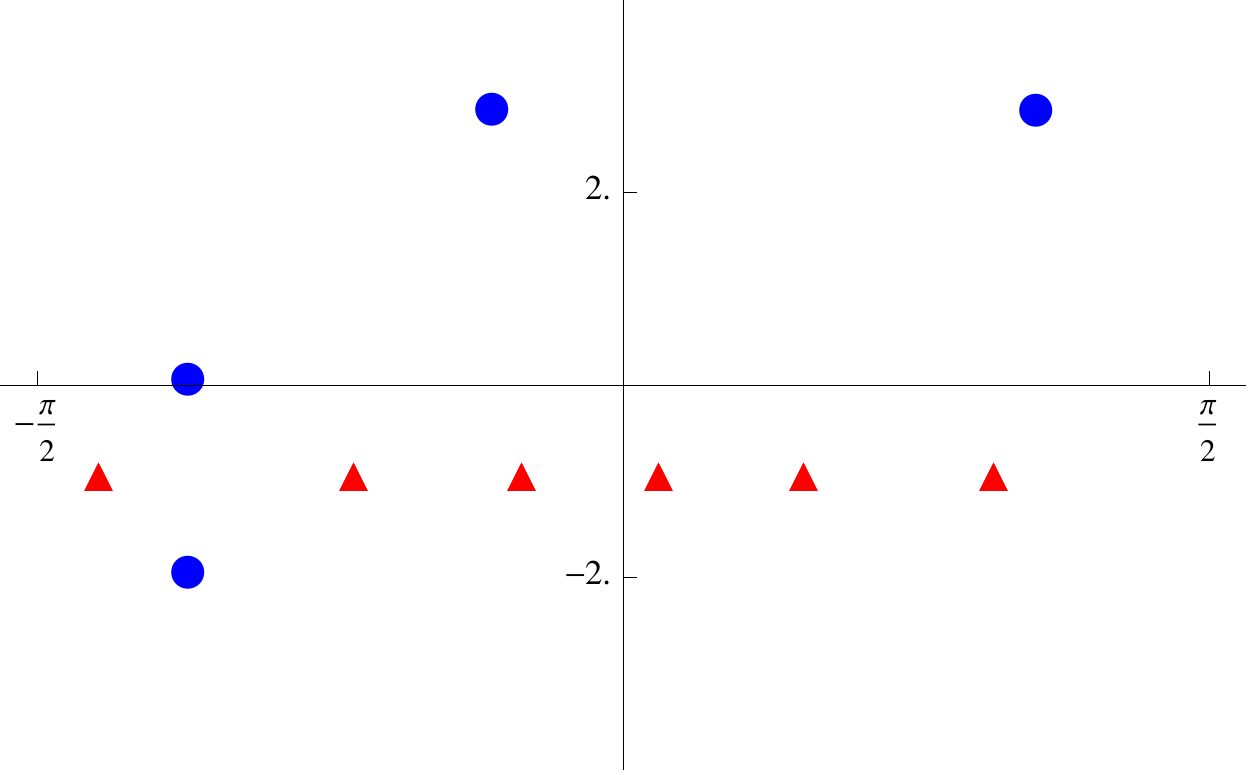}
           \end{center}
      \end{minipage}
       \begin{minipage}{0.33\hsize}
           \begin{center}
                \includegraphics[width=40mm]{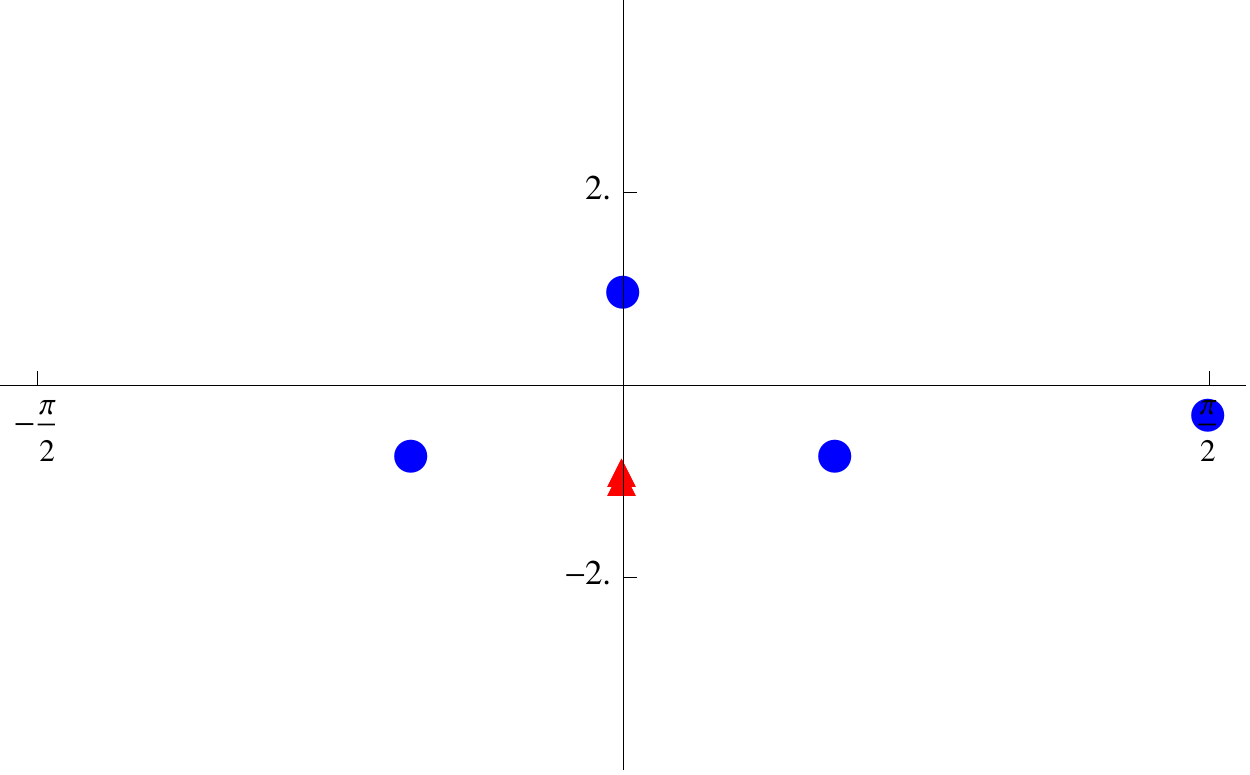}
           \end{center}
       \end{minipage}
    \end{tabular}
  \caption{`Worst 3' configurations at $T=0.46$.  }
   \label{fig:worst5T046}
 \end{center}
\end{figure}
\noindent
Figure~\ref{fig:worst5T046} shows the last three 
configurations at the same temperature. One still observes
2~strings, which may be remains of 4 strings, but no
particle-hole pairs.

The highly excited states do not affect the physical quantities.
They are physically irrelevant. All particle-hole pair excitations,
on the other hand, are low-lying and therefore physical.
To turn this into a quantitative argument we introduce a
quantity 
\begin{equation}
\mathfrak{r}= 
\frac{\sum_{\text{sum over particle-hole excitations}} |\Lambda_j |}{\sum_{\text{sum over all but the ground state}} |\Lambda_j |} \epp
\end{equation}
Clearly, $\mathfrak{r}$ measures the relative importance of the
particle-hole excitations. There is, of course, a certain
ambiguity in identifying the particle-hole excitations. The
explicit value of $\mathfrak{r}$ depends on the identification
criterion.\footnote{Here we adopt the following: $x^c$, $x^h$
is a pair if $|\Re{(x^c-x^h)/\gamma}|<0.05$, $0.93<\Im{(x^c-x^h)/\gamma}<1.07$.}
Its qualitative temperature behaviour, however, seems to be
independent of the choice.\footnote{If we require $0.95 <
\Im{(x^c-x^h)/\gamma} < 1.05$ then pairs are not present at
$T=0.8$ and $\mathfrak{r}$ at $T=0.6$ changes to 0.705, but
the other values in the table remain the same.}
We tabulate the values of  $\mathfrak{r}$ and the number of
pairs $n_{\text{pair}}$ in Table~\ref{table:valueofr}.
\begin{table}[htb] 
\begin{center}
\begin{tabular}{ |c|c|c|c|c|c|}
\hline
                      T  &         $0.9$  &       $0.8$&      $0.7$&      $0.6$&  $0.5$\\ \hline
 $n_{\text{pair}}$&              $0$&        $4$&          $10$&      $20$&       $29$\\ \hline 
 $\mathfrak{r}$  &                $0$&     $0.167$&  $0.335$&       $0.849$&       $0.917$  \\   
 \hline
\end{tabular}
 \caption{The numbers of the particle-hole pairs and values of $\mathfrak{r}$ 
 at various temperatures for 
 $N=8, \gamma=2, h=3.7$ and $s=0$.}
 \label{table:valueofr}
 \end{center}
\end{table}
Strictly speaking there are four states with accidental pairs at $T=0.9$.
We checked that they correspond to a situation in which the center of
a 3 string  and a hole come very close to  each other, hence, we neglect them.  
Even for $h=0$ there are such accidental pairs. We checked that their
contribution is independent of $T$ and very minor (less than 1\%).

To understand which levels contribute to $\mathfrak{r}$, we counted again
the number of states with particle-hole pairs in every 10 consecutive
`energy levels'. The corresponding histogram for $T=0.5$ is presented
in Figure~\ref{fig:histogram}.  
Table~\ref{table:valueofr} and the histogram clearly show that the
particle-hole pairs become the dominant excitations when the temperature
decreases at fixed finite $h$ in the $s=0$ sector.%
\footnote{To be precise, the 59th and 60th states seem more likely
to be accidental pairs. They are anyway very minor so we include them.}

\begin{figure}[!h]
\begin{center}
\includegraphics[width=.35\textwidth]{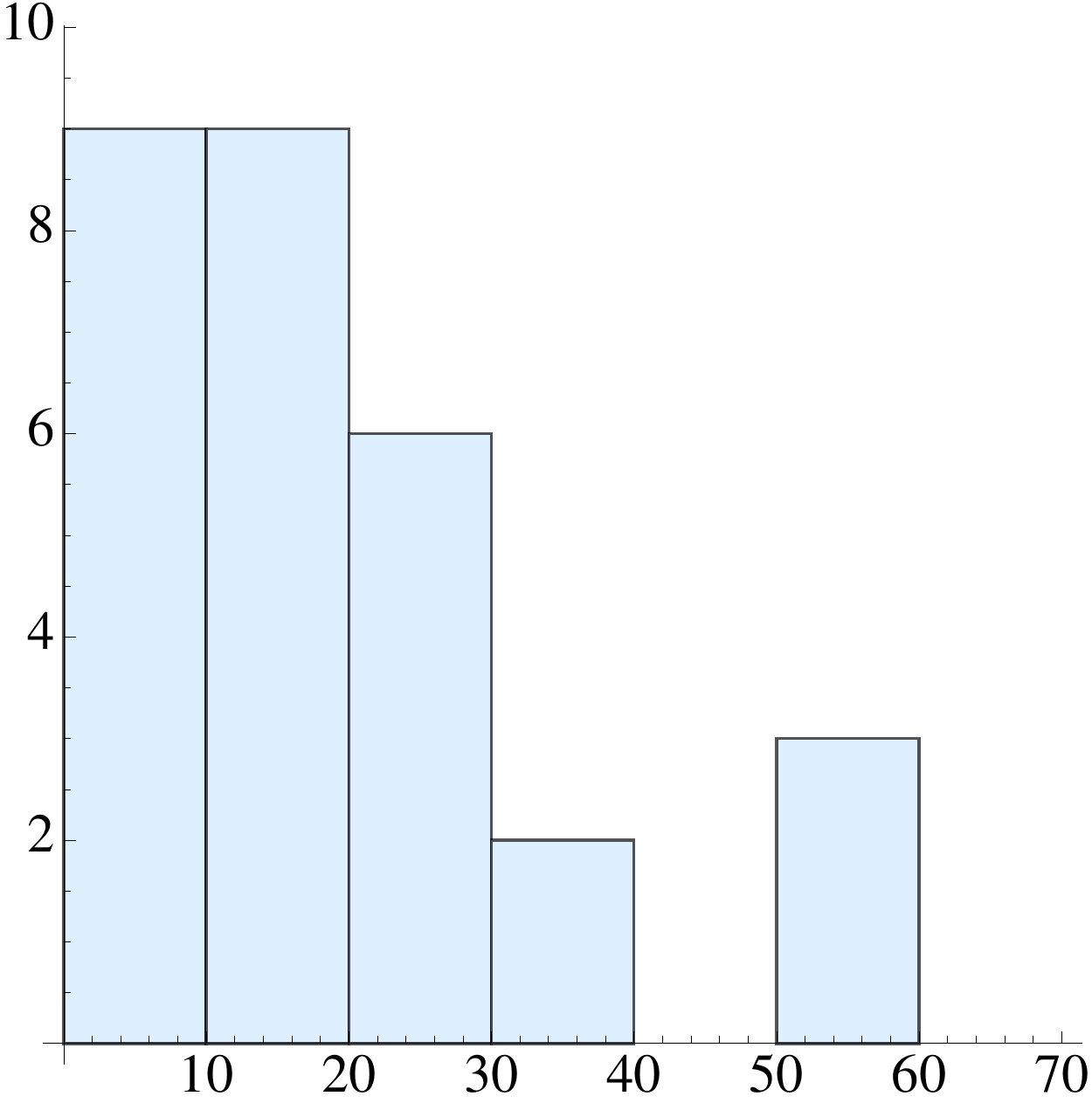}
\end{center}
\caption{The vertical axis represents the number of states 
with pairs in every 10 consecutive `energy levels'. In the
first bin, only the ground state is missing, while the 20th
level is missing in the 2nd bin and so on.
\label{fig:histogram}
}
\end{figure}

Next consider sectors with $s \ne 0$. We chose $\gamma=3$, fixed
$h=16$, and calculated the values of $\mathfrak{r}$ in several
sectors for $N=8,10$ and $N=12$. The results are summarized in
Table~\ref{table:valueofrII}. For this choice of the parameters
$\mathfrak{r}$ is equal to unity in the $s=0$ sector, that is, 
any but the ground state carries particle-hole pairs. The number
of particle-hole pairs and their contribution to $\mathfrak{r}$
decrease in the higher-$s$ sectors. Thus, particle-hole pairs
become less important. Instead, we find that low-lying excitations
in the $s \ne 0$ sectors are characterized by free holes. Take
$N=8$ and $s=1$ as an example. The first 10 `energy levels' do
not include particle-hole pairs, resulting in a small $\mathfrak{r}$.
We find numerically that there exist 5 available positions for
roots and holes near the real axis. In the present case, three
roots and two holes must be allocated there. This amounts to 10
possible configurations, and we checked that they generate
nothing but the first 10 states. Three examples are shown in
Figure~\ref{fig:freeholes}. Similarly we find that also the
most important contributions in other $s \ne 0$ sectors come
from free holes.

\begin{table}[htb] 
\begin{center}
\begin{tabular}{ |c||c|c||c|c||c|c|}
\hline
     &        \multicolumn{2}{|c||} { $N=8$}  &    \multicolumn{2}{c||}{$N=10$} &  \multicolumn{2}{c|}{$N=12$}   \\   \hline
 s  &         $  \mathfrak{r}$   &   $n_{\text{pairs}}/n_{\text{total}}$  &     $\mathfrak{r} $ &      $n_{\text{pairs}}/n_{\text{total}}$  &  $\mathfrak{r} $&   
 $n_{\text{pairs}}/n_{\text{total}}$ \\ \hline\hline
 1&                                         0.378&                                                 46/56&                                 0.351&                           178/210&        *  &  *   \\   \hline
 2&                                          0.115&                                                12/28&                                 0.035&                              46/120&          0&   0/66    \\  \hline 
3&                                         0.05 &                                                   1/8&                                         0 &                                  0/45&           0&    0/220\\ \hline 
\end{tabular}
 \caption{The values of $\mathfrak{r}$ in various sectors for $N=8,10$ and $12$.  
 The values at $\ast$ are not  available due to insufficient computer capacity.
 The symbol $n_{\text{total}}$ denotes the total number of states in a given sector.}
 \label{table:valueofrII}
 \end{center}
\end{table}

\begin{figure}[!h]
\begin{center}
\begin{tabular}{c}
 \begin{minipage}{0.33\hsize}
  \begin{center}
   \includegraphics[width=40mm]{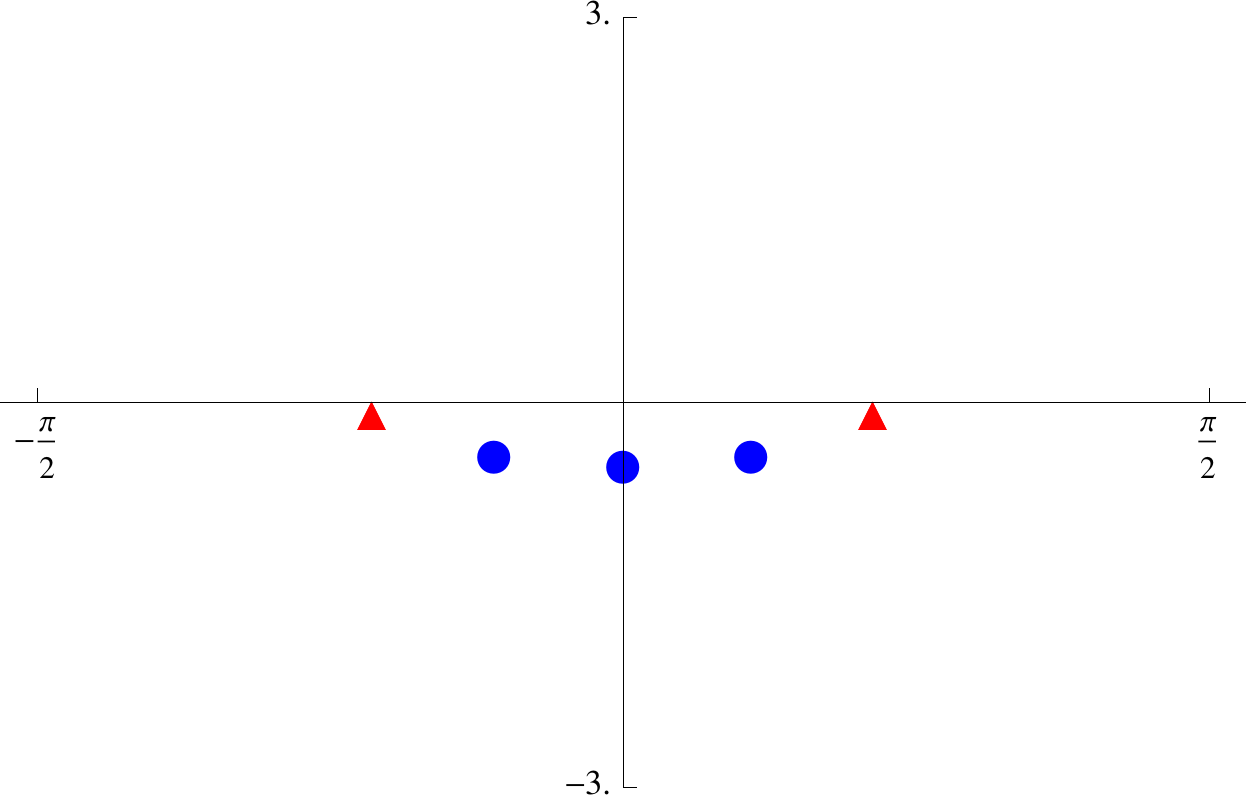}
  \end{center}
 \end{minipage}
 \begin{minipage}{0.33\hsize}
 \begin{center}
  \includegraphics[width=40mm]{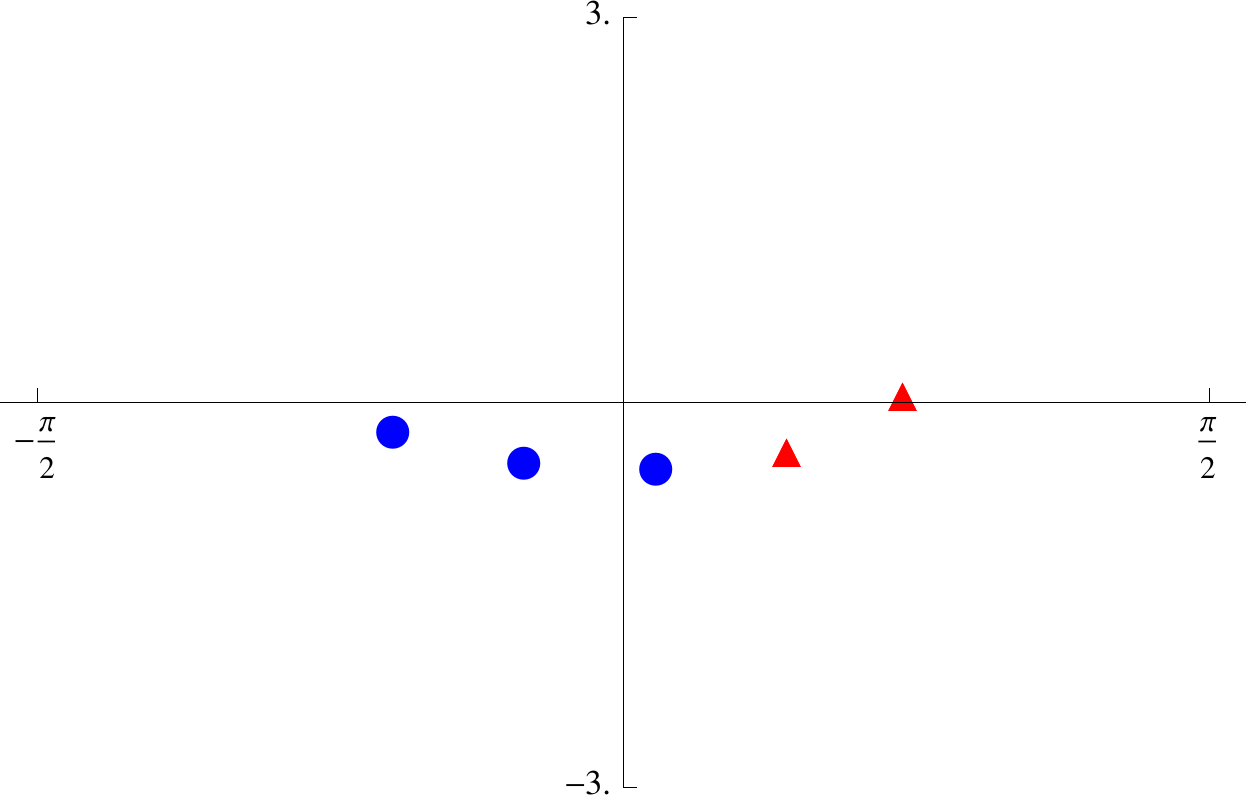}
 \end{center}
 \end{minipage}
 \begin{minipage}{0.33\hsize}
 \begin{center}
  \includegraphics[width=40mm]{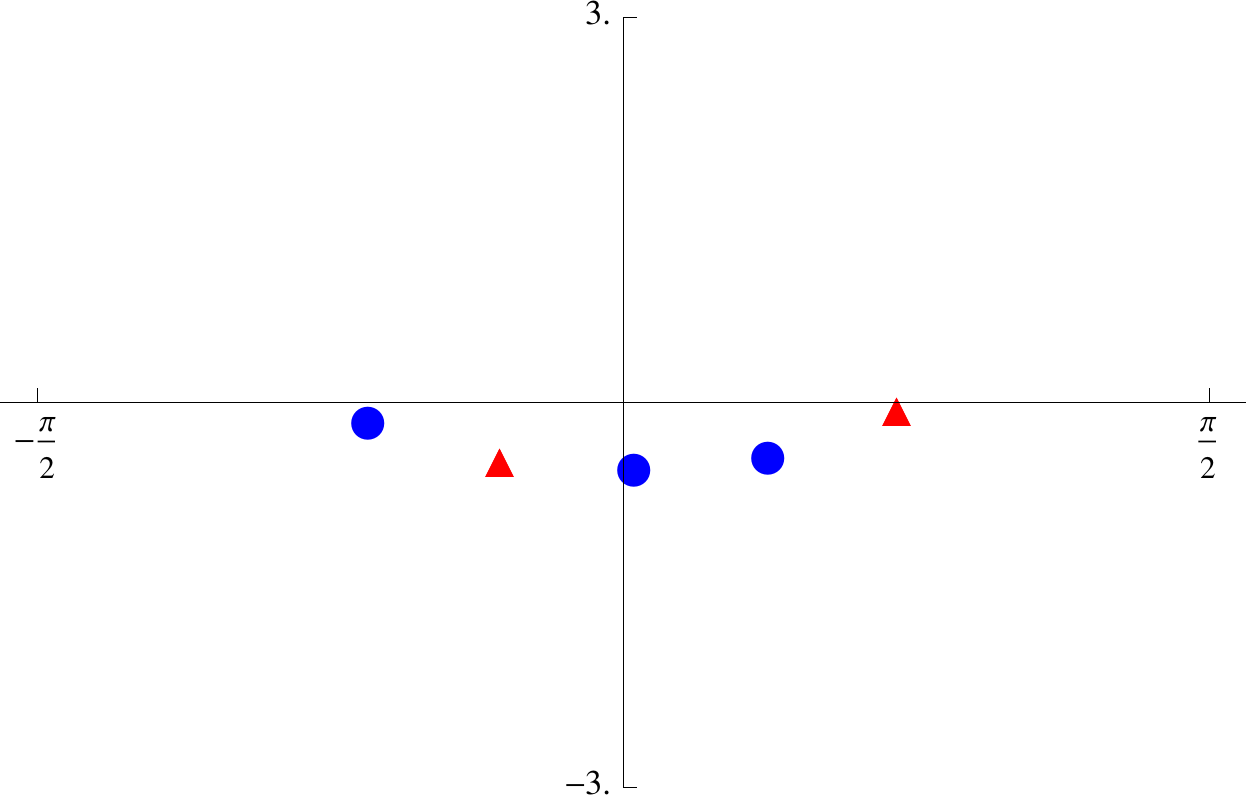}
 \end{center}
 \end{minipage}
 \end{tabular}
  \caption{Examples of configurations with free holes for  $N=8, M=3, \gamma=3, h=16$.   }
   \label{fig:freeholes}
 \end{center}
\end{figure}

The above observation suggests a better definition of $\mathfrak{r}$:
let $\mathfrak{r}'$ be the ratio of the sum of $|\Lambda_j|$ over
particle-hole excitations and free hole excitations to the sum of
all $|\Lambda_j|$ in a given $s \ne 0$ sector. A list of values of
$\mathfrak{r}'$ for $N=10$ and $ 12$ is shown in Table~\ref{table:valueofrIII}.
\begin{table}[htb] 
\begin{center}
\begin{tabular}{ |c||c|c|c|c| }
\hline
     &          \multicolumn{2}{c|}{$N=10$} &                 \multicolumn{2}{c|}{$N=12$}\\   \hline
 s  &       $\mathfrak{r}' $ &      $n_{\text{pf}}/n_{\text{total}}$&  $\mathfrak{r}' $ &      $n_{\text{pf}}/n_{\text{total}}$     \\ \hline\hline
  1&                                       0.999&              199/210&        *&        * \\   \hline
 2&                                          0.991&             97/120 &        0.972&      112/220  \\  \hline 
3&                                         0.989 &               35/45 &        0.992&       54/66  \\ \hline 
\end{tabular}
 \caption{The values of modified $\mathfrak{r}$  for $N=10$ and $ 12$.
 The symbol $n_{\text{pf}}$ means the sum of the number of particle-hole states
 plus that of the free hole states (avoiding double counts).}
 \label{table:valueofrIII}
 \end{center}
\end{table}
\enlargethispage{3ex}
The above finite-$N$ numerical investigations all justify the
claim in the main text: in the Trotter limit $N\rightarrow \infty$, 
in the presence of a finite magnetic field, the description of the
excitations in terms of free holes and particle-hole pairs
is complete at low temperatures.

\subsection{Higher-level Bethe Ansatz equation in finite-N
approximation}
We demonstrate the accuracy of the higher-level Bethe Ansatz
equations within the finite Trotter number approximation.
Consider the 10th largest eigenvalue state of the quantum
transfer matrix for $N=8$, $h=2$, $\gamma=2$ and $s=0$ at $T=0.46$.
The precise locations of the Bethe roots and holes are given in
Table~\ref{table:rootsandholes}. The root in the upper half
plane is identified as a close root $x^c$. The hole with
positive (negative) real part is labeled as $x^h_1$ ($x^h_2$).
Clearly $x^c$ and $x^h_1$ constitute a 2-string:$\frac{x^c-x^h_1}{\gamma}
= 1.11508 \i +0.0100447$.
\begin{table}[htb] 
\begin{center}
\begin{tabular}{ |c|c|}
\hline
             Bethe roots  &              holes            \\ \hline
$1.23901 + 1.09988 \i (x^c)$ &    $1.21892 - 1.13029 \i (x^h_1)$   \\
$0.585152 - 0.979686 \i$&     $-0.385543 - 0.942574 \i (x^h_2)$  \\
$0.0805622 - 0.956196 \i $&            \\
$-1.07883 - 0.940606 \i$&       \\    
 \hline
\end{tabular}
 \caption{The location of Bethe roots and holes in the complex plane
 for the 10th eigenvalue for $N=8$, $h=2$, $\gamma=2$ and $s=0$ at $T=0.46$.}
 \label{table:rootsandholes}
 \end{center}
\end{table}

To apply the higher-level Bethe Ansatz equation in finite $N$
approximation, we have to replace the dressed energy $\epsilon(x)$ by 
\begin{equation*}
\epsilon_N (x) = \frac{h}{2} + 
   \frac{NT}{2} \ln \Biggl( \frac{
       \cn \bigl( \frac{2K}{\pi} (x - \i \frac{\beta}{N}) \big| k \bigr)
       + \i \sn \bigl( \frac{2K}{\pi}(x - \i \frac{\beta}{N}) \big|k \bigr)}
      {\cn \bigl( \frac{2K}{\pi} (x + \i \frac{\beta}{N}) \big| k \bigr)
       + \i \sn \bigl( \frac{2K}{\pi}(x + \i \frac{\beta}{N}) \big|k \bigr)} \Biggr)
\end{equation*}
which is obtained by replacing $\e_0$ by $\e_0^{(N)}$  (see (\ref{defenulln}),
(\ref{defenull})) in (\ref{lineps}). With $\e_N$ instead of $\e$ in (\ref{a0})
(and by setting $n_+=n_-=0$), the result in the main body of the paper
claims that holes satisfy
$
 \fa^{(0)} (x^h_j) =-1\, (j=1,2) \epp
$
Note that, for the present case, we have independently checked that $k=0$.
The location of the close root is found to satisfy $\fa^{(0)} (x^c) +\fa^{(+)} (x^c) =-1$.
For $N$ being finite, the first term disappears and the subsidiary condition
simplifies to 
$
\fa^{(+)} (x^c) =-1 \epp
$
The numerical data in Table \ref{table:hBAE} indicate that they are satisfied
with reasonable accuracy despite the fact that $N$ is not really large.

\begin{table}[htb] 
\begin{center}
\begin{tabular}{ |c||c|}
\hline
$\fa^{(0)} (x^h_1) $ &    $-1.00126 - 0.00819979\i $  \\    
$\fa^{(0)} (x^h_2) $ &      $-0.99882 - 0.00656583\i$ \\
$ \fa^{(+)} (x^c)  $ &              $ -1.00025 - 0.0183492 \i$    \\
 \hline
\end{tabular}
 \caption{The values of the finite-$N$ auxiliary functions in low-$T$ approximation
 at the exact numerical holes' and particle's positions
given in Table \ref{table:rootsandholes}.
The parameters $N, h, \gamma, T$ are identical to those given there.
}
 \label{table:hBAE}
 \end{center}
\end{table}

}


\providecommand{\bysame}{\leavevmode\hbox to3em{\hrulefill}\thinspace}
\providecommand{\MR}{\relax\ifhmode\unskip\space\fi MR }
\providecommand{\MRhref}[2]{%
  \href{http://www.ams.org/mathscinet-getitem?mr=#1}{#2}
}
\providecommand{\href}[2]{#2}

\end{document}